\documentclass[traditabstract]{aa}
\usepackage{graphicx}
\usepackage{enumerate}
\usepackage{epsfig}
\usepackage{color}
\usepackage{txfonts}
\usepackage{natbib}
\usepackage{multirow}
\bibpunct{(}{)}{;}{a}{}{,} 

\begin{document}

\title{On the kinematic detection of accreted streams in the Gaia era:\\ a cautionary tale}
\titlerunning{On the kinematic detection of streams in the Gaia era}

\author{I.~Jean-Baptiste, \inst{1}       
 \and
P.~Di Matteo        \inst{1}
 \and
 M.~Haywood    \inst{1}
 \and
 A.~G{\'o}mez  \inst{1}
  \and 
  M.~Montuori  \inst{2}
  \and
   F.~Combes \inst{3, 4}
   \and 
   B.~Semelin \inst{3}
   }

\authorrunning{I.~Jean-Baptiste et al.}

\institute{GEPI, Observatoire de Paris, PSL Research University, CNRS, Univ Paris Diderot, Sorbonne Paris Cit\'e, Place Jules Janssen, 92195
Meudon, France\\
\email{paola.dimatteo@obspm.fr}
\and SMC-ISC-CNR and Dipartimento di Fisica,
Universit\`a ``La Sapienza'' Roma,
Ple. Aldo Moro 2, 
00185, Rome, Italy
\and Observatoire de Paris, LERMA, CNRS, PSL Univ., UPMC, Sorbonne Univ., 
F-75014, Paris, France
\and College de France, 11 Place Marcelin Berthelot, 75005, Paris, France
}

\date{Accepted, Received}

\abstract{The $\Lambda$CDM cosmological scenario predicts that our Galaxy should contain hundreds of stellar streams at the solar vicinity, fossil relics of the merging history of the Milky Way and more generally of the hierarchical growth of galaxies.  Because of the mixing time scales in the inner Galaxy,  it has been claimed that these streams should be difficult to detect in configuration space but can still be identifiable in kinematic-related  spaces like the energy/angular momenta spaces, $E-L_z$ and $L_\perp-L_z$, or spaces of orbital/velocity parameters.  
By means of  high-resolution, dissipationless N-body simulations, containing between 25$\times10^6$ and $35\times10^6$ particles, we model the accretion of a series of up to four 1:10 mass ratio satellites then up to eight 1:100 satellites and we search systematically for the signature of these accretions in these spaces. The novelty of this work with respect to the majority of those already available in the literature, is to analyze fully consistent models, where both the satellite(s) and the Milky Way galaxy are ``live" systems, which can react to the interaction, experience kinematical heating,  tidal effects and dynamical friction (the latter, a process often neglected in previous studies). 
We find that, in agreement with previous works, all spaces are rich in substructures, but that, contrary to previous works, the origin of these substructures -- accreted or in-situ -- cannot be determined, for the following reasons.
In all spaces considered (1) each satellite gives origin to several independent overdensities; (2) overdensities of multiple satellites overlap; (3) satellites of different masses can produce similar substructures; (4) the overlap between the in-situ and the accreted population is considerable everywhere; (5) in-situ stars also form substructures in response to the satellite(s) accretion. These points are valid even if the search is restricted to kinematically-selected halo stars only. 
As we are now entering the ``Gaia era", our results warn that an extreme caution must  be employed before interpreting overdensities in any of those spaces as evidence of relics of accreted satellites. Reconstructing the accretion history of our Galaxy will require a substantial amount of accurate spectroscopic data, that, complemented by the kinematic information, will possibly allow us to (chemically) identify accreted streams and measure their orbital properties.}


\maketitle

\section{Introduction}

Current spectroscopic surveys -- like APOGEE, APOGEE-2, GES, GALAH, RAVE \citep{allende08, majewski10, eisenstein11, sobeck14, gilmore12, anguiano14, desilva15, steinmetz06, zwitter08, siebert11, kordopatis13} -- are extending our knowledge of the chemical composition and radial velocities of several hundred  thousand stars in the Galaxy up to several kpc from the Sun. In the next years, starting from the first release in mid-September 2016, the ESA astrometric mission Gaia \citep{perryman01} will deliver parallaxes and proper motions for about 1 billion stars in the Galaxy and radial velocities for about one tenth of them. This unprecedented amount of data, coupled with Gaia follow-up spectroscopic surveys currently under definition --WEAVE, 4MOST, MOONS \citep{dalton12, dejong12, cirasuolo12} -- will potentially allow us to answer many simple, but still unraveled, questions of Galactic studies:  what are the characteristics of the different Milky Way stellar populations, how have they been shaped over time,  what are their evolutionary links and ultimately to what extent are they the result of in-situ star formation or rather the deposit of stars and structures accreted over time ? The Milky Way will constitute the natural testbed for any cosmological scenario, in particular for $\Lambda$CDM  cosmology, according to which galaxies grow hierarchically, from the formation of small subunits, that subsequently merge to evolve in the galaxies that we see nowadays \citep{white78}. \\
$\Lambda$CDM models predict indeed that a galaxy like the Milky Way should contain hundreds of stellar streams at the solar vicinity \citep{helmi99, helmi03, gomez13}, relics of the merging of other galactic systems, with masses comparable or significantly smaller than our own Galaxy at the time of their accretion \citep{bullock05, stewart08, delucia08, cooper10, font11, brook12, pillepich15, rodriguez16, deason16}. While we have evidence of recent and ongoing accretion events experienced by our Galaxy \citep{ibata94, ibata01, belokurov07} and some streams -- vestige of more ancient mergers -- have been possibly detected also at the solar vicinity \citep{helmi99, helmi06, klement09, smith09, nissen10},  we are still far from the numbers predicted by cosmological simulations. Is this the indication that we have  overestimated the importance of accretions in building  a Galaxy like our own over time, or is this rather a consequence of the difficulty in finding the remnants of accreted satellites, in particular of those related to the most ancient accretion events, once they are spatially mixed with in-situ stars in the Milky Way?\\

Numerous studies in the last fifteen years have suggested that the imprints of past merger events, even if dispersed in configuration space, can still be identified in kinematics-related spaces,  like the energy- angular momentum space (hereafter $E-L_z$, $L_z$ being $z$ component of the angular momentum, for an axisymmetric potential where $z$ is the symmetry axis), in the $L_\perp-L_z$ space, where $L_\perp$ is the angular momentum component in the $x-y$ plane,  or in spaces of orbital parameters, like the apocentre-pericentre-angular momentum space (hereafter APL) and its projections \citep{helmi99, helmi00, knebe05, brown05, helmi06, font06, choi07, morrison09, gomez10,  refiorentin15} -- see also \citet{klement10, smith16} for two recent and comprehensive reviews on the subject.\\ For example, distinct accreted satellites should appear as coherent structures in the $E-L_z$ space \citep{helmi00} and the shape of these structures should not significantly change during the accretion event, even in the case of a time-dependent potential \citep{knebe05, font06, gomez10}. From the number of clumps found in the $E-L_z$ space of several million Gaia stars, it should be potentially possible to set a lower limit to the number of past accretion events \citep{helmi00, gomez10}.  But is this search really feasible and meaningful ? What does an overdensity in the $E-L_z$ space really mean ? Can we realistically make use of the number of substructures found to set a lower limit to the number of merger events  that the Galaxy experienced in its past ? \\ Other spaces where the signature of accretion events may be found are the apocentre-pericentre space: here accreted stars should tend to cluster around a common value of eccentricity \citep{helmi06}, even if they can span different apocentre and pericentre distances. Is this coherence in the AP space really maintained ? And again, is it possible to separate accreted stars from the underlying, in-situ population, which -- as we will show -- can be also clustered in this space ? \\ And what about the $L_\perp-L_z$ space ? What are the regions in this space where the probability of finding accreted stars is the highest? \\

In this paper, we will address these questions, by means of high resolution N-body simulations of a Milky Way-type galaxy which accretes one or several  (up to four) satellites. The novelty of this work with respect to the majority of those already available in the literature, is to analyze fully consistent models, where both the satellite(s) and the Milky Way galaxy are ``live" systems, which can react to the interaction, experience kinematical heating, tidal effects and dynamical friction \citep[the latter, a process often neglected in previous studies, but see][]{knebe05, meza05, font06, refiorentin15}, exchange energy and angular momentum. We will show how, differently from previous findings, the energy-angular momentum space  -- and similarly, the $L_\perp-L_z$  and the apocentre-pericentre-angular momentum spaces --  become  hardly decipherable spaces when some of the most limiting assumptions which have affected previous works are removed (see Sect.~\ref{simubefore}).  We will show indeed that:
\begin{itemize}
\item because the energy and angular momentum of a satellite are not conserved quantities during an interaction,  each satellite gives origin to several independent overdensities;   
\item multiple satellites overlap in the $E-L_z$ space; 
\item in-situ stars affected by the interaction(s) tend to progressively populate a region of lower angular momentum and higher energy that the one initially (i.e. before the interaction) populated;
\item most of the accreted stars overlap with in-situ stars. 
\end{itemize}
We conclude from these findings that the search for the building blocks of our Galaxy in kinematic spaces is mostly inefficient. Similar conclusions are found for the APL and $L_\perp-L_z$ spaces. \\

In agreement with previous works \citep[see also][]{gould03}, we find that all these spaces are rich in substructures, but that the origin of these substructures cannot be determined with kinematics alone. As an example, we will show that  the in-situ stellar halo, formed as a result of the interaction, is neither smooth or non-rotating. This is different from what it  is usually assumed for in-situ halo stars in the Galaxy \citep{helmi99, kepley07, smith09}. Moreover, also in-situ halo stars can be clustered in kinematic spaces,  and  an extreme caution must be employed before interpreting overdensities in those spaces as evidence of relics of past accretion events \citep[see, for example,][]{morrison09, helmi06, klement09, refiorentin15}.  Detailed chemical abundances and/or ages will be definitely necessary to identify stellar streams, that should show up in those spaces as distinct patterns from those described by the in-situ stellar populations. \\

This paper is organized as follows: in Sect.~\ref{simu} we present the N-body simulations analyzed in this paper (Sect.~\ref{oursimu}), after discussing  the main numerical works done on the subject so far and the approach they have adopted (Sect.~\ref{simubefore}). In Sect.~\ref{results}, we will present the main results of our work, starting by analyzing the $E-L_z$ space (Sect.~\ref{ELzsect}) and then moving to show the results for the $L_\perp-L_z$ (Sect.~\ref{LzLperpsect_sun}) and the APL (Sect.~\ref{APLsect}) spaces. A discussion of our findings is given in Sect.~\ref{discussion}. Finally, in Sect.~\ref{conclusions}, the main conclusions of our work are presented.

\section{Numerical methods}\label{simu}
\subsection{Previous numerical modelling}\label{simubefore}
Before describing the N-body simulations analyzed in this paper, it is worth summarizing the main characteristics and the main dynamical processes modeled so far in the literature to study the kinematic signatures of accreted stars in the Galaxy. As we will see in the following of this paper, indeed, understanding these characteristics is essential to understand the reasons why the present work does not reach the main conclusions found in previous studies. \\ 

Schematically, the suggestion to detect stellar streams in the Milky Way by looking for (sub)-structures in kinematic spaces is mainly based on two kinds of models:
\begin{enumerate}
\item Test particle models
\item Self-consistent, N-body models
\end{enumerate}
\emph{Test particle models} have been developed since the pioneering works of \citet{helmi99, helmi00}. In this works \citep[see also][]{kepley07} the Milky Way is represented by a fixed, rigid potential and the satellites as a collection of  particles, usually gravitationally interacting. In most recent works, the Milky Way potential has been allowed to vary with time \citep[see, for example][]{gomez10a, gomez10}. All these models have confirmed Helmi's early suggestions: while the debris of early accreted satellites are very difficult to recover spatially, strong correlations and structures should be visible in velocity spaces ($E-L_z$, $L_\perp-L_z$, APL). Note that, because these models assume an analytic Galactic potential and no dynamical friction is taken into account, the energies and angular momenta of the accreted satellites will be overall conserved\footnote{For the angular momenta, $L$, this is true for a spherical potential, while in an axisymmetric one, it will be only the $z-$component to be strictly conserved.} by definition. Thus the finding that accreted satellites stand out as lumps in  integrals
of motion spaces and in particular that each lump corresponds to an accreted satellite \citep[see for example][for the $E-L_z$ space]{helmi00}, is, generally, a direct consequence of the assumptions made in these models, rather than an intrinsic feature of the accretion event. Unless satellites disperse before suffering substantial dynamical friction -- but note that in this case they would hardly attain the inner Galactic halo \citep[distances from the Galaxy center $\sim 15-20$~kpc, ][]{carollo07}, where kinematic detections of streams are currently mainly focused -- satellites lose their coherence in integrals of motion spaces during their accretion into the Galaxy (see Sect.~\ref{results}). \\ Caution should be also paid to the fact that these models generally do not include any in-situ (i.e. formed in the Galaxy) population, that act -- as we will see -- as a background where the signal of accretion events is mostly washed out. In all the above cited works, indeed, the stellar halo is exclusively made of accreted stars. This assumption has the natural consequence to maximize their signal, even when observational errors are taken into account, as in \citet{helmi00, gomez10}. An exception to this general approach is represented by the work by \citet{brown05}, who modeled also an in-situ population, 	concluding that a search for clumpy structures in the $E-L_z$ space is indeed very challenging for astrometric surveys like Gaia, one of the reasons being the presence of the background Galactic population. Note however that, in \citet{brown05} work, the in-situ halo population is modeled as a smooth distribution in phase-space, added \emph{a posteriori} to the model: as a consequence, this population cannot respond self-consistently to the interaction, experience neither kinematical heating nor any clustering  in kinematic spaces. The natural consequence is that the lumpiness of the accreted population is still overestimated, because it is modeled against an assumed smooth background.  We will show in the following pages that the situation is even more complicated than that discussed by \citet{brown05} and the search even more inefficient, when also the in-situ population is modeled self-consistently.\\
\emph{Self-consistent, "live" N-body models} remove the hypothesis of a rigid Galactic potential, since both the Galaxy and the satellite(s) are modeled as a collection of particles (dark matter and/or stars) which respond self-consistently to the merger. Dynamical friction is thus taken naturally into account and so should be also the response of the in-situ Galactic populations to the accretion events. To our knowledge,  the  importance of using this kind of approach for detecting streams kinematically has been firstly pointed out by \citet{knebe05}. They used cosmological, dark matter-only simulations and showed that the lumpy appearance of accreted satellites is significantly smeared out when a ``live" model is adopted. Unfortunately,  \citet{knebe05} do not discuss nor show  the  response of the  in-situ component to the interaction and, as a consequence, the efficiency of the ``integrals-of-motion" approach is still overestimated in their work. Many other simulations have investigated the $E-L_z$ space, or equivalent, by making use of self-consistent, N-body simulations \citep[see, for example,][]{meza05, helmi06, font06, gomez13, refiorentin15}. However, either they have not discussed the response of the in-situ population to the accretion event(s) \citep{meza05, helmi06, font06, gomez13}, or they have not taken into account the in-situ halo population that naturally forms in merger events \citep[as shown, for example, by][]{zolotov09, purcell10, font11, qu11, mccarthy12, cooper15}, replacing it with a smooth halo component added \emph{a posteriori} \citep{refiorentin15}. To our knowledge, the only works that have started to investigate the question of the overlap between in-situ and accreted stars are those by  \citet{ruchti14, ruchti15} -- however, they investigated mostly low mass mergers (mass ratio $\sim 1:100$). \\

The limitations that, in our opinion, nearly all previous works have suffered constitute the main motivation of our work. We consider it to be a first attempt to model in a more realistic way both the accreted and in-situ stellar populations during one or multiple accretion events. This is a necessary -- and not procrastinable -- step  to make predictions about the redistribution of in-situ and accreted stars in the Galaxy in view of Gaia, but also to caution the interpretation of current kinematic data where substructures are currently detected and an extra-galactic origin is often favored.

\subsection{Our simulations}\label{oursimu}

\begin{table}
\begin{center}
\begin{tabular}{lcccr}
\hline
\hline
	       & $M$ & $a$ & $h$ & $N$ \\ 
\hline	       
\\
MW galaxy: Thin disc & 11.11 & 4.7 & 0.3 & 10M\\
MW galaxy: Intermediate disc & 6.66 & 2.3 & 0.6 & 6M\\
MW galaxy: Thick disc & 4.44 & 2.3 & 0.9 & 4M\\
MW galaxy: GC system & 0.04 & 2.3 & 0.9 & 100\\
MW galaxy: Dark halo & 70.00 & 10 & -- & 5M\\
\\

Satellite: Thin disc & 1.11 & 1.48 & 0.09 & 1M\\
Satellite: Intermediate disc & 0.67 & 0.73 & 0.18 & 0.6M\\
Satellite: Thick disc & 0.44 & 0.73 & 0.27 & 0.4M\\
Satellite: GC system & 0.004 & 0.73 & 0.27 & 10\\
Satellite: Dark halo & 7.00 & 3.16 & -- & 0.5M\\
\hline
\end{tabular}
\caption{Masses, characteristic scale lengths and heights and number of particles, for the different components of the Milky Way-type galaxy and the satellite(s). All masses are in units of 2.3$\times10^9M_\odot$, distances in kpc.}\label{galparamtable}
\end{center}
\end{table}

\begin{figure}[h!]
\begin{center}
\includegraphics[clip=true, trim = 7mm 0mm 20mm 5mm, width=0.75\columnwidth]{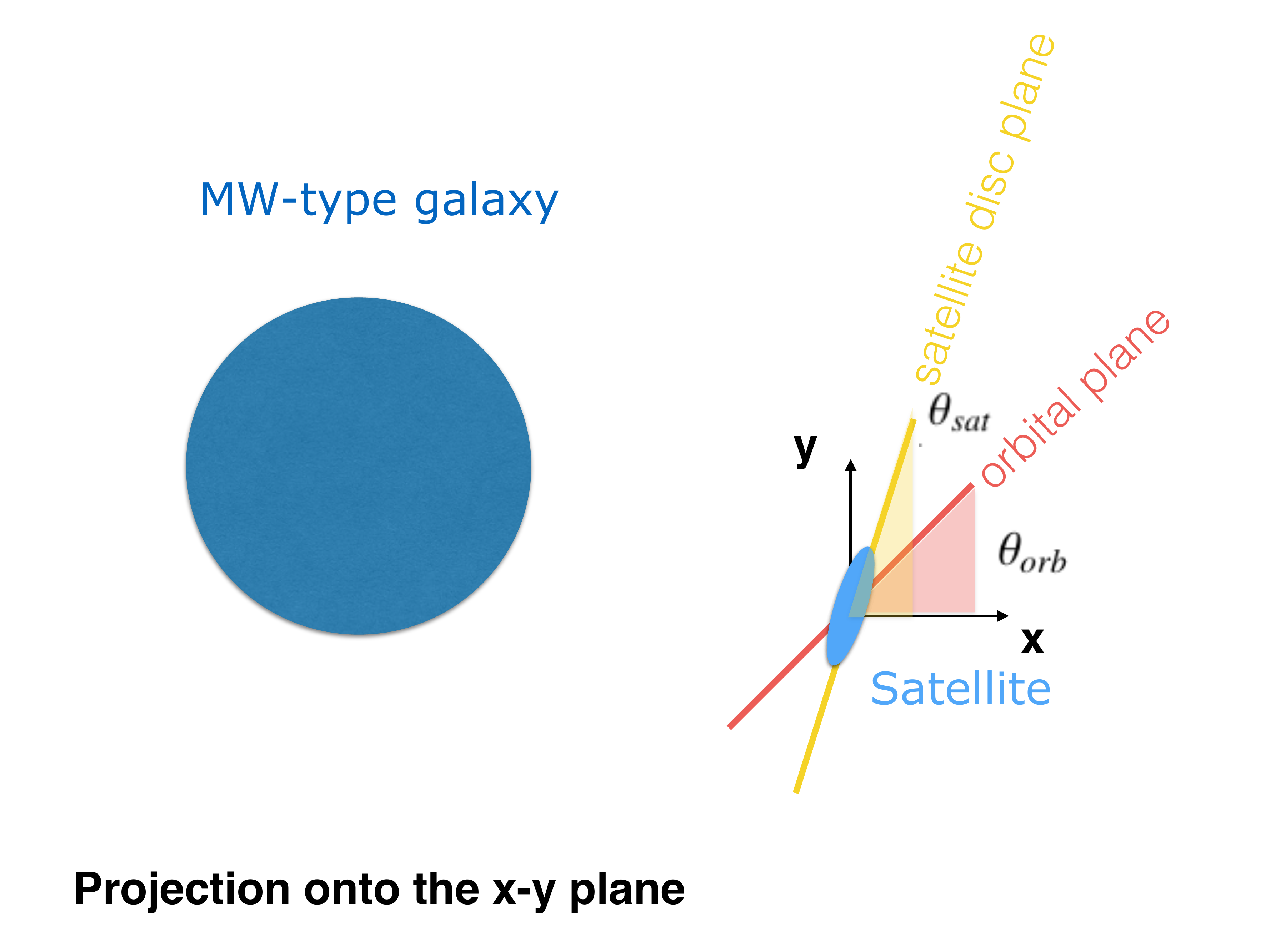}
\includegraphics[clip=true, trim = 7mm 0mm 20mm 5mm, width=0.75\columnwidth]{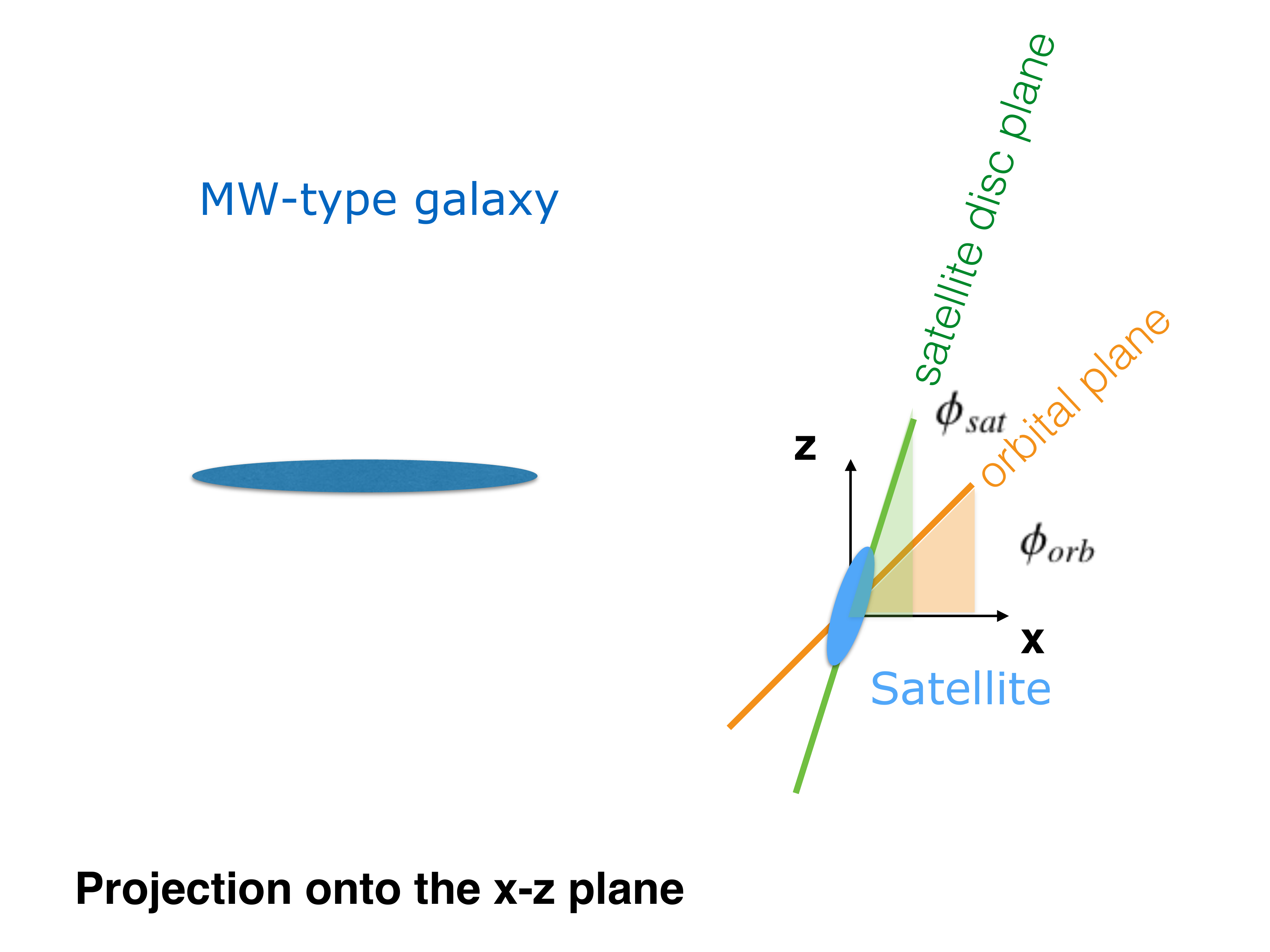}
\caption{Set-up of the satellite orbital plane and internal spin. In a reference frame with $xy-$plane coincident with the MW-disc plane and $z-$axis oriented as the spin of the MW-type galaxy, the spatial orientation of the satellite orbital plane is defined by two angles, $\theta_{orb}$ and $\phi_{orb}$. The former defines the angle between the intersection of the orbital plane with the $xy-$plane and the $x-axis$ (red line and area in the top panel), the latter defines the angle between the intersection of the orbital plane with the $xz-$plane and the $x-axis$ (orange line and area in the bottom panel). Similarly, the angles  $\theta_{sat}$ and $\phi_{sat}$, indicating the spatial orientation of the satellite spin, can be defined (yellow and green lines and areas, respectively, in the top and bottom panels.) }\label{orbitfig}
\end{center}
\end{figure}

\begin{table*}
\begin{center}
  \begin{tabular}{lccccccc}
      \hline
      \hline\\[0.01cm]
   \multicolumn{1}{c}{ } &
      \multicolumn{1}{c}{1$\times$(1:10)} &
      \multicolumn{2}{c}{2$\times$(1:10)} &
      \multicolumn{4}{c}{4$\times$(1:10)} \\
    \multicolumn{1}{c}{ } &
   \multicolumn{1}{c}{sat 1} &
      \multicolumn{1}{c}{sat 1} &
         \multicolumn{1}{c}{sat 2} &
   \multicolumn{1}{c}{sat 1} &
   \multicolumn{1}{c}{sat 2} &
      \multicolumn{1}{c}{sat 3} &
         \multicolumn{1}{c}{sat 4} \\[0.1cm]
         \hline\\[0.02cm]
           $x_{sat}$ & 83.86 & 83.86 & 92.38 & 83.86 & 92.38 & 42.48 & -9.08\\
       $y_{sat}$ & 0.00 & 0.00 & -21.98 & 0.00 & -21.98 & 11.16 & 75.00 \\
          $z_{sat}$ & 54.46 & 54.46 & -31.34 & 54.46 & -31.34 & -89.84 & -65.52 \\
	    $D_{sat}$ & 100.00 &100.00 & 100.00 & 100.00 & 100.00 & 100.00 & 100.00 \\[0.3cm]
 $v_{x,sat}$ & 1.22 & 1.22 & 1.46 & 1.22 & 1.46  & 0.81  & 0.07\\
       $v_{y,sat}$ & 0.30 & 0.30 & -0.05 & 0.30 & -0.05 & 0.40 & 1.38 \\
          $v_{z,sat}$ & 0.79 & 0.79 & -0.41 & 0.79 & -0.41  & -1.26 & -0.94 \\
	    $V_{sat}$ & 1.48 & 1.48 & 1.52 & 1.48 & 1.52 & 1.55 & 1.67 \\[0.3cm]
	    $L_{x, sat}$ & -16.16 & -16.16 & 7.53 & -16.16 & 7.53 &  21.63 & 20.34  \\
$L_{y, sat}$ & 0.00 & 0.00 & -7.79 &  0.00 & -7.79 & -18.83 & -12.93\\
$L_{z, sat}$ & 24.89& 24.89 & 27.66 & 24.89 & 27.66 & 7.89 & -17.61 \\[0.3cm]	        
	    $\theta_{orb}$ & 90.00 & 90.00 & 45.00 & 90.00 & 45.00  & 49.00 & 67.6 \\
 	    $\phi_{orb}$ & 33.00 & 33.00& -16.00 & 33.00 & -16.00 & -70.00 & 49.1 \\
	    $\theta_{sat}$ & 90.00 & 90.00 & 90.00 & 90.00 & 90.00 & 90.00 & 90.00 \\
 	    $\phi_{sat}$ & 83.00 & 83.00 & 83.00 & 83.00 & 83.00 & 83.00 & 83.00\\
    \hline
  \end{tabular}
  \caption{Initial positions, velocities and angular momenta for the different satellite galaxies. Alll quanttities are given in a reference frame whose origin coincides with the centre of the MW-type galaxy, whose $xy-$plane coincides with the MW disc and whose $z-$axis is oriented as the internal angular momentum of the MW-type galaxy. The angles $\theta_{orb}$ and $\phi_{orb}$, which indicate the orientation in this reference frame of the satellite orbital plane are given and expressed in degrees. The  angles $\theta_{sat}$ and $\phi_{sat}$ give the orientation of the spin of the satellite galaxies. See Fig.~\ref{orbitfig} for a schematic representation of all these angles. Distances are in kpc, velocities in units of 100km/s, angular momenta in units of 100kpc~km/s.  }\label{orbittable}
  \end{center}
\end{table*}

For this aim, we analyze in this paper three dissipationless, high-resolution, simulations of a Milky Way-type galaxy accreting one, two or four satellites\footnote{By including also a 4x(1:10) merger in the present paper we do not aim at suggesting that the Milky Way should contain such a high fraction of ex-situ stellar mass. But we consider that the analysis of this simulation brings an important element to the discussion and to the comprehension of the overlap of in-situ/ accreted stars. One may naively think that accreting more mass can increase the signal of accreted over in-situ stars in kinematic spaces. By adding an extreme case, such that of the 4x(1:10) merger, we show in the following sections that this is not the case : it is true that the remnant galaxy ends up with a larger fraction of accreted over in- situ material, but at the same time (1) this material redistributes over a larger portion of the phase-space; (2) the amount of in-situ stars heated by the interaction and found in the halo region also increases. The final result is that even a very extreme situation like that of a 4x(1:10) merger would not be more favorable to the kinematic detection of streams than a simple 1:10 or 2x(1:10). We consider this an important element of the following discussion.}, respectively. Each satellite has a mass which is one tenth of the mass of the Milky Way-like galaxy. The total number of particles used in these simulations varies between $N_{tot}=27\thinspace500\thinspace110$, for the case of a single accretion, to $N_{tot}=35\thinspace000\thinspace140$, when four accretions are modeled. 
The massive galaxy contains a thin, an intermediate and a thick stellar disc -- mimicking the Galactic thin disc, the young thick disc and the old thick disc, respectively \citep[see][]{haywood13, dimatteo16} --  a population of a hundred disc globular clusters, represented as point masses and is embedded in a dark matter halo. The total number of stellar (thin and thick disc) particles in the main galaxy is $20\thinspace000\thinspace000$, the number of globular cluster particles is 100 and the number of dark matter particles is $5\thinspace000\thinspace000$. The discs are modeled with Miyamoto-Nagai density distributions, of the form 
\begin{eqnarray*}
\rho_*(R,z)=\Big(\frac{{h_*}^2M_*}{4\pi} \Big)\frac{a_*R^2+(a_*+3\sqrt{z^2+{h_*}^2}) (a_*+\sqrt{z^2+{h_*}^2})^2}{\Big[{a_*}^2+\Big(a_*+\sqrt{z^2+{h_*}^2}\Big)^2\Big]^{5/2}(z^2+{h_*}^2)^{3/2}},
\end{eqnarray*}
with $M_*, a_*$ and $h_*$ masses, characteristic lengths and heights which vary for the thin, intermediate and thick disc populations (see Table~\ref{galparamtable} for all the values); 
the system of disc globular clusters has scale length and scale height equal to those of the thick disc and the dark matter halo is modeled as a Plummer sphere, whose density is
\begin{eqnarray*}
\rho_{halo}(r)=\Big( \frac{3M_{halo}}{4\pi {a_{halo}}^3}\Big)\Big(1+\frac{r^2}{{a_{halo}}^2}\Big)^{-5/2}
\end{eqnarray*}  
with $M_{halo}$ and $a_{halo}$ characteristic mass and radius, respectively (see Table~\ref{galparamtable} for all the values). The satellite galaxies are rescaled versions of the main galaxy, with masses and total number of particles 10 times smaller and sizes reduced by a factor $\sqrt{10}$ (again, see Table~\ref{galparamtable}).  
It has been shown that most of the initial dark halo mass is rapidly  lost by the satellite galaxy as it is placed in the host potential -- \citet{villalobos08}, for example, estimate that
after few $10^8$ Myrs, that is before the first pericentre passage, only $\sim30\%$ of the initial dark halo mass is still bound to the system.  We have assumed an initial halo mass for the satellite of $M_{halo}=1.6\times10^{10}M_\odot$, a visible mass of $M_\star=5\times10^9M_\odot$, which implies a $M_\star/M_{halo}$ ratio = 0.3, compatible with the $M_\star/M_{halo}$ ratio $\sim0.2$ found in the simulations by \citet{villalobos08}, before the first close encounter between the satellite and their modeled Milky Way-type galaxy.
Note also that our choice to use a core dark matter halo for both the Milky Way-type galaxy and the satellite(s) comes from a number of observational evidence that seem to be more
consistent with a dark halo profile with a nearly flat density core \citep{flores94, deblok01a, deblok01b, marchesini02, gentile05, kuzio06, kuzio08}.\\
The Milky Way disc initially coincides with the $x-y$ plane of the reference frame, the spin of the Milky Way-type galaxy being oriented as the $z-$axis and positive.  In a reference frame with the origin at the centre of the Milky Way-type galaxy, the $x-y$ plane coinciding with its disc and the $z-$axis being oriented as the spin of the Milky Way-type galaxy, satellites positions are  ($x_{sat}, y_{sat}, z_{sat}$) and their distances from the centre of the Milky-Way type galaxy are equal to $D_{sat}$. Their initial 3D-velocities, relative to the Milky Way-like galaxy  are  ($v_{x,sat}, v_{y,sat}, v_{z,sat}$), $V_{sat}$ being their absolute value.  Their orbital planes are inclined of ($\theta_{orb}, \phi_{orb}$) with respect to the Milky Way-type galaxy, $\theta_{orb}$ being the angle between the intersection of the satellite orbital plane with the $xy-$plane and the $x-$axis,  $\phi_{orb}$ being the angle between the projection of the satellite orbital plane onto the $xz-$plane and the $x-$axis. Each satellite has an internal angular momentum whose orientation in the space is described by the angles ($\theta_{sat}, \phi_{sat}$), $\theta_{sat}$ being the angle between the projection of the satellite spin onto the $xy-$plane and the $x-$axis, $\phi_{sat}$ being the angle between the projection of the satellite spin onto the $xz-$plane and the $x-$axis. A schematic representation of all these angles is given in Fig.~\ref{orbitfig}. Their values are reported in Table~\ref{orbittable}, together with positions, velocities and initial orbital angular momenta. Note, in particular,  that:
\begin{itemize} 
\item all satellites are initially on direct orbits ($L_{z,sat}$, the $z-$component of their orbital angular momentum, being positive), except for satellite 4 of the 4$\times$(1:10) merger, which is initially placed on a retrograde orbit, $L_{z,sat}$ being negative, see Table~\ref{orbittable}. 
\item Initial orbital velocities of the satellite galaxy correspond to that of a parabolic orbit
for a 1:10 mass ratio with a MW-type galaxy mass of 100 (in our
mass units). This value for the total mass is
slightly higher ($\sim10\%$) than the total mass initially in the Milky Way-type
galaxy. But since the Milky Way-type system accretes in our models not only one,
but also 2 or 4 satellites, we have taken an initially slightly higher
mass to take into account - at least partially - also this mass growth.
The choice of exploring parabolic orbits, in particular, is in agreement
with cosmological predictions \citep{khochfar06}.
\end{itemize}
We will comment further on the choice of our initial conditions in Sect.~\ref{our limits}.\\
Initial conditions have been generated adopting the iterative method described in \citet{rodionov09}. All simulations have been run by making use of the TreeSPH code described in \citet{semelin02}, which has been already extensively used by our group over the last ten years to simulate the secular evolution of galaxies and accretions and mergers as well. Gravitational forces are calculated using a tolerance parameter  $\theta=0.7$ and include terms up to the quadrupole order in the multiple expansion. A Plummer potential is used to soften gravitational forces, with constant softening lengths for different species of particles. In all the simulations described here, we adopt $\epsilon=50~{\rm pc}$. The equations of motion are integrated using a leapfrog algorithm with a fixed time step of  $\Delta t=2.5$ $\times $   $10^5~{\rm yr}$.\\

In this work, we make use of the following set of units: distances are given in kpc, masses in units of $2.3\times10^9M_{\odot}$, velocities in units of $100$~km/s and $G=1$. Energies are thus given in units of $10^4 \rm{km^2/s^2}$ and time is in unit of $10^7$~years. With this choice of units, the stellar mass of the Milky Way-type galaxy, at the beginning of the simulation, is 5.1$\times10^{10}M_{\odot}$. Since all the  N-body models presented in this work are dissipationless, units can be easily rescaled and - for example - masses reduced to mimic a Milky Way-type galaxy at higher redshifts\footnote{Such a rescaling would still be within estimated uncertainties, even for dark matter halos with masses below $10^{12} M_\odot$, that is in the steep part of the $M_\star/M_{halo}$ relation, at least between $z=0$ and $z=1$. Indeed, for a galaxy with a halo mass $M_{ halo}\sim~2\times10^{11}$ at z=0, and a corresponding $M_\star/M_{halo}\sim0.01$ at that redshift, if one decreases its dark matter mass by a factor of 2 between z=0 and z=1 and at the same time maintains the $M_\star/M_{halo}$ ratio constant and equal to 0.01, this ratio would be between 2 and 3 sigma from the mean relation at z=1  \citep[see Figs. 7 and 8 from][]{behroozi13}.}.

\begin{figure*}
\begin{flushleft}
\includegraphics[clip=true, trim = 15mm 0mm 45mm 5mm, angle=270,width=2.\columnwidth]{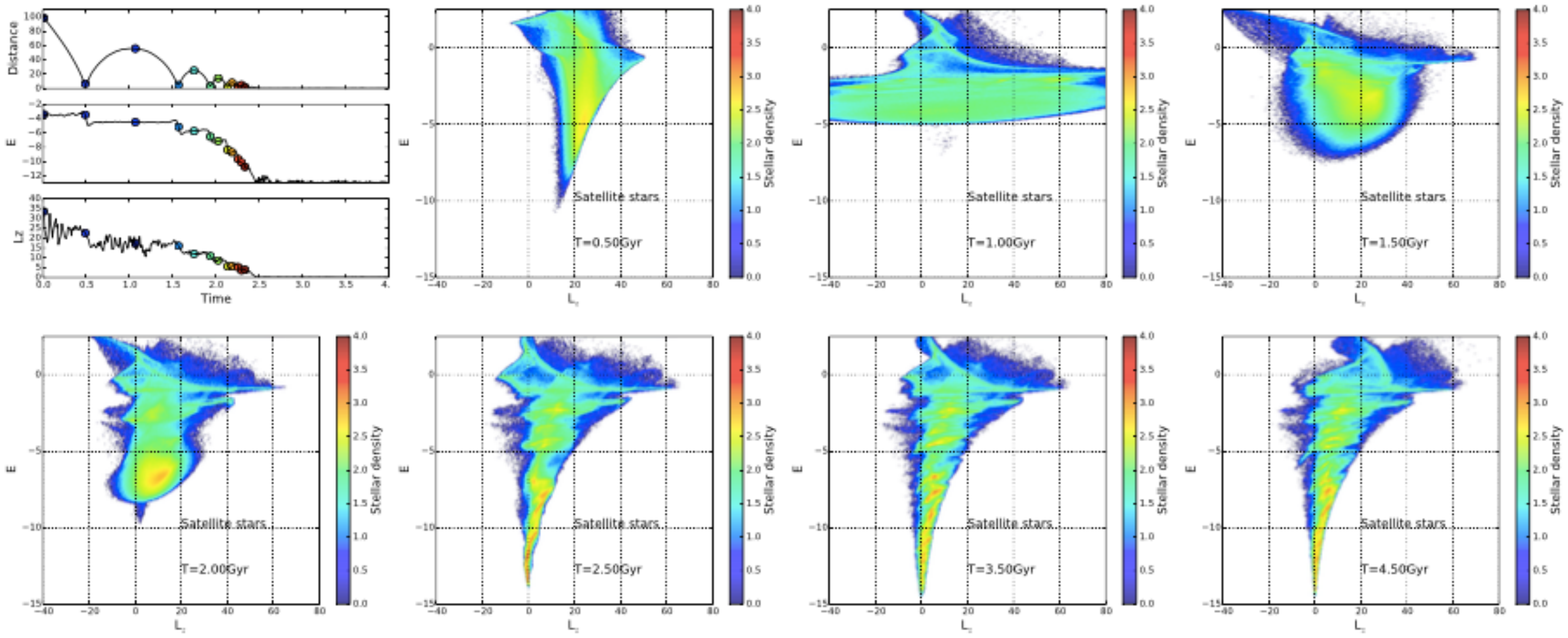}
\end{flushleft}
\begin{center}
\caption{Simulation of a 1:10 merger on a Milky Way-type galaxy. \emph{Leftmost panels in the top row: }Time evolution on the satellite distance from the centre of the Milky Way-type galaxy, evolution of its total energy $E$,  and of the $z-$component of the angular momentum $L_z$. The colored dots indicate different epochs of pericentre and apocentre passage of the satellite during its orbit around the Milky Way-type galaxy. \emph{From left to right and from top to bottom: } Distribution in the $E-L_z$ plane of stars belonging to the satellite, at different epochs during the accretion event. All stars belonging to the satellite, both the unbound and the
     bound population, are shown. The satellite merges at about 2.5~Gyr from the beginning of the simulation (t=0). In all the $E-L_z$ plots, colors code the stellar density, in logarithmic scale,  as indicated by the color bar.}\label{ElzsatGLOBtime}
\end{center}
\end{figure*}

\begin{figure*}
\begin{flushleft}
\includegraphics[clip=true, trim = 7mm 0mm 10mm 25mm, angle=270,width=2.\columnwidth]{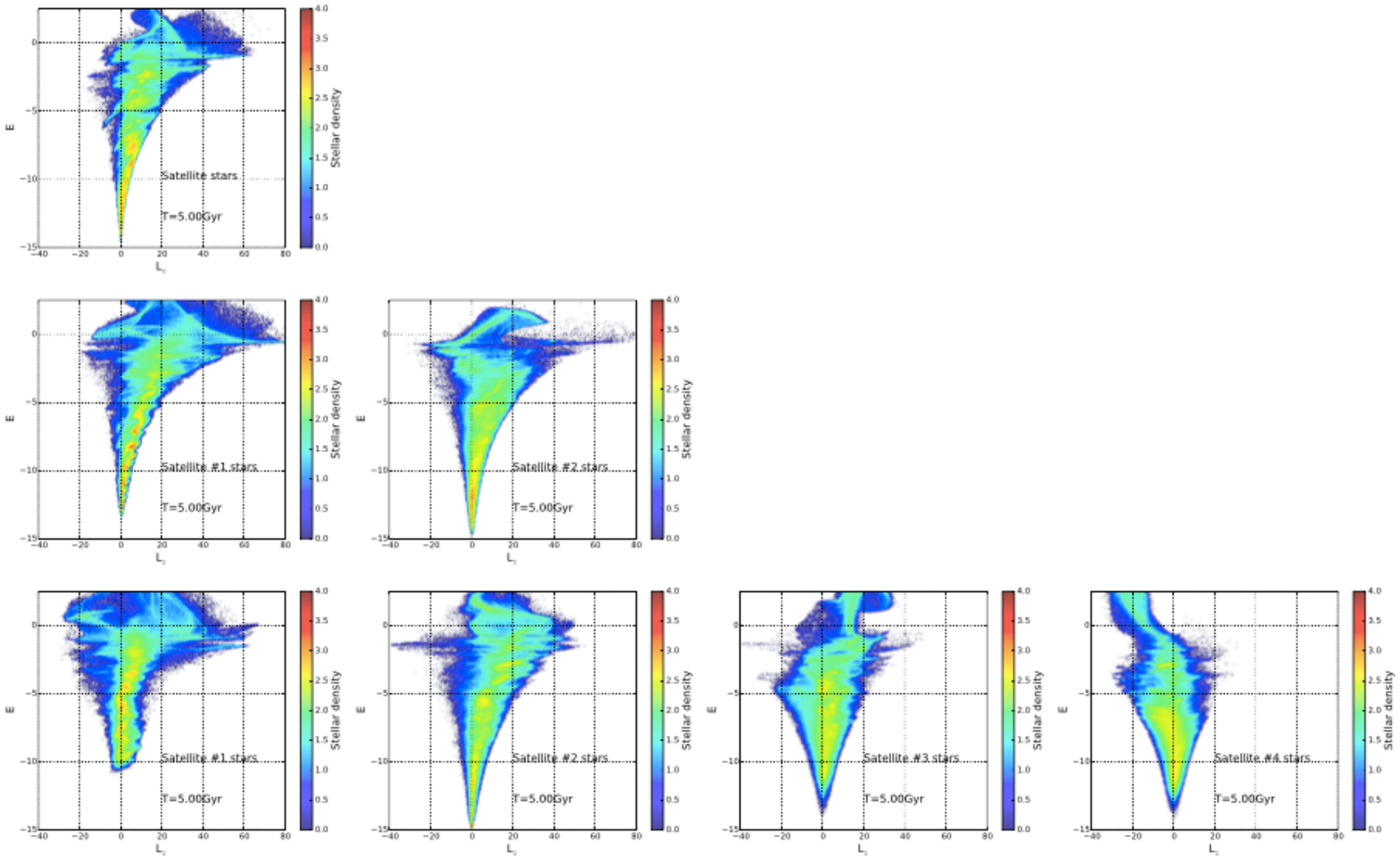}
\end{flushleft}
\begin{center}
\caption{\emph{From top to bottom: } Distribution in the $E-L_z$ space of satellite stars for the 1$\times$(1:10) merger (\emph{top row}), for the 2$\times$(1:10) merger (\emph{middle row}) and for the 4$\times$(1:10) merger (\emph{bottom row}). For the 2$\times$(1:10) and 4$\times$(1:10) mergers, the contribution of each satellite is shown from left to right. All the distributions are given at the final time of the simulation, corresponding to $t=5$~Gyr. In all the plots, colors code the stellar density, in logarithmic scale, as indicated by the color bar. }\label{ElzsatGLOB}
\end{center}
\end{figure*}

\begin{figure*}
\begin{flushleft}
\includegraphics[clip=true, trim = 0mm 0mm 0mm 5mm, angle=270,width=2.\columnwidth]{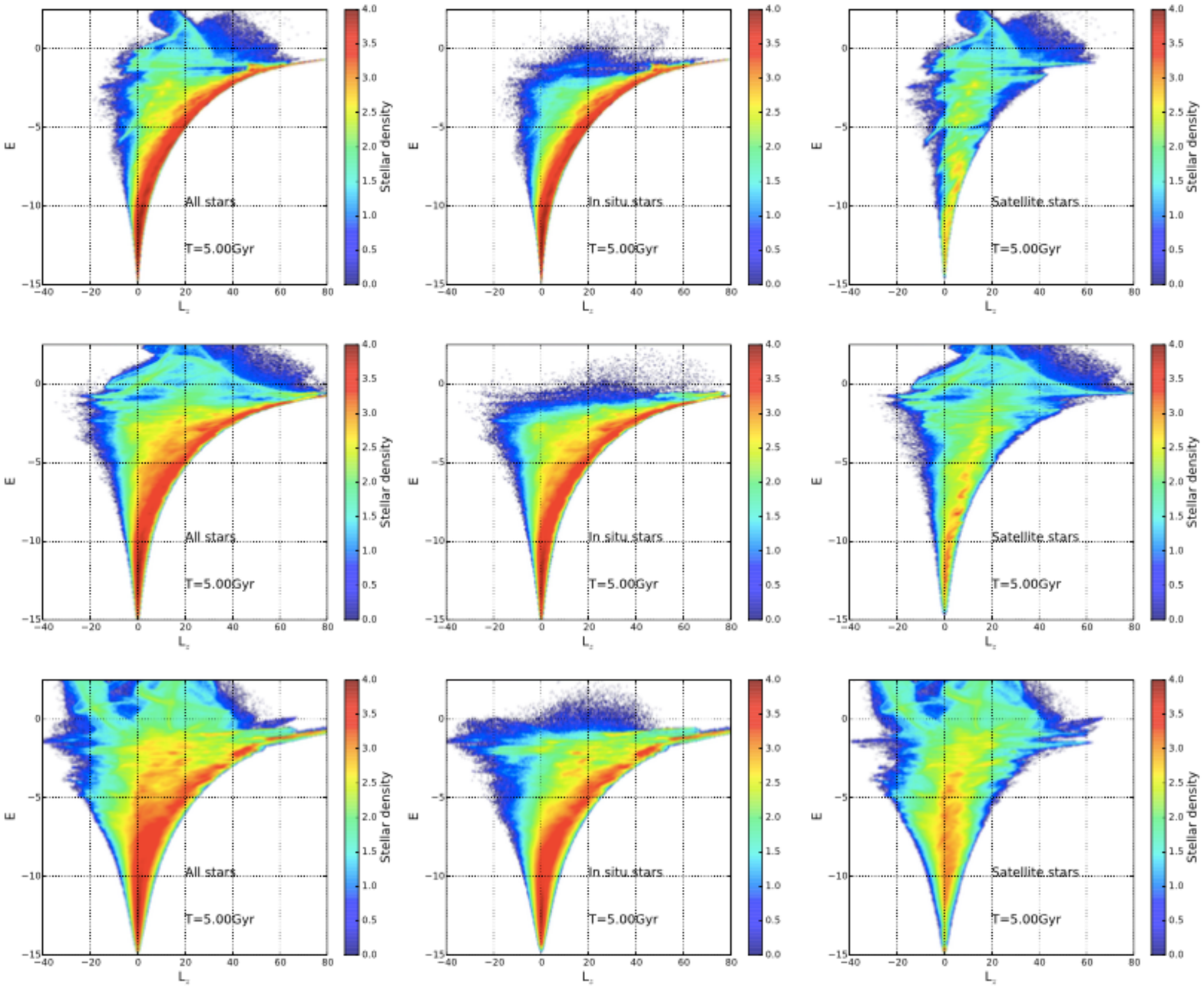}
\end{flushleft}
\begin{center}
\caption{\emph{From left to right: } Distribution in the $E-L_z$ space of all stars (\emph{leftmost panel}), in-situ stars (\emph{middle panel}) and accreted stars (\emph{rightmost panel}). \emph{From top to bottom: }Different rows show, respectively, the case of a 1$\times$(1:10),  2$\times$(1:10) and 4$\times$(1:10) merger, respectively.  All the distributions are shown at the final time of the simulation, $t=5$~Gyr.  In all the plots, colors code the stellar density, in logarithmic scale, as indicated by the color bar.}\label{ElzGLOB}
\end{center}
\end{figure*}

\section{Results}\label{results}

In the following of this analysis, unless explicitly  stated, all quantities are evaluated in a reference frame whose origin is at the centre of the Milky Way-type galaxy.  This centre is evaluated, at each snapshot of the simulations, as the density centre, following the method described in \citet{casertano05}. The $x-y$ plane coincides with its disc and the $z-$axis is perpendicular to it. All the results presented in this Section concern the stellar distribution in integrals-of-motion and kinematic spaces, in the case of a Milky Way-type galaxy accreting one or several satellites. We refer the reader to Appendix~\ref{isoMW} for the distributions obtained when the Milky Way is evolved in isolation, without experiencing any accretion. 

\subsection{On the $E-L_z$ space}\label{ELzsect}

We start our investigation by looking at the $E-L_z$ space, which has been proposed as a natural space where to look for the signatures of past accretion events \citep{helmi00}.
The results are structured in the following way: first we discuss how one or several  satellites redistribute their stars in this space, during their accretion into the galactic potential; then we show the predictions of our models about the overlap between accreted and in-situ stars;  finally we restrict our analysis at a ``solar vicinity" and discuss the difficulty and overall efficiency of searching stellar streams in the $E-L_z$ space.
 
\subsubsection{Coherence of accreted structures}

The redistribution in the $E-L_z$ space of satellite stars during their accretion into our Milky Way-type galaxy is shown in Fig.~\ref{ElzsatGLOBtime}, at different times during a 1:10 merger. The satellite represents, at the beginning of the interaction, a clump in the $E-L_z$ space, characterized by high energy and a large spread in the $z-$component of the angular momentum. As it approaches the first pericentre passage, at $t=0.5$~Gyr, the distribution of satellite stars in the $E-L_z$ space changes: stars are redistributed over a much larger interval of energies, from slightly positive (for stars which become unbound to the system), to significantly negative (for those stars which approach the innermost galactic regions). At the same time, the satellite becomes overall closer to the galactic centre and dynamical friction becomes stronger: as a consequence  the angular momentum tends to diminish -- hence the more compact distribution in $L_z$ found at this epoch. This strong elongation in the redistribution of stars in the $E-L_z$ space  is reminiscent of the strong spatial elongation of satellite galaxies (and even less massive systems, like globular clusters) when they approach the pericentre of their orbit:  tidal tails are particularly elongated and relatively thin at this epoch and can span up to tens of kpc around the satellite bound mass. As the satellite reaches its next apocentre passage, its spatial distribution becomes more compact, so does its energy distribution (see Fig.~\ref{ElzsatGLOBtime} a time $t=1$~Gyr), while the $z-$component of its angular momentum becomes broader again\footnote{At t=1~Gyr one can note that part of the satellite stars have $L_z >
     L_{z, irc}(E)$, with $L_{z,circ}(E)$ being the $z$-component of the angular momentum of a circular orbit of energy $E$. This comes from the fact that in Fig.~\ref{ElzsatGLOBtime} we are plotting
     \emph{all} stars belonging to the satellite, both the unbound and the
     bound population. By t=2.5~Gyr all satellite particles are unbound
     since this is the merger epoch for the 1x(1:10) interaction. But before
     that time, and in particular at t=1~Gyr, a large fraction of the
     satellite stars is still gravitationally bound to the system. This
     means that, together with the motion of the centre of mass of the
     satellite in the Milky Way potential, one needs to take into account also
     the motion of satellite stars in the satellite potential.  The
     angular momentum $L$ is thus $L= R\times(v_{sys}+v_{pec})=L_{sys}+L_{pec}$, with
     $v_{sys}$ the velocity of the satellite barycenter in the Milky Way reference
     frame, and $v_{pec}$ the peculiar velocities of satellite stars with
     respect to the satellite centre. Satellite stars rotate initially
     as fast as $\sim100~$km/s around its centre. When the satellite is
     located at $R=40~kpc$ from the Milky Way centre (as it is the case at t=1~
     Gyr) this implies a $L_{pec}$ that can be as high as $\sim40$ (in units of
     100km/s.kpc). At t=1~Gyr, the energy of the centre of mass of the
     satellite is E= -5 and the z-component of its angular momentum is
     Lz=20. By adding the contribution of $L_{z,pec}$ to $L_{z,sys}$, one can
     explain the range of $L_z$ found at this time. Note that since $L$
     depends on $R$, the effect of the peculiar velocities is particularly
     evident at large distances from the Milky Way centre, and it is reduced or
     it vanishes when the satellite is at its pericentre (see for ex the
     time t=2~Gyr). }. At each orbit, satellite stars go through this phase of compression and expansion in the $E-L_z$ space. Globally, as an effect of dynamical friction, the energy and the angular momentum decrease and the satellite penetrates deeper and deeper in the potential well of the main galaxy. During the interaction, part of the stars  become unbound, leaves the satellite and goes to populate the tidal tails which develop around it. Because stars lost at different passages are characterized by different values of energy and angular momenta (at first approximation, their energy and angular momenta at the moment they escape the satellite are those of the satellite centre), escaped stars pack into different substructures, depending on the time they escaped the satellite. Stars lost at early epochs populate the upper part of the $E-L_z$ diagram, while stars lost at more advanced epochs of the interaction are preferentially grouped into substructures of lower (i.e. more bound) energies. Thus, in general, if dynamical friction has time to act on the satellite before it becomes a gravitational unbound set of stars, \emph{satellite stars loose their coherence in the $E-L_z$ space} \citep[as also found by][]{meza05}: a satellite gives rise to several streams or horizontal ripples, which are not homogenous all along their length and  can become clumpy in the middle. In the following, we will use the terms ``ripples" or ``streams" to identify the large scale overdensities in the $E-L_z$ plane and we will refer to ``clumps" to identify either the high density regions of those streams, or small-scale overdensities. 
The number and density of these ripples depend on the number of passages the satellite experienced around the main galaxy and on the mass loss it experienced at each passage. Note that the loss of coherence of satellite stars in $E-L_z$ space is not dependent on the particular choice of orbital parameters: in all simulated cases, from the case of a single 1:10 accretion, to the case of  2x(1:10) and 4x(1:10) mergers, each satellite contributes to several ripples and clumps in the $E-L_z$ space, redistributing its stars over a large extent, both in energy and angular momentum (see Fig.~\ref{ElzsatGLOB}). Note also that when multiple satellites are accreted, even if their initial energies and/or angular momenta are different, once the merger is completed their stars tend to redistribute over a similar portion of the $E-L_z$ space, that is the overlap between stars initially belonging to different satellites is not negligible. This indicates that, even in the ideal case of the absence of an in-situ population, a given region of the $E-L_z$ space and in particular any given stream in this space, can be the result of the overlap of different accreted structures. Finally, it is worth emphasizing that accreted stars do not redistribute only in clumps: part of them is more diffusely distributed and cannot be associated with any clear overdensity  (see Fig.~\ref{ElzsatGLOB}).

\subsubsection{Overlap with in-situ stars}\label{overlap}

In the previous part of this section, we have seen how one or several satellites accreted onto a Milky Way-type galaxy redistribute in the $E-L_z$ space. We have seen that ripples and clumps in this space cannot be associated to a single accretion event: a satellite, during its accretion into the Galaxy, redistributes in several streams and different satellites -- with initial different orbital parameters -- show a not negligible overlap in this space.  However, there is another element that complicates the research of substructures in the $E-L_z$ space: the presence of in-situ stars which respond to the interaction and redistribute in the energy-angular momentum plane, occupying a region similar to those of accreted stars, as we detail in the following.\\

That in-situ disc stars - kinematically heated during accretion events -- redistribute in a thicker disc and inner halo  has been known since two decades \citep{quinn93, walker96, villalobos08, villalobos09, zolotov09, purcell10, font11, dimatteo11, qu11, mccarthy12, cooper15}. As a result of this kinematic heating, in-situ stars tend to gain some energy and lose part of their initial angular momentum, since their orbits tend to  become more radially and vertically elongated. This change in the spatial and kinematic properties of in-situ disc stars during an interaction has naturally also some consequences on their redistribution in the $E-L_z$ space, as shown in the central column of Fig.~\ref{ElzGLOB}. Stars that -- in the absence of an interaction -- would remain confined in a relatively thin and elongated region of the $E-L_z$ space, corresponding to that occupied by our initial thin/thick disc (see Fig.~\ref{ELziso} in Appendix~\ref{isoMW}), tend to redistribute towards lower level of angular momentum and higher energies (see Fig.~\ref{ElzGLOB}). The higher the number of accreted satellites (and thus the larger the accreted mass), the broader the distribution of in-situ stars in $E-L_z$ space is (Fig.~\ref{ElzGLOB}). Moreover, also the distribution of in-situ stars is clumpy in this space. In the case of a single accretion, as well as in the case of the accretion of four satellites, clumps appear not only in the region occupied by the stellar disc, as already pointed out by \citet{gomez12}, but also in the less bound region of the diagram -- the region naturally occupied by halo stars (an example of the spatial distribution of stars belonging to some of these clumps, for the case of the accretion of two satellites, is given and discussed in Appendix~\ref{spatialclumps}). When all stars are taken into account, without any differentiation on their origin -- in-situ or accreted -- one sees clearly that this space is hardly decipherable: part of the lumps found (see for ex, those in the bottom-right panel of Fig.~\ref{ElzGLOB}, at $0 < L_z\le 20$ and $-5\le E < 0$) have not an extragalactic origin, but are made of in-situ stars. This finding is valid globally, but also locally -- i.e. when the search is restricted to ``solar vicinity" volumes, as we show in the following. \\
Before moving to local searches for accreted streams, however, we want to drive the reader's attention to another point that seems to us worth emphasizing:  as a consequence of the angular momentum redistribution taking place during mergers, part of the in-situ disc stars -- that initially rotate in a direct sense, with a positive $L_z$ -- tend to redistribute on retrograde orbits. The larger the accreted mass, the higher the number of in-situ stars that end up with negative $L_z$ and the greater the maximum value of the negative angular momentum attained. In the case of $4\times(1:10)$ mergers, for example, some high energy, in-situ stars are on retrograde orbits with angular momenta lower than $L_z=-20$ (in our units, see Sect.~\ref{oursimu}). As a consequence, we caution that \emph{the presence of stars on retrograde orbits in the Galaxy, even at high energies, is not necessarily evidence of an accreted, extragalactic origin}.

\subsubsection{Looking for streams in the solar vicinity}\label{Elzstreams_sun}

\begin{figure}
\begin{center}
\includegraphics[clip=true, trim = 20mm 90mm 20mm 90mm, width=1.1\columnwidth]{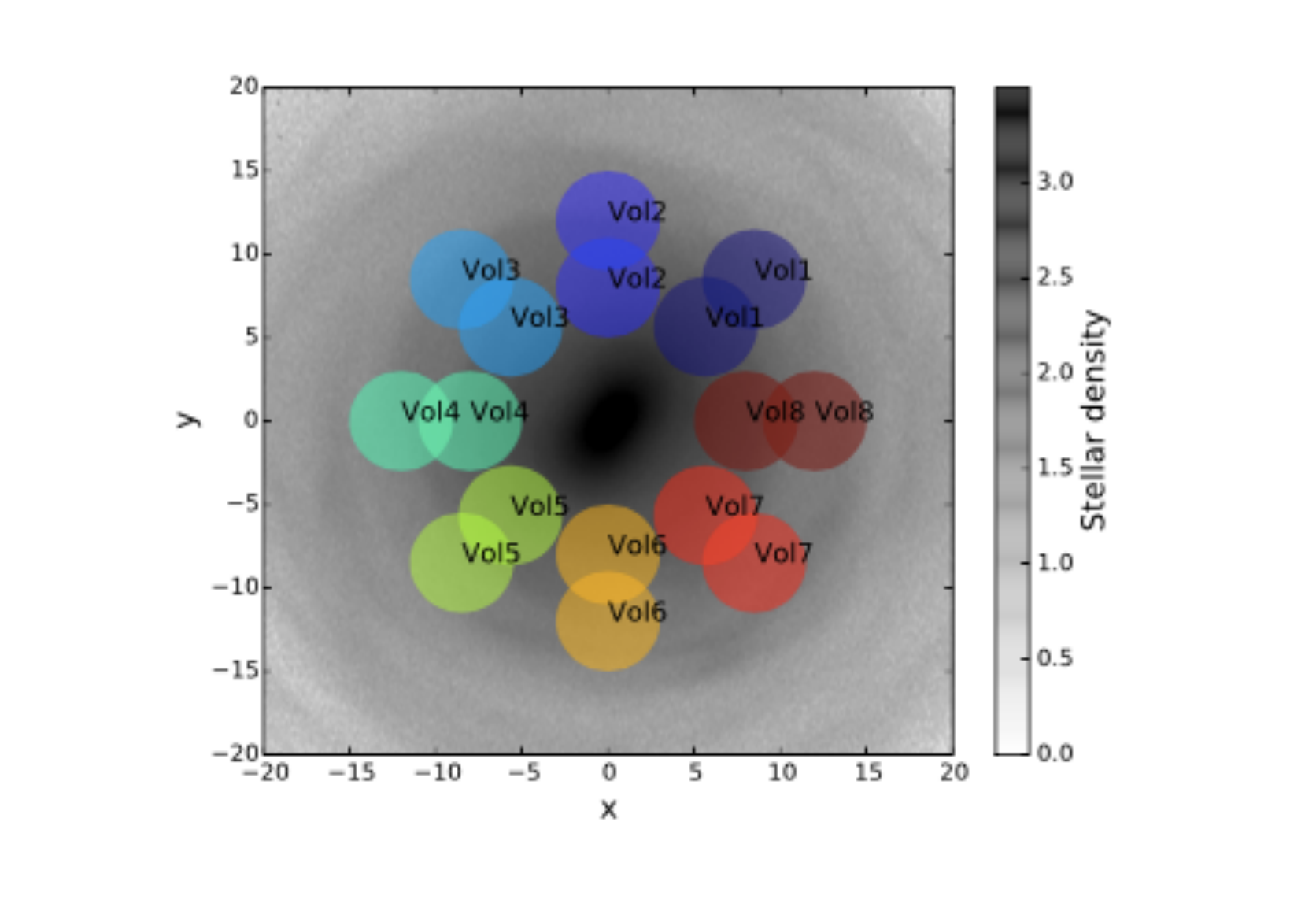}
\end{center}
\caption{The sixteen solar vicinity volumes chosen for the analysis performed in Sects.~\ref{Elzstreams_sun} and \ref{LzLperpsect_sun}. Each spherical volume has a radius of 3~kpc. Volume are located at 8~kpc and 12~kpc from the galaxy centre and are homogeneously distributed in azimuth. The grey map in foreground is simply used to indicate the location of the volumes, for one of the simulations analyzed in this paper. }\label{sunvol}
\end{figure}

\begin{figure*}
\begin{flushleft}
\includegraphics[clip=true, trim = 5mm 0mm 20mm 10mm, angle=270,width=2\columnwidth]{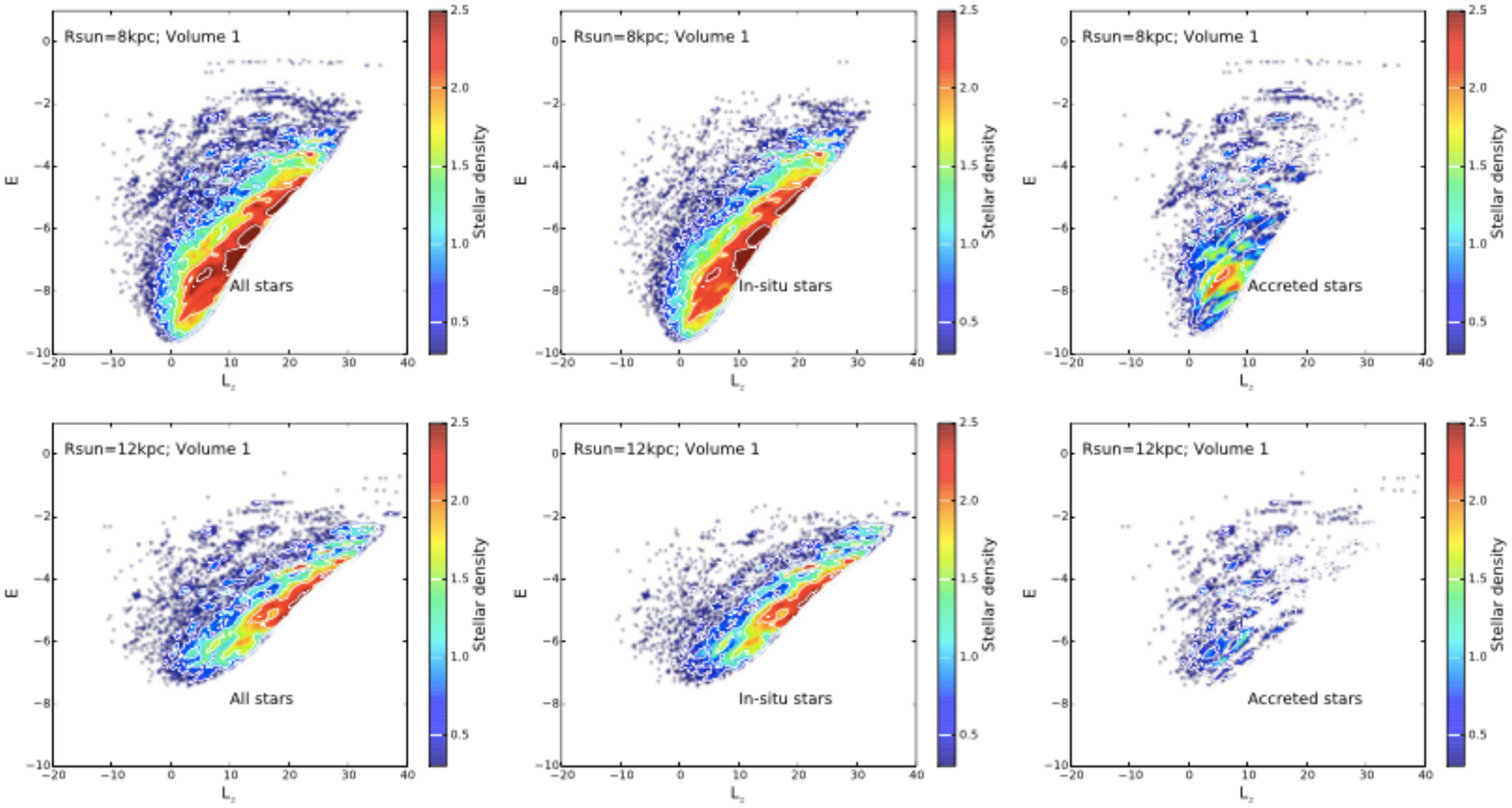}
\end{flushleft}
\begin{center}
\caption{The $E-L_z$ space, for the 1$\times$(1:10) simulation and for two different ``solar vicinity" volumes, one located at 8~kpc (top row) and one at 12~kpc (bottom row) from the galaxy centre. \emph{From left to right:} all stars, in-situ stars and accreted stars in the volume are shown. No additional selection has been operated. Colors code the stellar density, in logarithmic scale, as indicated by the color bar.}\label{ELz1satRsun8part1}
\end{center}
\end{figure*}

\begin{figure*}
\begin{flushleft}
\includegraphics[clip=true, trim = 2mm 2mm 18mm 15mm, angle=270,width=2\columnwidth]{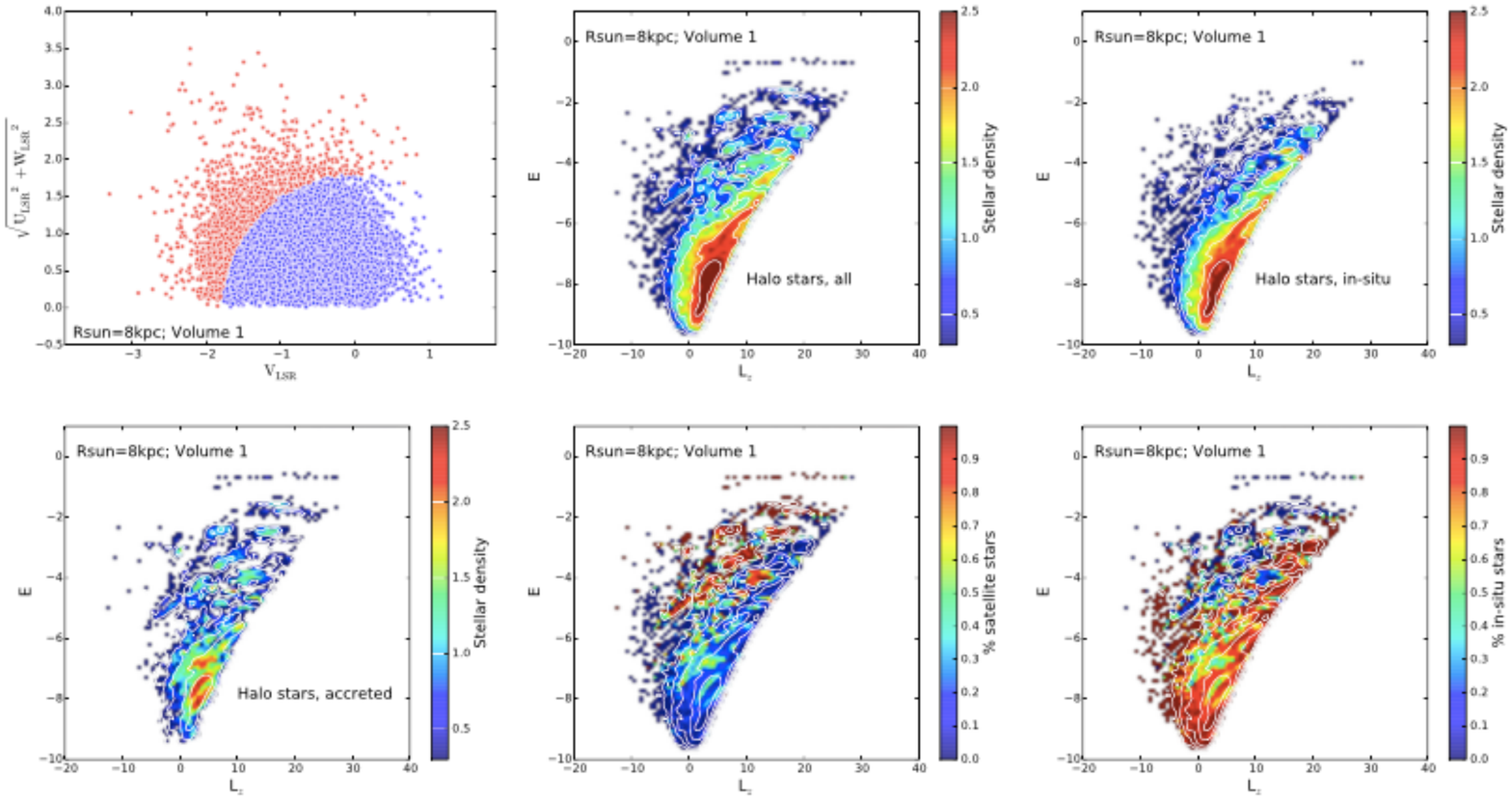}
\end{flushleft}
\caption{Similar to Fig.~\ref{ELz1satRsun8part1}, for one of the volumes at 8~kpc, as indicated in the legend, but this time only halo stars in the selected volume are shown. Halo stars are selected on the basis of their location on the Toomre diagram (\emph{upper-left panel}), as those with $\sqrt{{U_{LSR}}^2+{V_{LSR}}^2+{W_{LSR}}^2}\ge 1.8$. Red points indicate the distribution of one out of fifty halo stars in the volume, blue points indicate thin and thick disc stars in the volume. The following figures, from left to right and from top to bottom, indicate the distribution in the $E-L_z$ space of all halo stars in the selected volume, of in-situ halo stars and of accreted halo stars. The middle and right panels in the bottom row indicate, respectively, the fractional contribution of satellite stars to the total stellar distribution in the $E-L_z$space and the  complementary fractional contribution of in-situ stars. Velocities are in units of 100~km/s, angular momenta in units of 100~km/s.kpc, energies in units of 100~${km/s}^2$.   }\label{ELz1satRsun8part1_halo}
\end{figure*}

\begin{table}
\caption{Fractions $f_{90}$ and $f_{60}$ of the area of the $E-L_z$ space dominated by accreted stars, relative to the area occupied by the whole distribution of halo stars (see Sect.~\ref{Elzstreams_sun} for their definition), in the case of the 1$\times(1:10)$ simulation.  These fractions are reported here for all the different ``solar vicinity" volumes shown in Fig.~\ref{sunvol}.}           
\label{tablefrac1}      
\centering                                    
\begin{tabular}{c c c c}          
\hline\hline                      
$\mathrm{ R_{sun}}$ & Volume $\#$  & $f_{90}$ & $f_{60}$\\    
\hline                                 
	8. & 1 & 0.16 & 0.25\\
    	8. & 2 &  0.16 & 0.25 \\
    	8. & 3 & 0.18 & 0.28 \\
    	8. & 4 & 0.15 & 0.26 \\
    	8. & 5 &0.13 & 0.22 \\
    	8. & 6 & 0.16 & 0.26 \\
    	8. & 7 &  0.16 & 0.27 \\
        8. & 8 & 0.14 & 0.25 \\
        
        	12. & 1 & 0.21 & 0.33\\
    	12. & 2 & 0.20 & 0.31 \\
    	12. & 3 & 0.27 & 0.40 \\
    	12. & 4 & 0.21 & 0.32 \\
    	12. & 5 & 0.19 & 0.30 \\
    	12. & 6 & 0.22 & 0.36 \\
    	12. & 7 & 0.21 & 0.34 \\
        12. & 8 & 0.22 & 0.35 \\
\hline                                             
\end{tabular}
\end{table}

After discussing the global behavior of accreted and in-situ stars in the $E-L_z$ space, we now investigate the feasibility of detecting stellar streams in this space, in local volumes. Defining ``solar vicinity" volumes in a N-body simulation is not trivial. One can choose to place the local volume  either  at a distance from the  galaxy centre comparable to the Sun--Galactic centre distance (i.e. 8--8.5~kpc), or to relate its position to the length of the stellar bar, or to the bar's resonances.  Moreover, because accreted stars do not necessarily redistribute homogeneously in configuration space, the fraction of accreted stars in a particular ``solar volume" can significantly vary depending on the particular choice of that volume. To try to make our analysis as more generic as possible and not dependent on the particular choice of our ``solar vicinity" volume, we have  chosen to identify several regions in our simulated galaxies, as schematically represented in Fig.~\ref{sunvol}: for each simulation, we have indeed defined 16 spherical volumes, all with radii of 3~kpc, centred at a distance of 8~kpc or 12~kpc from the galaxy centre and distributed homogeneously in azimuth. For all stars in each of these volumes at the final time of the simulation, we have analyzed their distribution in the $E-L_z$ space. For two of these volumes -- one placed at 8~kpc and one at 12~kpc from the galaxy centre -- the corresponding distribution is shown in Fig.~\ref{ELz1satRsun8part1}, for the case  of a single merger (i.e. 1$\times$(1:10)). Despite some differences in the values attained by the stellar distribution in the different regions -- one can note, for example, that in the volume at 12~kpc, for any given value of energy, stars attain larger values of $L_z$, as expected -- some points are common to all examined cases, not reported in the figure for a seek of synthesis:
\begin{enumerate}
\item in a 3~kpc--wide region around the Sun, the $E-L_z$ space is rich of substructures;  
\item these substructures are present in the in-situ population and in the accreted population as well; 
\item a single satellite gives rise to a multitude of substructures, of different extent and sizes;  
\item the overlap between the in-situ population and the accreted one is substantial, to the point that no clear and evident distinction can be made between the two on the basis of the analysis of the $E-L_z$ space alone;
\item  in none of these volumes, the presence of a substructure can be easily and unambigously  interpreted as evidence of an extragalactic origin for the stars that compose it: several stellar clumps, also some of those at moderate or negative $L_z$ and relatively high energies, have indeed an in-situ origin, being made of  stars initially in the disc, then heated by the interaction. 
\end{enumerate}

\begin{figure*}
\begin{flushleft}
\includegraphics[clip=true, trim = 2mm 2mm 0mm 10mm, angle=270,width=2\columnwidth]{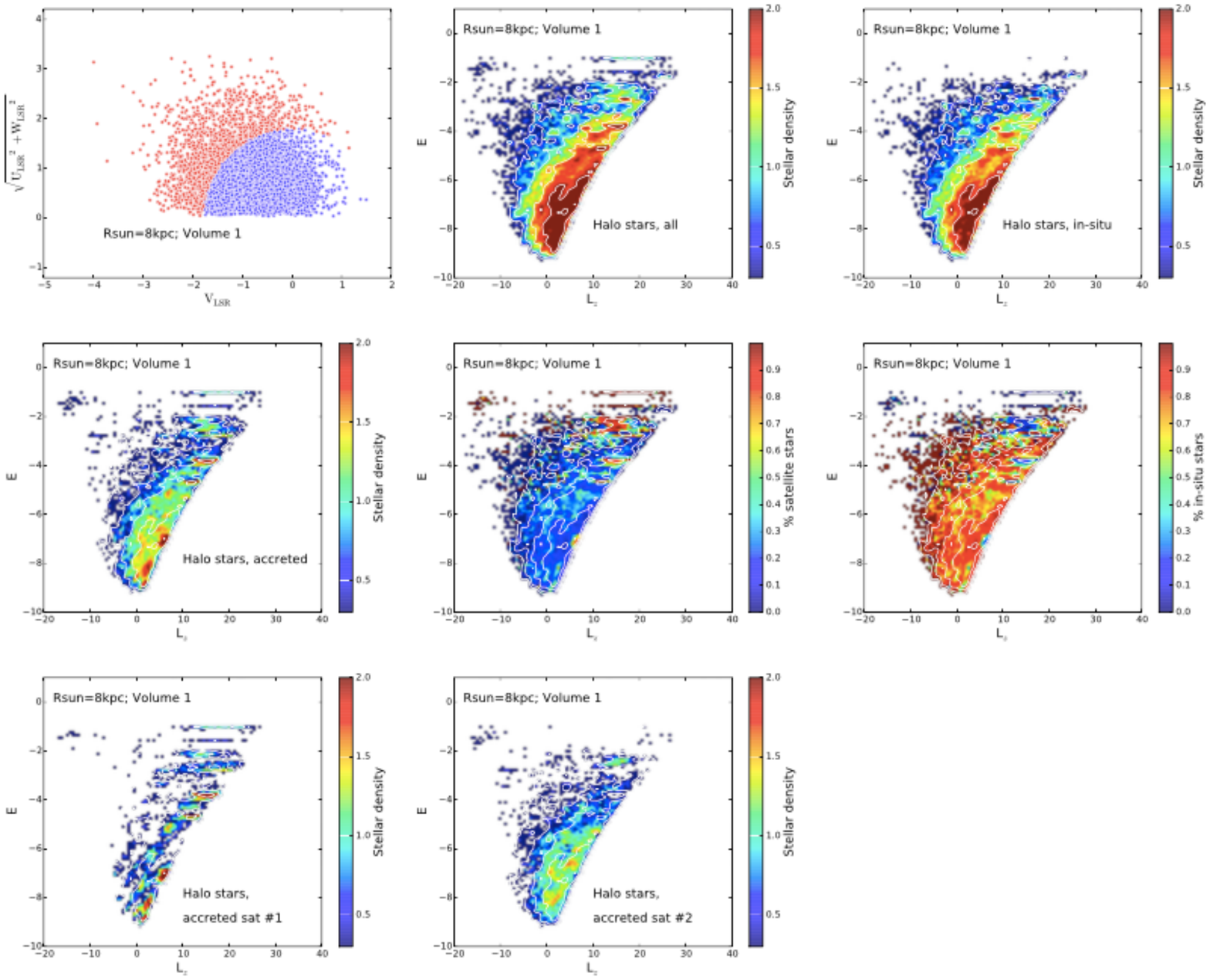}
\end{flushleft}
\caption{Same as Fig.~\ref{ELz1satRsun8part1_halo}, but for the case of the 2$\times$(1:10) merger. The distribution in the $E-L_z$ space of each of the two satellites is also shown in the bottom row. }\label{ELz2satRsun8part1_halo}
\end{figure*}

\begin{figure*}
\begin{flushleft}
\includegraphics[clip=true, trim = 2mm 35mm 5mm 0mm, width=2.\columnwidth]{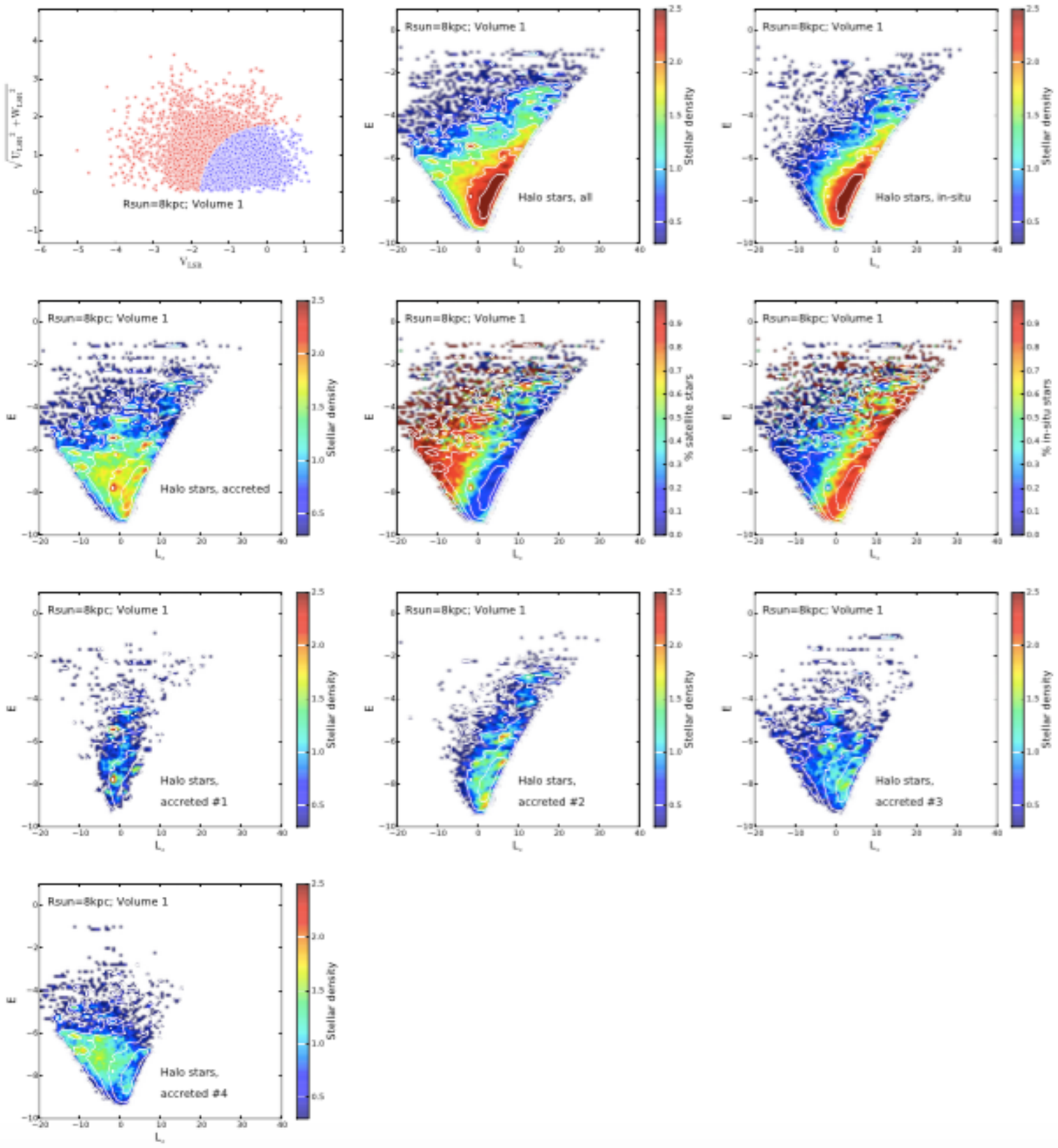}
\end{flushleft}
\caption{Same as Fig.~\ref{ELz1satRsun8part1_halo}, but for the case of the 4$\times$(1:10) merger.  The distribution in the $E-L_z$ space of each of the four satellites is also shown in the third and fourth rows..}\label{ELz4satRsun8part1_halo}
\end{figure*}

\begin{table}
\caption{Fractions $f_{90}$ and $f_{60}$ of the area of the $E-L_z$ space dominated by accreted stars, relative to the area occupied by the whole distribution of halo stars (see Sect.~\ref{Elzstreams_sun} for their definition), in the case of the 2$\times(1:10)$ simulation.  These fractions are reported here for all the different ``solar vicinity" volumes shown in Fig.~\ref{sunvol}.}              
\label{tablefrac2}      
\centering                                      
\begin{tabular}{c c c c}          
\hline\hline                        
$\mathrm{ R_{sun}}$ & Volume $\#$  & $f_{90}$ & $f_{60}$\\    
\hline                                   
	8. & 1 & 0.09 & 0.14\\
    	8. & 2 &  0.08 & 0.14 \\
    	8. & 3 & 0.07 & 0.10 \\
    	8. & 4 & 0.08 & 0.14 \\
    	8. & 5 &0.06 & 0.09 \\
    	8. & 6 & 0.09 & 0.15 \\
    	8. & 7 &  0.07 & 0.12 \\
        8. & 8 & 0.06 & 0.10 \\
        
        	12. & 1 & 0.09 & 0.14\\
    	12. & 2 & 0.12 & 0.17 \\
    	12. & 3 & 0.10 & 0.13 \\
    	12. & 4 & 0.11 & 0.16 \\
    	12. & 5 & 0.06 & 0.10 \\
    	12. & 6 & 0.13 & 0.19 \\
    	12. & 7 & 0.07 & 0.12 \\
        12. & 8 & 0.08 & 0.13 \\
\hline                                             
\end{tabular}
\end{table}

\begin{table}
\caption{Fractions $f_{90}$ and $f_{60}$ of the area of the $E-L_z$ space dominated by accreted stars, relative to the area occupied by the whole distribution of halo stars (see Sect.~\ref{Elzstreams_sun} for their definition), in the case of the 4$\times(1:10)$ simulation.  These fractions are reported here for all the different ``solar vicinity" volumes shown in Fig.~\ref{sunvol}.}              
\label{tablefrac4}      
\centering                                      
\begin{tabular}{c c c c}          
\hline\hline                        
$\mathrm{ R_{sun}}$ & Volume $\#$  & $f_{90}$ & $f_{60}$\\    
\hline                                   
	8. & 1 & 0.25 & 0.45\\
    	8. & 2 &  0.25 & 0.47 \\
    	8. & 3 & 0.24 & 0.47 \\
    	8. & 4 & 0.22 & 0.44 \\
    	8. & 5 &0.22 & 0.43 \\
    	8. & 6 & 0.22 & 0.45 \\
    	8. & 7 &  0.24 & 0.47 \\
        8. & 8 & 0.26 & 0.47 \\
        
        	12. & 1 & 0.30 & 0.46\\
    	12. & 2 & 0.29 & 0.47 \\
    	12. & 3 & 0.30 & 0.50 \\
    	12. & 4 & 0.30 & 0.49 \\
    	12. & 5 & 0.30 & 0.47 \\
    	12. & 6 & 0.31 & 0.51 \\
    	12. & 7 & 0.29 & 0.49 \\
        12. & 8 & 0.32 & 0.52 \\
\hline                                             
\end{tabular}
\end{table}

The analysis presented in Fig.~\ref{ELz1satRsun8part1} has been made considering all stars (i.e. disc(s), as well as halo stars) in a given stellar volume. One may thus think to alleviate the problem of the contamination of the in-situ component by kinematically selecting halo stars only. This is indeed a common strategy, supported by the idea that it is among the oldest populations in our Galaxy -- currently found in the stellar halo -- that the remnants of past accretion events can be found. We have thus repeated the analysis presented in Fig.~\ref{ELz1satRsun8part1}, this time by applying it only to kinematically defined halo stars. To this aim, for each ``solar vicinity" volume we have selected only stars whose velocities $(U_{LSR}, V_{LSR}, W_{LSR})$  in the Local Standard of Rest (i.e. LSR) reference frame\footnote{Here  $U_{LSR}, V_{LSR}$ and $W_{LSR}$ denote, respectively, the radial, tangential and vertical velocity in a reference frame whose origin is in the chosen stellar particle position, $R_{\star}$ and which rotates around the galaxy centre at a constant circular velocity, $V_{Sun}$, equal to $V_C(R=R_{Sun})$, i.e. equal to the value of the rotation curve of the modeled galaxy at  $R=R_{Sun}$.} satisfy the condition $\sqrt{{V_{LSR}}^2+{U_{LSR}}^2+{W_{LSR}}^2} \ge 180$~km/s.\footnote{The threshold of 180~km/s  is commonly used to select halo stars in observational samples at the solar vicinity \citep[see, for example][]{nissen10}. Moreover, because in the selected ``solar vicinity" volumes, our modeled galaxies attain values of the rotation curve similar to those estimated in our Galaxy at the solar radius (i.e. between 200 and 250 km/s), we feel comfortable in applying the same selection criterion also to our modeled data.}  This selection is shown in the top-left panel of Fig.~\ref{ELz1satRsun8part1_halo}, for the same volume considered in the upper row of Fig.~\ref{ELz1satRsun8part1}. As it can be appreciated from this figure, even when the analysis is restricted to kinematically selected halo stars only, the two main problems affecting the analysis presented in Fig.~\ref{ELz1satRsun8part1} are not alleviated: 
\begin{enumerate}
\item the contamination of in-situ stars is still severe; 
\item  the clumpy redistribution of stars in the $E-L_z$ space is not a characteristics of the accreted population only, since in-situ halo stars show a inhomogeneous and lumpy distribution as well. 
\end{enumerate}
In fact accreted stars dominate only a marginal region of the $E-L_z$ space of halo stars (see bottom-middle panel of Fig.~\ref{ELz1satRsun8part1_halo}). Most of this space -- which is all but smoothly distributed -- is indeed dominated by the in-situ halo population  (see bottom-right panel of Fig.~\ref{ELz1satRsun8part1_halo}). To better quantify the importance of accreted stars, we have evaluated the area of the region, in the $E-L_z$ space, occupied by accreted stars. To this aim, we have selected only regions where the percentage  of accreted stars in greater or equal to 90\% of the total (this corresponds, for example, to the red-brown regions in Fig.~\ref{ELz1satRsun8part1_halo}, bottom-middle panel) -- and we have divided this area by the area of the $E-L_z$ space occupied by all halo stars. We refer to this fraction as $f_{90}$. For the volume shown in Fig.~\ref{ELz1satRsun8part1_halo}, $f_{90}=0.16$. Thus, in this volume, accreted stars dominate only a marginal region of the overall distribution.  We can relax the threshold and look for the area where the percentage of accreted stars is greater or equal to 60\% of the total distribution and divide -- as before -- this area by that occupied by the whole distribution. We call this fraction $f_{60}$. For the volume examined in  Fig.~\ref{ELz1satRsun8part1_halo}, $f_{60}=0.25$. Thus, even relaxing the criterion, accreted stars turn out to dominate only a limited region of the $E-L_z$ space. For this specific volume and this specific simulation, the probability that a given region of the $E-L_z$ space is mostly made of in-situ stars is thus  very high, about 75\%. We have repeated this calculation for all the volumes defined in Fig.~\ref{sunvol}, for the remnant of the 1$\times$(1:10) simulation. The corresponding values, $f_{90}$ and $f_{60}$ are reported in Table~\ref{tablefrac1}. On average, for the volumes located at 8~kpc, $f_{90}=0.16$ and $f_{60}=0.26$. For those located at 12~kpc, the fraction of the $E-L_z$ space dominated by accreted stars increases slightly ($f_{90}=0.21$ and $f_{60}=0.34$), this mainly because of the decreasing density of  the in-situ halo population.\\

All the analysis presented in the first part of this section concerns the case of a single 1:10 accretion. It is thus important to understand how the results are sensitive to the accretion history of the galaxy and in particular to the number of accreted satellites. For this, we need to examine the cases of 2$\times$(1:10) and 4$\times$(1:10) accretions. By examining these two cases, we do not think, of course, to have explored the full parameter space. But we can, in any case, derive some lessons, as detailed in the following.
Similarly to the analysis done before, Figs.~\ref{ELz2satRsun8part1_halo} and \ref{ELz4satRsun8part1_halo} show, respectively, the distribution in $E-L_z$ space of all halo stars in a ``solar vicinity" volume, that of all in-situ and accreted stars (separating also the contribution of the different satellites) and their relative contributions to the total  distribution. As before, halo stars have been selected as those satisfying the kinematic condition $\sqrt{{V_{LSR}}^2+{U_{LSR}}^2+{W_{LSR}}^2} \ge 180$~km/s. There are some points which is worth retaining from this analysis:
\begin{enumerate}
\item in all cases, neither the total distribution of halo stars, nor that of accreted stars, nor that of in-situ stars is smooth  in the $E-L_z$ space; 
\item  the higher the number of accretion events -- and thus the accreted stellar mass --  the larger the distribution of in-situ stars in $E-L_z$ space; 
\item  each satellite redistributes over a large extent of the $E-L_z$ space at the solar vicinity, giving rise to several substructures; 
\item stars initially belonging to different satellites -- even with significantly different initial orbits, as it is the case of the 4$\times$(1:10) merger -- can significantly overlap in the $E-L_z$ space, to the point that the global distribution of accreted stars is hardly decipherable. 
\end{enumerate}

As a consequence, 
\begin{itemize}
\item because of Point 1, substructures in this space cannot be interpreted as an indication of the extragalactic origin of the stars that make it;
\item because of Point 2, it is not possible to define \emph{a priori} what the extension of in-situ halo stars in the $E-L_z$ space should be: depending on the specific accretion history of the galaxy, in-situ stars heated by the interaction can occupy a variable extent of this space. Any cut at any particular value of $E$ and $L_z$ -- as tentatively suggested by \citet{ruchti14, ruchti15} -- is arbitrary. Moreover, as we discuss more extensively in Sect.~\ref{discussion}, in our modelling the in-situ halo stellar population is made only of stars initially in the disc and then kinematically heated by the interaction to form the halo. In other words, in our models an initial population of in-situ halo stars is missing. If this population was taken into account, the distribution of the in-situ population in the $E-L_z$ space may become even more extended, depending -- among other factors --  on the intrinsic amount of rotation this initial in-situ halo population would have. Modelling this initial population is beyond the scope of this paper, but it is worth taking into account that in all this analysis we may probably still underestimate the role of in-situ halo stars. 
\item because of Point 3, from the number of substructures found in this space we cannot trace back the number of accretion events experienced by the galaxy. 
\item because of Point 4, it is not possible to separate the different satellite progenitors in this space.  
\end{itemize}

Finally, we can repeat the analysis made for the 1$\times$(1:10) merger and ask ourselves which fraction the $E-L_z$ space is dominated by accreted stars in the case of our 2$\times$(1:10) and 4$\times$(1:10) mergers. The results of this analysis are reported in Tables~\ref{tablefrac2} and \ref{tablefrac4}, respectively. It is only for the 4$\times$(1:10) simulation that the fraction of the $E-L_z$ space is significant, with a $f_{60}\sim0.5$ and $f_{90}\sim0.3$, on average. This higher fractions are mostly due to the presence of a significant fraction of accreted stars on retrograde orbits  -- due to the accretion of satellite $\#$4 and partially also to the satellite $\#3$ (see also Fig.~\ref{ELz4satRsun8part1_halo}). Even in this case, however, the role of the in-situ population is still important, half of the $E-L_z$ space being still dominated by in-situ stars. \\

\subsection{On the $L_\perp-L_z$ space}\label{LzLperpsect_sun}

Differently from the energy, the determination of the angular momentum components does not require any knowledge of the Galactic potential. Moreover, $L_z$ is conserved in an axisymmetric potential and $L_\perp$, even if not strictly conserved, may vary only marginally \citep[see Sect.~2, Chapter~3 in][]{binney87}.  This is the reason why the $L_z-L_\perp$ space  has been used several times as a natural space where to look for the presence of stellar streams, since the seminal work by  \citet{helmi99}. By studying a sample of about a hundred metal-poor red giants and RR Lyrae stars within 1~kpc from the Sun, they pointed out that while at values of $L_\perp\le 1000$~kpc~km/s and $|L_z|\le 1000$~kpc~km/s the distribution appears quite symmetric, with stars spanning all values of $L_z$, for larger $L_\perp$ the distribution appears asymmetric with respect to $L_z$, with an excess of stars on prograde (i.e. in the direction of Galactic rotation) orbits, showing up as an overdensity. According to their work, indeed, for such high values of $L_\perp$, very few stars appear on retrograde orbits ($L_z<0$) and there is a lack of stars on polar orbits ($L_z=0$). By comparing the observed distribution of stars in the $L_z-L_\perp$ space  with Monte Carlo simulations of smooth, non rotating stellar halos, the overdensity found at ($L_z, L_\perp$)$\simeq$(1000,2000)~kpc.km/s  was shown to be significant and interpreted as evidence of an accreted stream, known as the ``Helmi stream" after this discovery. Other studies have subsequently confirmed the presence of this overdensity in the $L_z-L_\perp$ space, confirming its extra-galactic nature \citep{chiba00, refiorentin05, dettbarn07, kepley07, klement09, smith09, refiorentin15} and concluding that about 5\% of the local stellar halo should be made of stars belonging to this stream. 
Apart from the Helmi's one, other possible streams have been potentially detected in the $L_z-L_\perp$ plane \citep[see, for example][]{kepley07, smith09, refiorentin15}.  A point that is particularly critical in the search for streams and substructures in the $L_z-L_\perp$ space is the comparison that it is often made with what it is supposed to be the distribution of in-situ halo stars in this space \citep[see, for example][]{helmi99, kepley07, smith09}: \emph{in-situ halo stars are assumed to have a smooth distribution in the $L_z-L_\perp$ space and to not rotate. This is a critical assumption, which affects the significance of any possible detection. } 

We thus concentrate the following analysis to investigate the locus occupied by in-situ halo stars in this plane and to understand and quantify their overlap with accreted stars. Once again, we recall the reader that the stellar halo, in our simulations, is entirely made of disc stars kinematically heated by the interaction. No modelling of a stellar halo pre-existing the accretion event is included in this work. 

\subsubsection{Looking for streams at the ``solar vicinity": accreted, in-situ stars and their overlap}\label{LzLperpsect}

\begin{figure*}
\begin{flushleft}
\includegraphics[clip=true, trim = 10mm 45mm 15mm 40mm, width=2.\columnwidth]{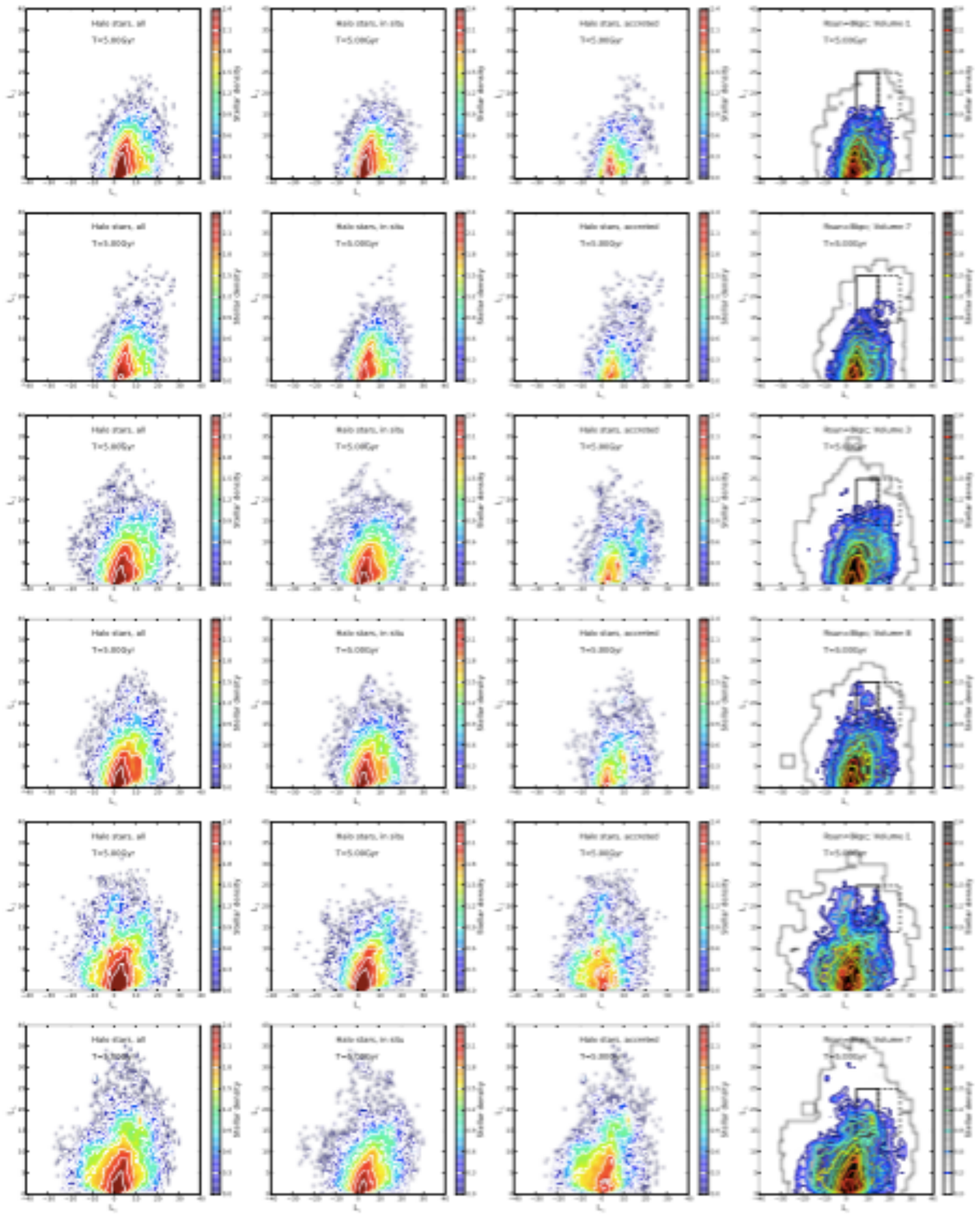}
\end{flushleft}
\begin{center}
\caption{The distribution in the $L_\perp-L_z$ space of stars in some ``solar vicinity" volumes, for the 1$\times$(1:10) simulation (\emph{first and second row}), for the 2$\times$(1:10) simulation (\emph{third and fourth row}) and for the 4$\times$(1:10) simulation (\emph{last two rows}). \emph{From left to right:} Distribution, in the selected volume, of all stars, in-situ stars and accreted stars,respectively.  \emph{Rightmost column:} only the distribution of accreted stars is shown (great scale), with the contours representing respectively the accreted population (thick lines) and the in-situ one (thin lines). The two rectangular areas correspond to the Helmi region and to the `extreme prograde region" (see text for details). }\label{LzLperp1satRsun8part1}
\end{center}
\end{figure*}

In Fig.~\ref{LzLperp1satRsun8part1}, we show the distribution of halo stars, for some of the solar vicinity volumes defined in Fig.~\ref{sunvol} and for the case of the 1$\times$(1:10) accretion, the 2$\times$(1:10) and the 4$\times$(1:10), respectively. Halo stars have been selected kinematically, using the same criterion already adopted in Sect.~\ref{Elzstreams_sun} and in Figs.~\ref{ELz1satRsun8part1_halo}, \ref{ELz2satRsun8part1_halo} and \ref{ELz4satRsun8part1_halo}: among all stars in the chosen solar volume, halo stars are those satisfying the condition $\sqrt{V_{LSR}^2+U_{LSR}^2+W_{LSR}^2} \ge 180$~km/s.\\

The main points we want to make from this analysis are the following:
\begin{enumerate}
\item Halo stars occupy a large portion of the $L_z-L_\perp$  space, whose extension depends on the number of accretion events experienced by the galaxy,  and on the orbital characteristics of the accreted satellites (prograde or retrograde orbits). In our models, it is in the case of the 4$\times$(1:10) accretion, that halo stars redistribute over the largest extent in the $L_z-L_\perp$  space.
\item In general, the distribution is asymmetric with respect to the axis $L_z=0$, towards positive $L_z$ values (i.e. prograde motions). The strength of this asymmetry depends on the  accreted mass -- the higher the number  of accretion events, the more symmetric the distribution becomes -- and on the presence of satellites on retrograde orbits. The cause of this asymmetry cannot be attributed only to the presence of accreted stars. Indeed, as Fig.~\ref{LzLperp1satRsun8part1} shows, also  in-situ halo stars  show an asymmetric distribution, skewed towards positive $L_z$. Again, the strength of this asymmetry depends on the number of accretion events and on their characteristics -- also for in-situ stars, the distribution tends to become more symmetric with respect to the $L_z=0$ axis in the case of the 4$\times(1:10)$ interaction -- but in all cases, at high values of $L_\perp$, the distribution of in-situ  halo stars becomes skewed towards more prograde, inclined orbits. Even if the accretion event causes  an important redistribution of the initial angular momentum of the main galaxy and in particular of its $z-$component, to the point that some in-situ stars end up on retrograde orbits, overall in-situ halo stars retain part of their initial rotation, forming a slow-rotating stellar halo. 
This finding is fundamental: it implies that \emph{any observational evidence of an asymmetric distribution in the $L_z-L_\perp$  space skewed towards prograde, inclined orbits is not in itself an indication that this region of the space is dominated by accreted stars}. Any accretion event indeed generates an in-situ halo population (made of pre-existing disc stars heated by the interaction), whose distribution, at large values of $L_\perp$, is skewed towards large values of $L_z$.   
\item Overall the overlap between the in-situ and accreted population in the $L_z-L_\perp$ plane is considerable everywhere. There is no particular region where the contribution of accreted stars appears dominant.  For values of   $(L_z-L_\perp)$ approaching $(0,0)$, the in-situ population dominates, while moving towards the periphery of the distribution, the two populations show overall comparable stellar densities. This is visible, for example, in the rightmost panels of Fig.~\ref{LzLperp1satRsun8part1}, where the density contours of the in-situ and accreted populations are reported:  they show a striking similarity everywhere, suggesting that even in the periphery of the distribution (for large values of $L_z$ and $L_\perp$, for example), the probability to find in-situ stars is high. To better quantify this point, for each solar vicinity volume, we have selected two regions of the $L_z-L_\perp$ plane: one corresponding to the region where the Helmi stream has been found  observationally (i.e., in units of 100~kpc.km/s, $5<L_z<15$ and $14<L_\perp<25$, hereafter called the "Helmi region", see \citet{kepley07}) and one characterized by similar  values of $L_\perp$  but more extreme values of  $L_z$  ($15 < L_z < 25$,  hereafter called ``extreme prograde region"). For each of these two regions, for each solar vicinity volume and for all the three simulations, we have evaluated the number of satellite stars contained in that region and compared it to the total number of stars found in the same region. These values are reported in Tables~\ref{tablefrachelmi1}, \ref{tablefrachelmi2} and \ref{tablefrachelmi4}, for the 1$\times$(1:10), 2$\times$(1:10) and 4$\times$(1:10) simulations, respectively. While the fraction of satellite stars depends on the volume under consideration, we found that -- on average -- there is no clear dependence of this value on the number of accreted satellites and most importantly  that in all cases \emph{a significant fraction of in-situ stars is present in the selected regions}, even when the Milky Way-type galaxy experiences multiple accretion events. Averaging over all the volumes, in the Helmi region the average fraction of accreted stars is 0.39 in the case of the 1$\times(1:10)$ simulation, 0.37 for the 2$\times(1:10)$ simulation and 0.5 for the 4$\times(1:10)$ simulation. For the extreme prograde region, these fractions are, respectively, 0.63, 0.46 and 0.22. 
\item Finally, some words on the smoothness of the $L_z-L_\perp$ space. The overall distribution does not appear to be smooth, thus confirming observational findings \citep[see, for example][]{helmi99, smith09, refiorentin15}. The presence of substructures is visible among accreted halo stars, in agreement with previous models \citep{helmi99, kepley07, refiorentin15}. But most importantly -- a result to our knowledge never pointed out before -- also the in-situ halo population is not smoothly redistributed in the $L_z-L_\perp$ space. Some examples of this clumpy, in-situ halo distribution can be appreciated in Fig.~\ref{LzLperp1satRsun8part1}. Overdensities 
 of in-situ stars appear not only at low values of $L_\perp$, thus for low inclination orbits, but also at values as large as $L_\perp\sim 20$. This suggests that an extreme caution must be taken in interpreting the origin of substructures and clumps in the $L_z-L_\perp$ space: \emph{the presence of isolated/dense groups of stars  in some regions of this space, or of groups with extreme prograde or retrograde orbits, it is not in itself a probe of an extragalactic origin of these stars.}
\end{enumerate}

\begin{table}
\caption{Fractional contribution of accreted stars to the ``Helmi region" and to the ``extreme prograde region" for the $1\times(1:10)$ simulation. For each solar vicinity volume, the total number of stellar particles in the ``Helmi region" ($N_{star, Reg1}$) and in the ``extreme prograde region" ($N_{star, Reg2}$) is shown, together with the respective fractions of satellite stars ($f_{sat, Reg1}$ and $f_{sat, Reg2}$). See Sect.~\ref{LzLperpsect} for details. }              
\label{tablefrachelmi1}      
\centering                                      
\begin{tabular}{c c c c c c}          
\hline\hline                        
$\mathrm{ R_{sun}}$ & Volume $\#$ &  $N_{star, Reg1}$& $f_{sat, Reg1}$ & $N_{star, Reg2}$ & $f_{sat, Reg2}$\\    
\hline                                   
	8. & 1 & 343 & 0.41 & 115 & 0.67\\
    	8. & 2 &  264 & 0.28 & 127 & 0.75 \\
    	8. & 3 & 283 & 0.44 & 88 & 0.78 \\
    	8. & 4 & 241 & 0.34  & 50 &  0.34 \\
    	8. & 5 & 275 & 0.18 & 65 & 0.52 \\
    	8. & 6 & 277 & 0.35 & 153 & 0.78\\
    	8. & 7 &  354 & 0.48 &  152 & 0.76\\
        8. & 8 & 284 & 0.43 & 137 & 0.80 \\
        
        	12. & 1 & 167 & 0.51 & 125 & 0.73\\
    	12. & 2 & 140 & 0.29 & 137 & 0.67 \\
    	12. & 3 & 169 & 0.55 & 84 & 0.62 \\
    	12. & 4 & 150 & 0.33 & 53 & 0.30 \\
    	12. & 5 & 131 & 0.29 & 43 & 0.35 \\
    	12. & 6 & 177& 0.42 & 286 & 0.89 \\
    	12. & 7 & 229 & 0.56 & 110 & 0.63 \\
        12. & 8 & 163 & 0.39 & 71 & 0.55  \\
\hline                                             
\end{tabular}
\end{table}

\begin{table}
\caption{Same as Table~\ref{tablefrachelmi1}, but for the $2\times(1:10)$ satellite.}              
\label{tablefrachelmi2}      
\centering                                      
\begin{tabular}{c c c c c c}          
\hline\hline                        
$\mathrm{ R_{sun}}$ & Volume $\#$ &  $N_{star, Reg1}$& $f_{sat, Reg1}$ & $N_{star, Reg2}$ & $f_{sat, Reg2}$\\    
\hline                                   
	8. & 1 & 1018 & 0.49 & 418 & 0.68\\
    	8. & 2 &  785 & 0.37 & 424 & 0.72 \\
    	8. & 3 & 744 & 0.28 & 474 & 0.60 \\
    	8. & 4 & 1114 & 0.27  & 486 &  0.37 \\
    	8. & 5 & 1124 & 0.31 & 440 & 0.44 \\
    	8. & 6 & 1105 & 0.47 & 369 & 0.47\\
    	8. & 7 &  769 & 0.34 &  234 & 0.29\\
        8. & 8 & 859 & 0.39 & 282 & 0.42 \\
        
        	12. & 1 & 696 & 0.38 & 234 & 0.40\\
    	12. & 2 & 439 & 0.34 & 394 & 0.72 \\
    	12. & 3 & 521 & 0.33 & 355 & 0.55 \\
    	12. & 4 & 789 & 0.35 & 53 & 0.30 \\
    	12. & 5 & 785 & 0.40 & 353 & 0.33 \\
    	12. & 6 & 671& 0.47 & 415 & 0.33 \\
    	12. & 7 & 436 & 0.36 & 216 & 0.26 \\
        12. & 8 & 478 & 0.33 & 288 & 0.48  \\
\hline                                             
\end{tabular}
\end{table}

\begin{table}
\caption{.Same as Table~\ref{tablefrachelmi1}, but for the $4\times(1:10)$ satellite.}              
\label{tablefrachelmi4}      
\centering                                      
\begin{tabular}{c c c c c c}          
\hline\hline                        
$\mathrm{ R_{sun}}$ & Volume $\#$ &  $N_{star, Reg1}$& $f_{sat, Reg1}$ & $N_{star, Reg2}$ & $f_{sat, Reg2}$\\    
\hline                                   
	8. & 1 & 1303 & 0.25 & 385 & 0.20\\
    	8. & 2 & 1593 & 0.42 & 616 & 0.13 \\
    	8. & 3 & 2530 & 0.68 & 277 & 0.17 \\
    	8. & 4 & 2465 & 0.49 & 661 & 0.37 \\
    	8. & 5 & 2778 & 0.29 & 1146 & 0.22 \\
    	8. & 6 & 3856 & 0.69 & 950 & 0.20 \\
    	8. & 7 & 3377 & 0.76 & 543 & 0.20 \\
        8. & 8 & 1007 & 0.47 & 311 & 0.24 \\
        
        	12. & 1 & 601 & 0.23 & 197 & 0.28 \\
    	12. & 2 & 1232 & 0.53 & 292 & 0.15 \\
    	12. & 3 & 1310 & 0.62 & 158 & 0.23 \\
    	12. & 4 & 1730 & 0.57 & 427 & 0.26 \\
    	12. & 5 & 1705 & 0.19 & 902 & 0.12 \\
    	12. & 6 & 1848 & 0.61 & 543 & 0.19 \\
    	12. & 7 & 1925 & 0.74 & 352 & 0.32 \\
        12. & 8 & 519 & 0.49 & 239 & 0.32 \\
\hline                                             
\end{tabular}
\end{table}

\subsubsection{Some words about in-situ and accreted stars in velocity spaces}

\begin{figure*}
\begin{center}
\includegraphics[clip=true, trim = 0mm 14mm 5mm 9mm, angle=270,width=2\columnwidth]{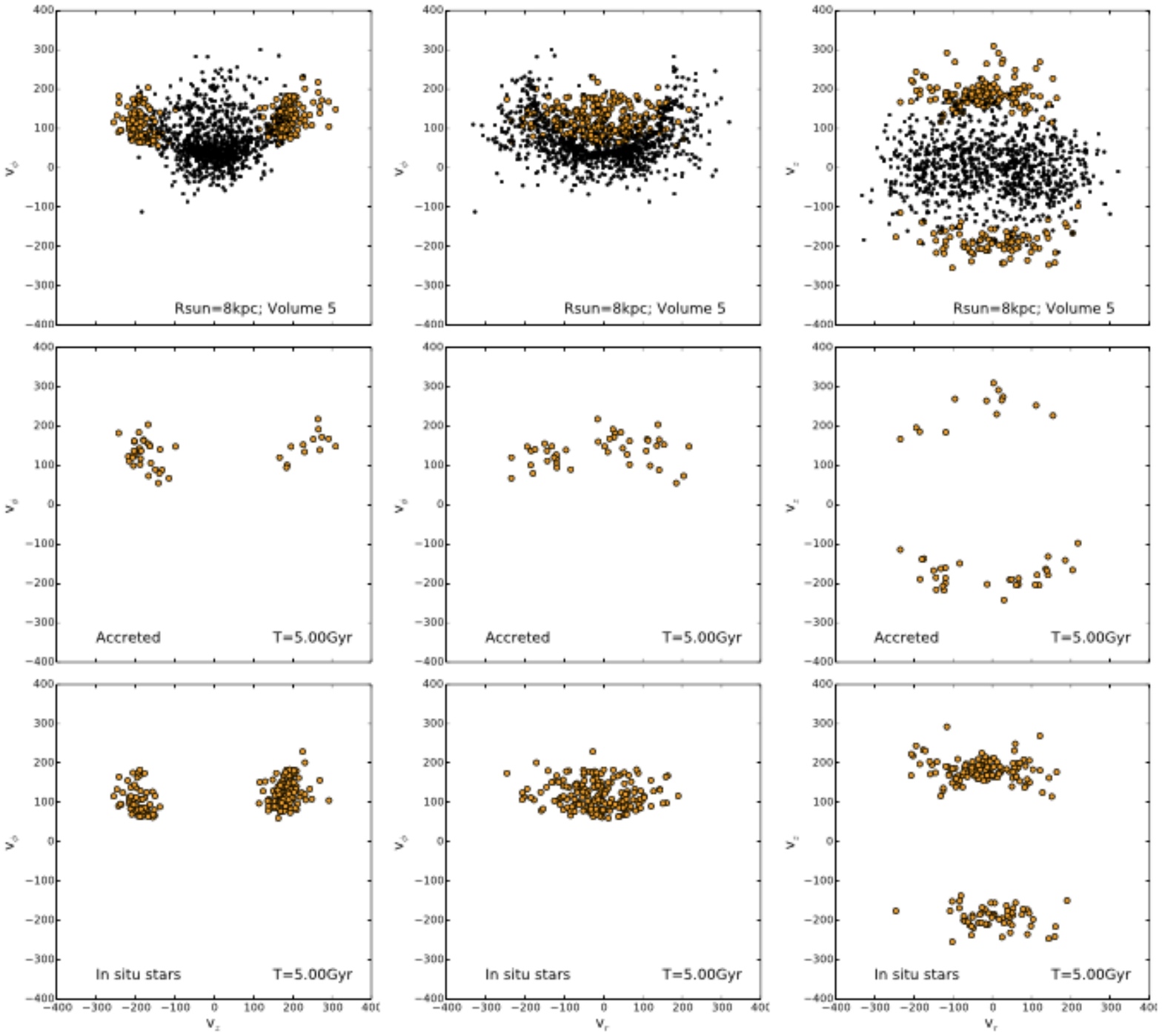}
\caption{1$\times$(1:10) simulation: Velocities of stars belonging to one of the ``solar vicinity" volumes studied in this paper. \emph{From left to right:}Projection of the velocities in the $v_z-v_\phi$, $v_r-v_\phi$ and $v_r-v_z$ planes. \emph{From top to bottom:} all stars,  accreted stars and in-situ stars in the selected volume are shown. In all the panels, the black points represents the velocities of all stars in the volume (only 1 particle over 100 is shown), while the orange points indicate only those stars in the volume that belong to  the Helmi region. Velocities are in units of km/s.  }\label{velhelmi2}
\end{center}
\end{figure*}

\begin{figure*}
\begin{center}
\includegraphics[clip=true, trim = 15mm 35mm 10mm 9mm, width=1.7\columnwidth]{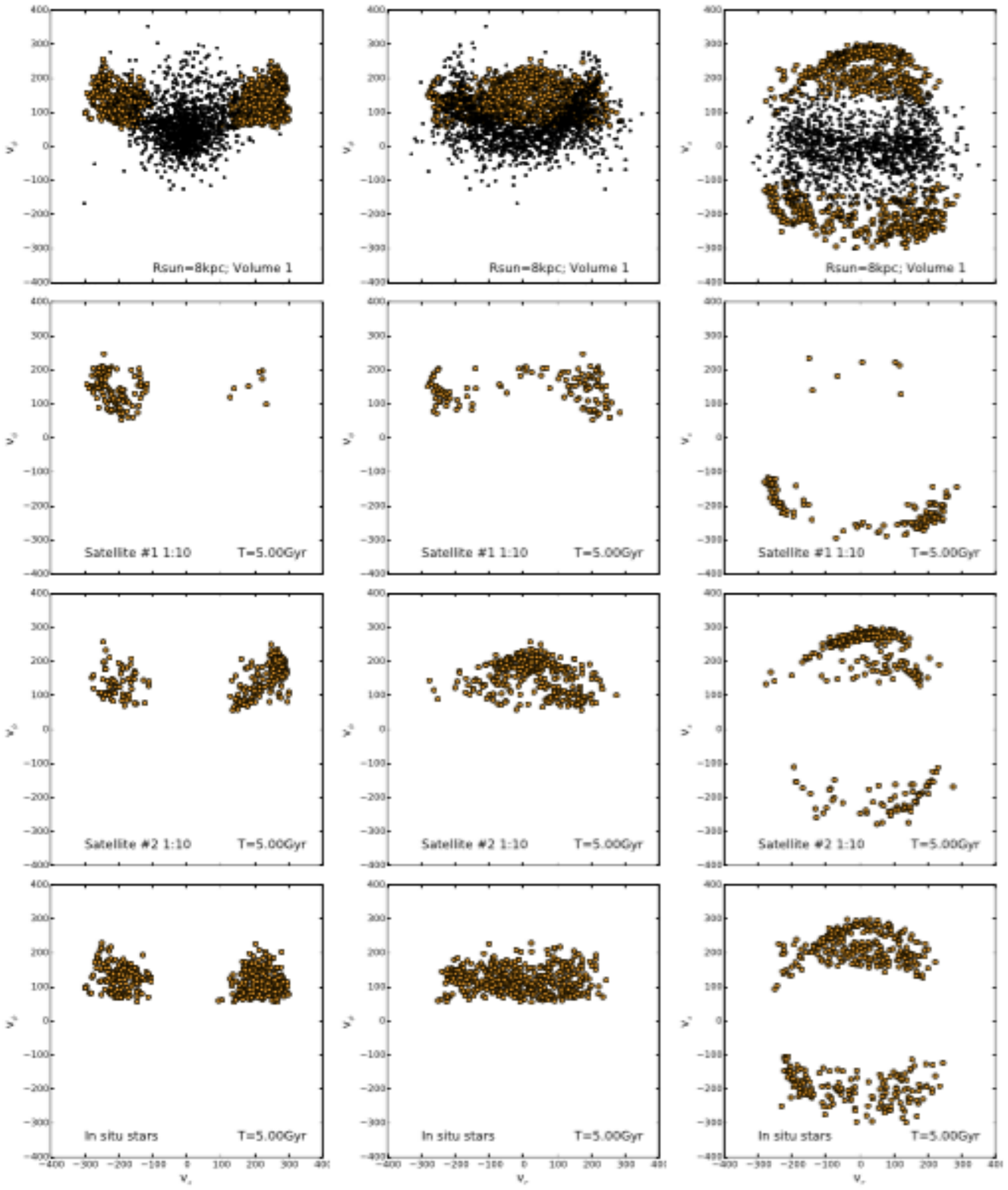}
\caption{Same as Fig.~\ref{velhelmi2}, but for the 2$\times$(1:10) merger. Velocities of accreted stars in the $v_z-v_\phi$, $v_r-v_\phi$ and $v_r-v_z$ planes are also shown for each of the two satellites, separately (second and third row).}\label{velhelmi3}
\end{center}
\end{figure*}

\begin{figure*}
\begin{center}
\includegraphics[clip=true, trim = 15mm 10mm 10mm 9mm, width=1.7\columnwidth]{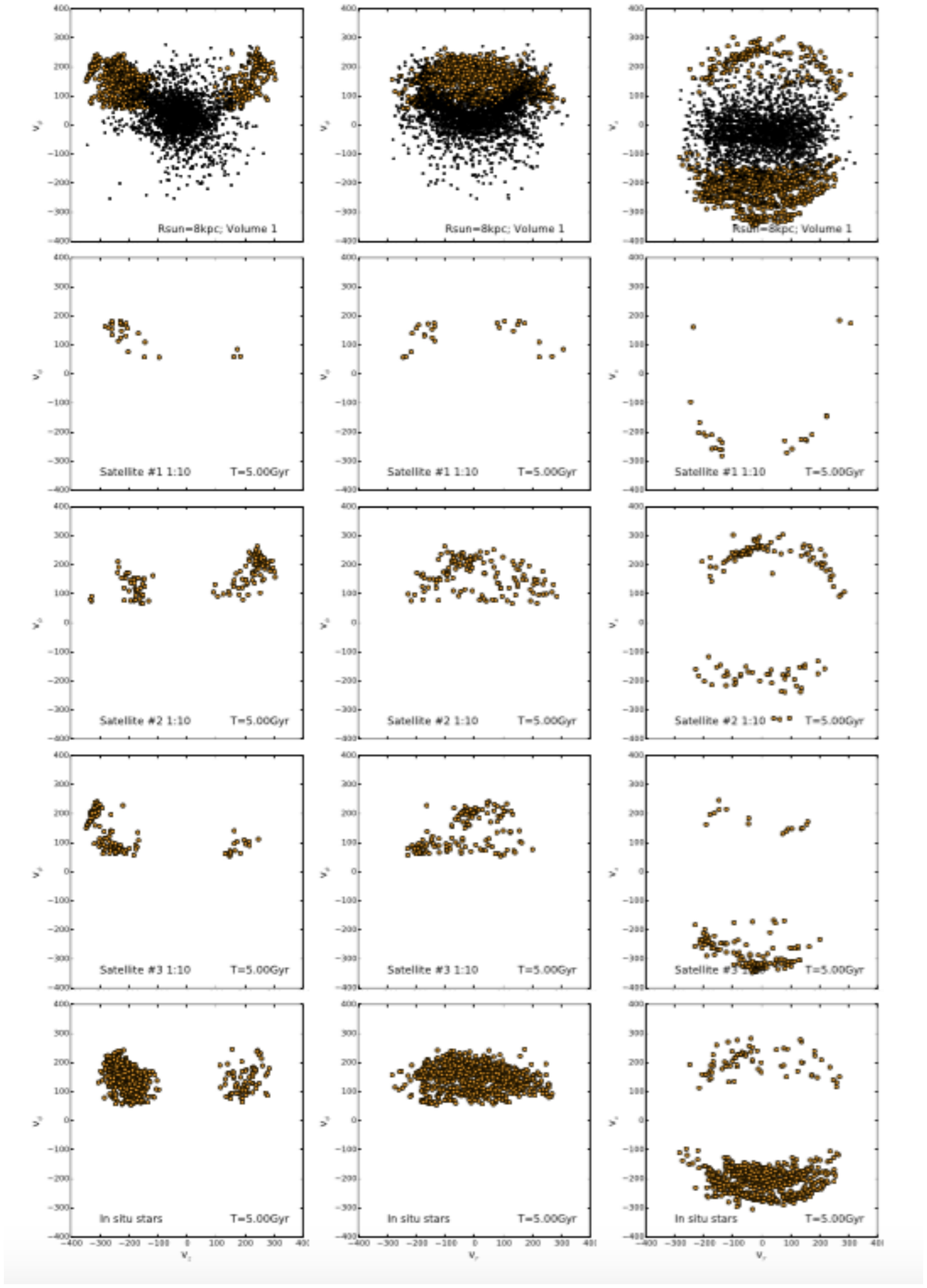}
\caption{Same as Fig.~\ref{velhelmi2}, but for the 4$\times$(1:10) merger. Velocities of accreted stars in the $v_z-v_\phi$, $v_r-v_\phi$ and $v_r-v_z$ planes are also shown for each of the satellites, separately (second, third and fourth rows). Note that the satellite 4 does not contribute with stars to the ``solar vicinity" volume shown in this figure and for this reason it is not shown}\label{velhelmi4}
\end{center}
\end{figure*}

\begin{figure*}[h!]
\begin{center}
\includegraphics[clip=true, trim = 0mm 30mm 0mm 0mm, width=2.\columnwidth]{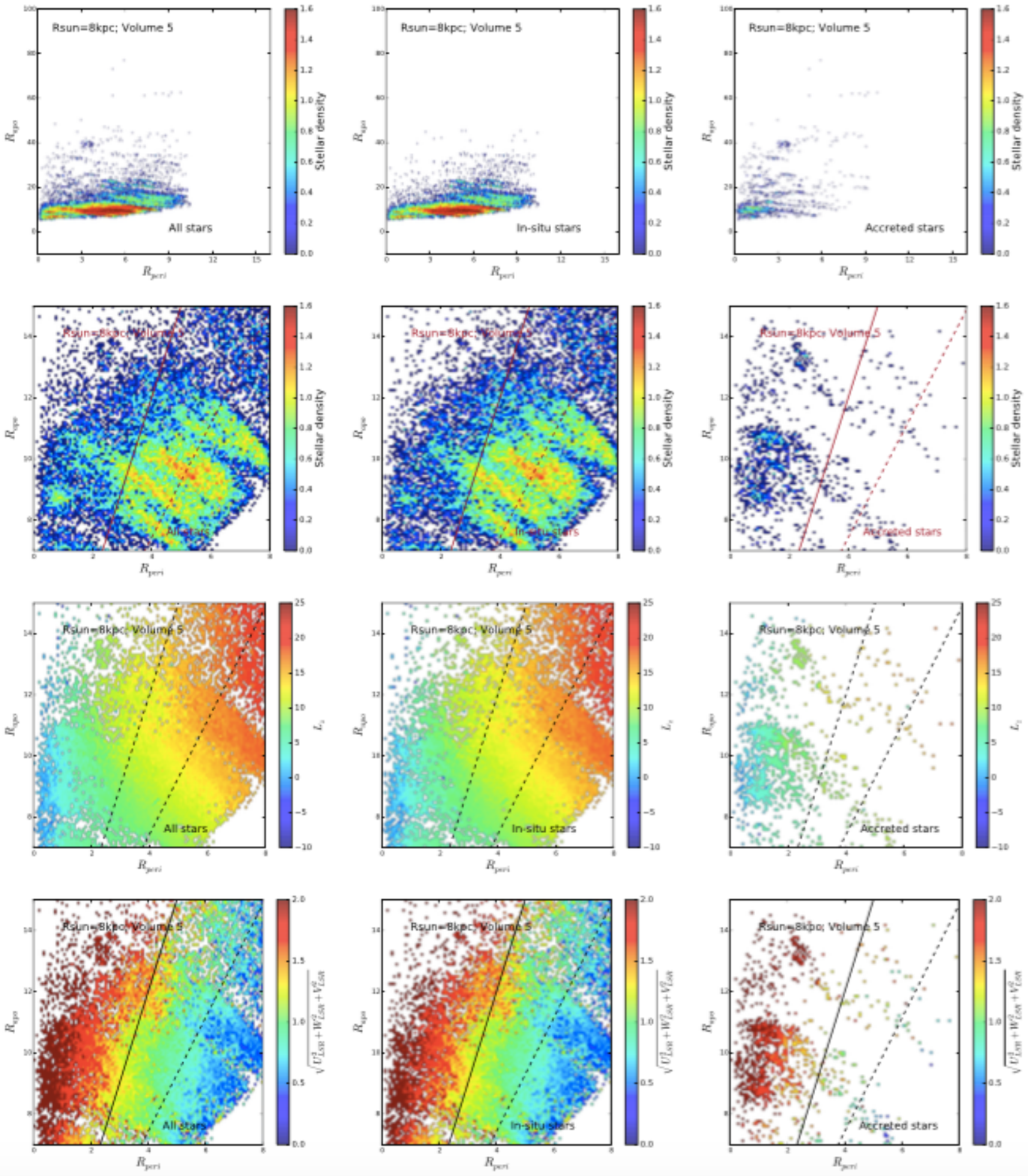}
\caption{The $R_{apo}-R_{peri}$ space for the 1$\times$(1:10) simulation and for one of the ``solar vicinity" volume studied in this paper. \emph{From left to right:} all stars, in-situ stars and accreted stars in the volume are shown, respectively. \emph{First row:} The whole extent of the  $R_{apo}-R_{peri}$ space is shown, colors code the stellar density, in logarithms scale, as indicated by the color bar. \emph{Second row:} zoom on a region with $0\le R_{peri}\le8$~kpc and $0\le R_{apo}\le 15$~kpc. The solid and dashed diagonal lines show, respectively, the line of constant eccentricity $e=0.5$ and $e=0.3$, corresponding to the region where possible accreted streams have been found in observational data of stars at the solar vicinity \citep[see][]{helmi06}.Colors code the stellar density, in logarithms scale, as indicated by the color bar. \emph{Third  row:} Same as the second row, but this time the colors code $L_z$, the $z-$component of the angular momentum. \emph{Fourth row:} same as the second row, but this time the colors code the total velocity of the stars, in the $LSR$ reference frame.}\label{MW1APL}
\end{center}
\end{figure*}

\begin{figure*}[h!]
\begin{center}
\includegraphics[clip=true, trim = 0mm 30mm 0mm 0mm, width=2.\columnwidth]{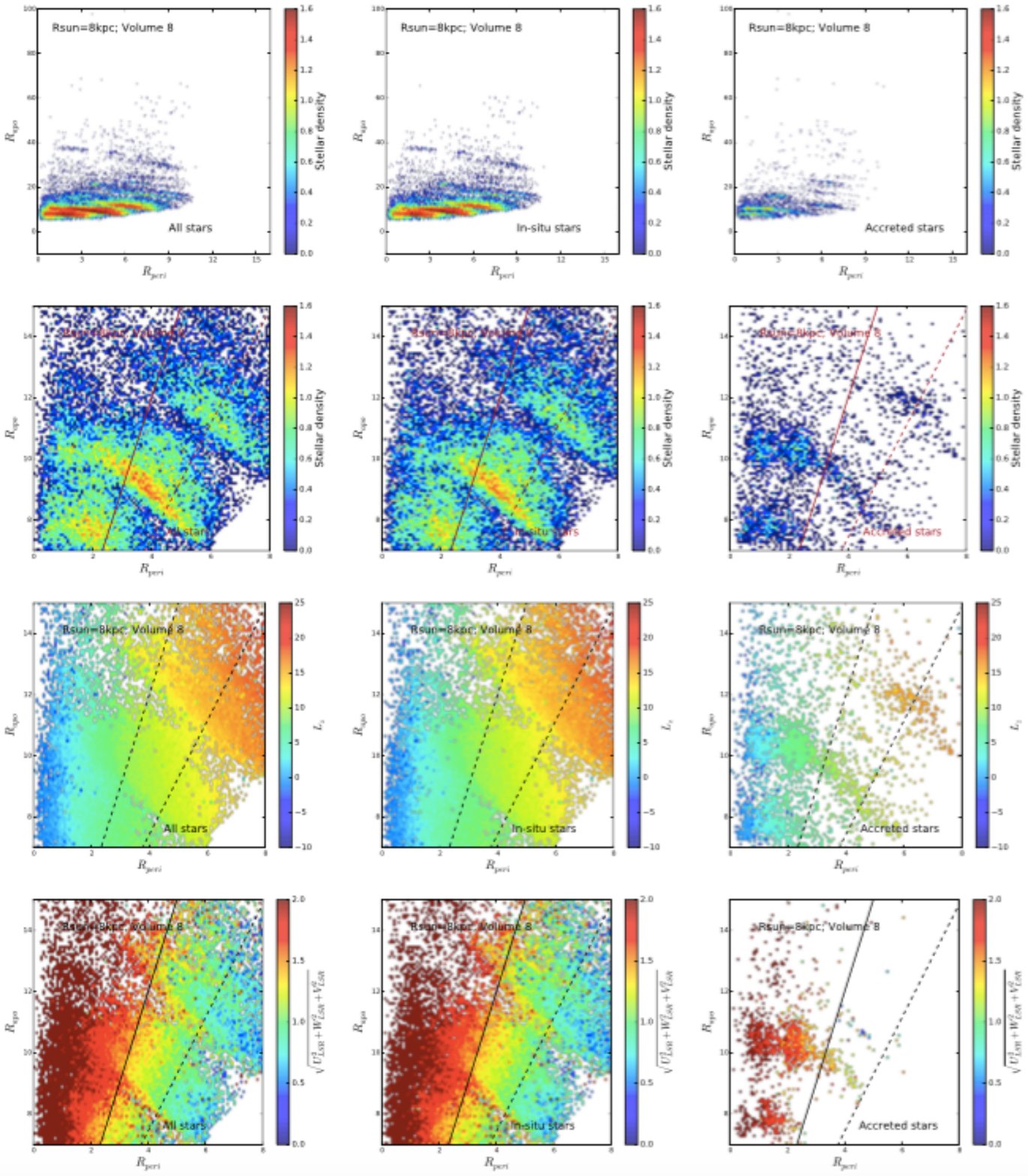}
\caption{Same as Fig.~\ref{MW1APL}, but for the 2$\times$(1:10)simulation.}\label{MW2APL}
\end{center}
\end{figure*}

\begin{figure*}
\begin{center}
\includegraphics[clip=true, trim = 0mm 30mm 10mm 9mm, width=1.7\columnwidth]{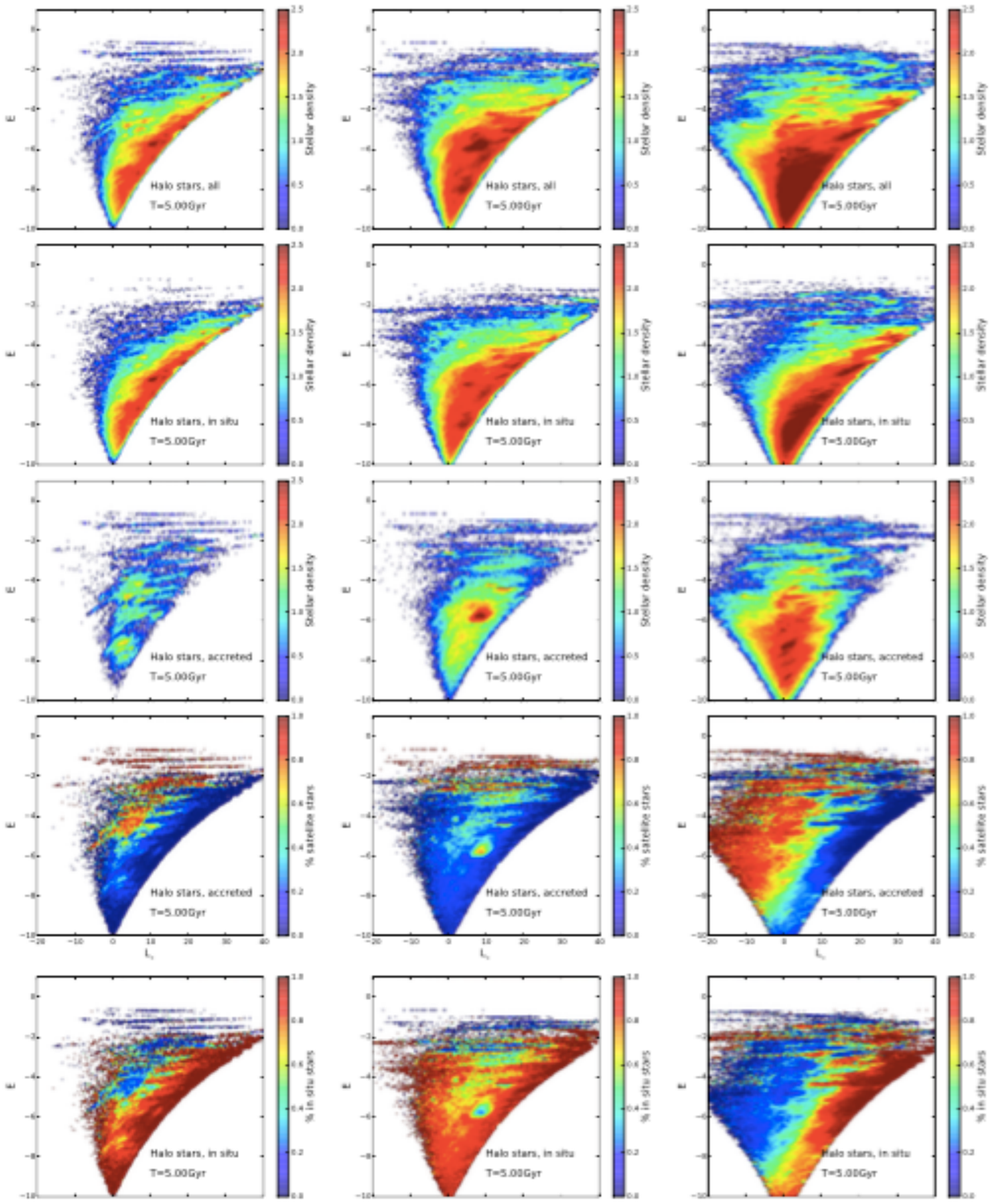}
\caption{\emph{From left to right: }Distribution in the $E -L_z$ space of halo stars in a 10~kpc volume around the Sun, for the 1$\times$(1:10) merger (\emph{left column}), for the 2$\times$(1:10) merger (\emph{middle column}) and for the 4$\times$(1:10) merger (\emph{right column}). The first, second and third row show, respectively, all stars in the selected volume, in-situ stars only and accreted stars, only. Colors code the stellar density, as indicated in the color bar. The fourth and last row show, respectively, the fractional contribution of in-situ stars and accreted stars, to the total. In all panels, to minimize the contribution of disc stars, only stars at vertical distances $z$ from the disc plane greater than 3~kpc have been selected. }\label{ELzWEAVE}
\end{center}
\end{figure*}

\begin{figure*}
\begin{center}
\includegraphics[clip=true, trim = 0mm 25mm 10mm 9mm, width=1.7\columnwidth]{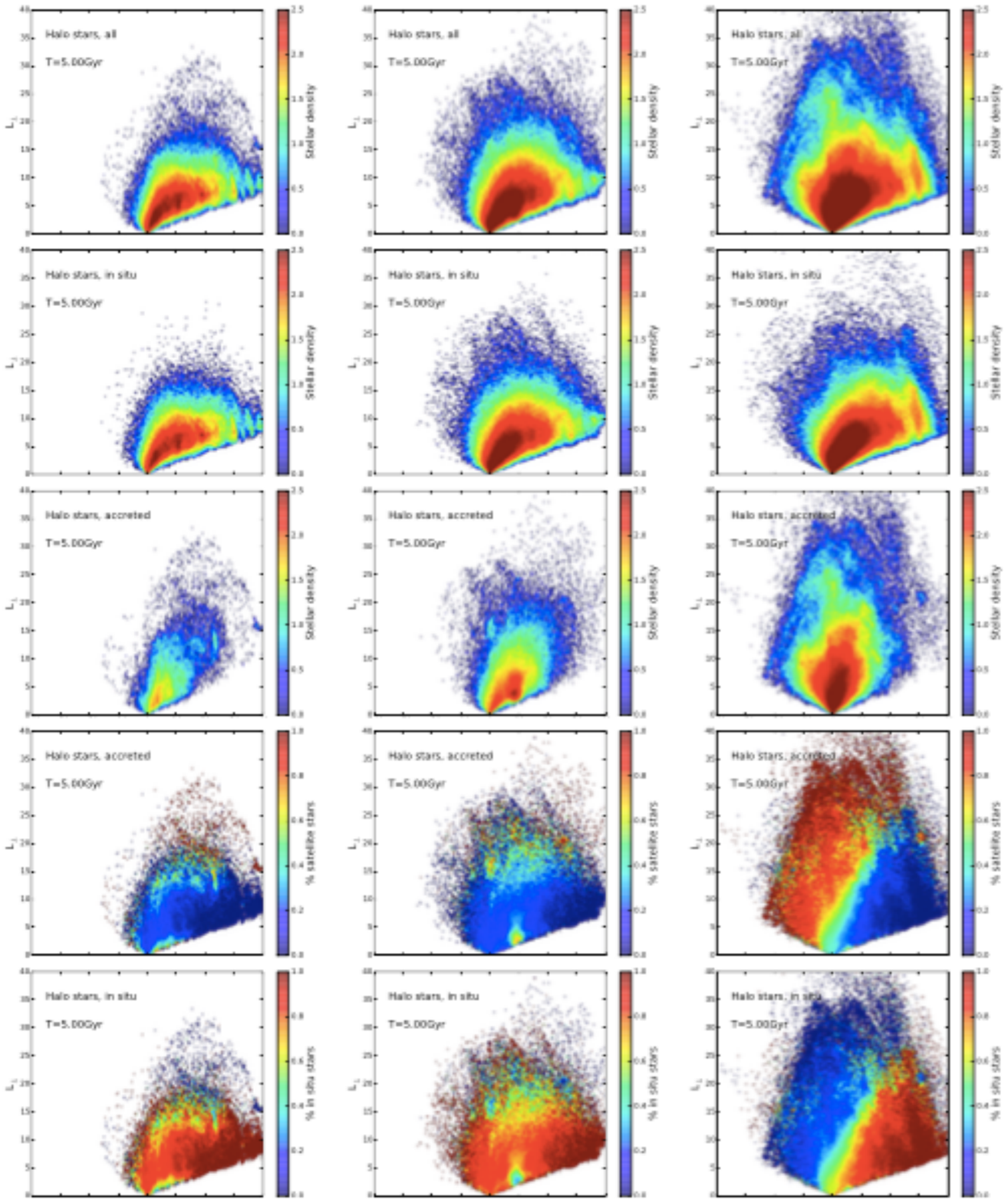}
\caption{\emph{From left to right: }Distribution in the $L_\perp -L_z$ space of halo stars in a 10~kpc volume around the Sun, for the 1$\times$(1:10) merger (\emph{left column}), for the 2$\times$(1:10) merger (\emph{middle column}) and for the 4$\times$(1:10) merger (\emph{right column}). The first, second and third row show, respectively, all stars in the selected volume, in-situ stars only and accreted stars, only. Colors code the stellar density, as indicated in the color bar. The fourth and last row show, respectively, the fractional contribution of in-situ stars and accreted stars, to the total. As in Fig.~\ref{ELzWEAVE}, in all panels, to minimize the contribution of disc stars, only stars at vertical distances $z$ from the disc plane greater than 3~kpc have been selected. }\label{LzLperpWEAVE}
\end{center}
\end{figure*}

Before moving to discuss the signature of accretion events in the APL space, it is worth commenting on the distribution of in-situ and accreted stars in velocity spaces. \\
The velocity of stars belonging to the Helmi stream was investigated in \citet{helmi99, kepley07} \citep[see also][]{klement08, helmi08, smith16}. 
In the $v_r-v_z$ (i.e. Galactocentric radial velocity -- vertical velocity) plane, stars in the stream appear as two separate groups, with $v_z$ clumping respectively at about $\pm 200\rm{km/s}$ and radial velocities distributed in the interval  $-200\rm{km/s}\lesssim v_r \lesssim 200\rm{km/s}$. The split found in $v_z$ and visible also in the $v_z-v_\phi$ plane, where $v_\phi$ is the tangential velocity, can be understood with the presence of streams of stars having similar $L_\perp$, but moving on opposite directions (i.e. inward and upward) with respect to the Galactic plane. These observations were  compared to  N-body  simulations aimed to model the accretion of a satellite galaxy in an analytic Milky Way potential in \citet{helmi99, kepley07, helmi08}. It was found that the properties of stars in the Helmi stream could be reproduced if the stream was the remnant of a satellite with an initial internal velocity dispersion of $12-18$~km/s, probably resembling the Fornax or the Sagittarius dwarf galaxy \citep{helmi99} and accreted between 6 and 9~Gyr ago \citep{kepley07}.  \\
The message one can retain from these works is that a number of information can be retrieved by looking at the kinematics alone of stars at the solar vicinity: their in-situ or extragalactic origin and - in the latter case - the mass of the progenitor satellite which generated the stream. Our models delineate a more complex scenario.\\
The $v_r, v_\phi$ and $v_z$ velocities of simulated halo stars in some of the solar vicinity volumes previously studied are shown in Figs.~\ref{velhelmi2}, \ref{velhelmi3} and \ref{velhelmi4}. As in Sects.~\ref{Elzstreams_sun} and \ref{LzLperpsect_sun}, for each solar vicinity volume in Fig.~\ref{sunvol}, we have selected halo stars as those satisfying the condition $\sqrt{V_{LSR}^2+U_{LSR}^2+W_{LSR}^2} \ge 180$~km/s. and, among those latter, stars in the Helmi region  are defined as those with  $5<L_z<15$ and $14<L_\perp<25$ (in units of 100~kpc.km/s, see previous section and the rightmost panels of Fig.~\ref{LzLperp1satRsun8part1}). Their velocities are reported in Figs.~\ref{velhelmi2}, \ref{velhelmi3}, \ref{velhelmi4} for some of the examined solar vicinity volumes and for all the simulations. The velocities of all simulated stars occupying the ``Helmi region"  is reported in the top panels of each of these figures and compared to the velocities of all stellar particles in the selected volume. One can see that the kinematic characteristics found for the Helmi stream at the solar vicinity \citep{helmi99} are recovered: our modeled ``Helmi stars" show extreme values of $v_z$ and $v_\phi$ when compared to the whole set of halo stars at the solar vicinity, they show the split in the $v_z$ distribution also found in the observations (see panels in first and last columns of Figs.~\ref{velhelmi2}, \ref{velhelmi3} and \ref{velhelmi4}) and their $v_r$ and $v_\phi$ values are similar to those attained  by the observed sample. However, the finding that the modeled stars reproduce the kinematic distribution of observed ones does not tell anything about the nature of these stars (in-situ or accreted). The central and bottom panels of Figs.~\ref{velhelmi2}, \ref{velhelmi3} and \ref{velhelmi4} indeed show that in-situ and accreted stars in the Helmi region have the same velocities: similar $v_r$, $v_\phi$ and $v_z$. In particular also in-situ stars show the characteristic split in $v_z$ observed for stars in the Helmi stream. \\ That in-situ and accreted stars in the Helmi region have the same velocities is not surprising. They belong to the same region of the $L_z-L_\perp$ plane -- as a consequence they have similar angular momenta -- and they have been selected in a limited volume around the Sun -- this limits the range of distances they span. The angular momenta and the distances being fixed and similar, this naturally leads also to similar velocities, this independently on their in-situ or accreted nature. The conclusion that we draw from this analysis is that all stars in the Helmi region should share the same kinematics, independent on  their in-situ our accreted origin. The fact that, under certain conditions, N-body models can reproduce the velocities observed in the Helmi region by means of an accreted stream is not a probe in itself that the stars observed in that region are accreted:  once the spatial volume is fixed, all stars in that volume with similar angular momenta will have similar velocities, independent on their accreted or in-situ nature. \\
Not only our models suggest that  the use of these velocity spaces cannot solve the question of the in-situ or accreted origin of the stars that make them, but also that it is not possible to recover the mass of the progenitor satellite -- if any -- from them. While \citet{helmi99} and \citet{kepley07} suggest that the progenitor of the Helmi stream may have been a galaxy similar to Fornax or Sagittarius, our models show that the kinematic properties observed for this stream can be reproduced with one (or multiple) satellite(s) significantly more massive -- each of them having an initial stellar mass  about 100 times larger than the current stellar mass in Fornax \citep{deboer12}.

\subsection{On the $R_{apo}-R_{peri}$ space}\label{APLsect}

We end our quest for accreted stars by exploring the APL, apocentre-pericentre-angular momentum, space. The distribution of solar vicinity stars from the Nordstr\"om catalogue in this space has been already explored by \citet{helmi06}, who compared the observed distribution with predictions of N-body models of satellites accreted in a fixed analytic Milky Way potential and with cosmological simulations. Since in both cases the distribution of in-situ stars -- and their possible overlap with accreted populations -- has not been investigated, we discuss the issue in this section. In the following, the apocentres and pericentres, indicated as $R_{apo}$ and $R_{peri}$ respectively, are the true apocentre and pericentre distances spanned by the particles in the last Gyr of evolution, that is, no assumption on the gravitational potential is done \emph{a posteriori} to reconstruct the orbital parameters. \\
Figs.~\ref{MW1APL}, \ref{MW2APL} show, respectively, the APL space for the 1$\times$(1:10) and 2$\times$(1:10) simulations, for all stellar particles  in a given solar vicinity volume. From these plots we can deduce the following results:
\begin{itemize}
\item as for the $E-L_z$ and $L_\perp-L_z$ spaces discussed in the previous sections, the overlap of in-situ and accreted stars is substantial also in the $R_{apo}-R_{peri}$ space, at the point that no clear distinction can be made among in-situ and accreted stars on the basis of this space, only. Similar conclusions are found for the $R_{apo}-L_z$ space (not shown).
\item for some of the volumes explored, some clumps of stars are found at large apocentre values ($R_{apo}\sim40$~kpc), but the origin of these clumps is not always extragalactic. For example, while in Fig.~\ref{MW1APL} (top row), the overdensity found at ($R_{peri}, R_{apo}$)=(3, 40)~kpc is made of accreted stars, in Fig.~\ref{MW2APL} the two evident overdensities at ($R_{peri}, R_{apo}$)=(6, 45)~kpc and ($R_{peri}, R_{apo}$)=(8.5, 38)~kpc are mostly made of in-situ stars. 
\item The global distribution of all the stars in the volume in the apocentre-pericentre space reveals the presence of several streaks, already noticed by \citet{helmi06} in their models. Each of these streaks spans a large range in eccentricities, $e$,  from $e\le 0.1$ to $e\ge 0.7$. They are visible both in the accreted and in-situ component, both on a large scale (top row panels of Figs.~\ref{MW1APL} and \ref{MW2APL}) and on a zoom-in region of the $R_{apo}-R_{peri}$ space, defined by $0\le R_{peri}\le8$~kpc and $7 \le R_{apo}\le15$~kpc, whose extent is similar to that studied by \citet{helmi06}. As we checked, these streaks are made of stars with similar energies, but which span a large extent in the vertical component of the angular momentum  $L_z$ (see Figs.~\ref{MW1APL}, \ref{MW2APL}). They are associated to spiral-like features, or rings of different extent, induced by the merger event, similarly to the structures discussed in \citet{gomez12}. These streaks are not only visible in the stellar (thin and thick) discs, as already noticed by \citet{gomez12}, but also among halo stars -- those that in the bottom rows of Figs.~\ref{MW1APL}, \ref{MW2APL} show the highest velocities ($\sqrt{V_{LSR}^2+U_{LSR}^2+W_{LSR}^2} \ge 180$~km/s) in the $LSR$ reference frame.
\item In the zoom-in region, according to our models in-situ stars occupy most of the $R_{apo}-R_{peri}$ space, independent on the number of accreted satellites and on the ``solar vicinity" volume under study. This is  true also for the region lying between the lines of constant eccentricities $e=0.3$ and $e=0.5$ (see central rows in Figs.~\ref{MW1APL}, \ref{MW2APL}), which corresponds to the region where overdensities found in in the solar vicinity data from the Nordstr\"om catalogue have been associated to accreted streams \citep{helmi06}. Our models suggest that in this region  the contribution of in-situ stars heated by the interaction and gone to populate the thick disc-inner stellar halo, is substantial.
\item Finally, for the overdensities found in the $R_{apo}-R_{peri}$ distribution of stars in the Nordstr\"om catalogue, that follow a diagonal pattern at eccentricities $0.3\le e \le 0.5$ and for which an extragalactic origin has been suggested, a trend has been reported by \citet{helmi06}: these stars tend to have progressively larger $L_z$, as the $R_{apo}$ and $R_{peri}$ increase. We notice from the bottom panels of Figs.~\ref{MW1APL}, \ref{MW2APL} that in our models this trend -- an increase of $L_z$ with $R_{apo}$ and $R_{peri}$ -- is found among accreted stars and among the in-situ population, as well. 
 \end{itemize}
To conclude, both the diagonal distribution of stars in the $R_{apo}-R_{peri}$ space and the presence of streaks or overdensities and the trend observed for the angular momentum $L_z$ with $R_{apo}$ and $R_{peri}$, all these features are common both to accreted and in-situ stars. Without detailed chemical abundances, the differentiation of the extragalactic or in-situ origin of stars in this space is --according to these results-- unfeasible. \\
Similar results are found also for the 4$\times(1:10)$ simulation.

\subsection{Leaving the solar volume: kinematic detection of streams on a 10~kpc scale}

Most of the previous analysis, except that presented in Figs.~\ref{ElzsatGLOBtime}, \ref{ElzsatGLOB} and  \ref{ElzGLOB}, has been performed selecting  stellar particles in restricted solar volumes (3~kpc in radius). These are the typical volumes over which the kinematic search of stellar streams has been focused until now \citep{helmi99, kepley07, morrison09, smith09, refiorentin05, refiorentin15}.  In the next years, with Gaia and spectroscopic related surveys like GALAH, WEAVE, 4MOST, we will have an almost complete view of stars up to  $\sim$3~kpc from the Sun: distances, kinematics, ages and detailed abundances. This full set of quantities should thus provide enough information to separate patterns and reconstruct the origin and the formation sites of ``local" stars. However, beyond the $\sim3$~kpc-sphere, at larger distances from the Sun, uncertainties in dating stars will possibly be too large to determine ages with sufficient accuracy\footnote{Under the hypothesis that the atmospheric parameters will not be the limiting factor in dating stars, 10\% errors or better in age estimates are expected to be achieved within $\sim3$~kpc from the Sun}, detailed abundances  from large spectroscopic surveys with precision  $\sim$0.03--0.05~dex   will also be difficult to obtain\footnote{Abundances with precision of $\lesssim$0.1~dex will be probably achievable only inside a sphere of few kpc from the Sun. We require a precision of 0.03--0.05~dex, because we need to detect abundance differences 
on the order of 0.1~dex  \citep[see][]{nissen10}.}. It is thus at those scales (typically between 3 and 10~kpc from the Sun), that it is fundamental to understand if the search for the fossil records of the Galaxy by kinematics alone is feasible and meaningful. \\
In Fig.~\ref{ELzWEAVE}, the analysis of these extended solar volumes, up to distances of 10~kpc from the Sun, is shown, for the $E-L_z$ plane.  For each simulation, we have selected all stars with distances inside 10~kpc from the Sun, where four different Sun positions have been chosen, corresponding to the centres of volumes 1 and 5 in Fig.~\ref{sunvol}, at distances of 8~kpc and 12~kpc from the galaxy centre. From this analysis we have also excluded stars at vertical distances $z$ from the plane below 3~kpc, to maximize the fraction of halo stars in the examined samples. In Fig.~\ref{ELzWEAVE} only one of these volumes is shown, the conclusion of the analysis being the same for all the four volumes examined. This figure shows that all the problems that affect the local samples are also present in a 10~kpc--extended solar volumes: significant overlap of in-situ and accreted stars, in-situ stars that dominate a large part of the $E-L_z$ spaces, non-smooth distribution for both populations. Making the blind test of looking at any of the distributions shown in the top row of Fig.~\ref{ELzWEAVE}, it appears impossible to establish how many different satellites have contributed to determine those distributions, the masses of the progenitor systems and which fraction of the lumpy regions is made of in-situ and accreted stars, respectively. Similar conclusions are reached for the $L_\perp-L_z$ space (see Fig.~\ref{LzLperpWEAVE}).

\section{Discussion}\label{discussion}
\subsection{In-situ halo stars, the elephant in the room}

All the results presented in the previous sections strongly point to a problem: the kinematic search for streams of accreted satellites in the Milky Way cannot be achieved neglecting the presence of the in-situ stellar population. In the simulations presented in this work -- as recalled several times in this manuscript -- there is no in-situ stellar halo before the interaction(s). The in-situ halo found at the end of the simulations is only the result of the heating of a pre-existing stellar disc and no other channel for the formation of the in-situ halo -- as those described by \citet{cooper15} -- has been taken into account. That heating from satellite accretions is effective in forming -- or contributing to form -- a stellar halo has been already pointed out in a number of papers \citet{zolotov09, purcell10, font11, qu11, mccarthy12, cooper15}. But here we show that the consequence of this heating is important for any search in integrals--of--motion or kinematic spaces: 
\begin{itemize}
\item \emph{heated halo stars are not smoothly distributed:} this result is in contradiction with the usual assumption made in the literature that the in-situ halo should be smooth in those spaces. At least the part of the in-situ halo that results from satellite heating is, on the contrary, structured, even several Gyrs after the accretion has taken place. Halo stars with a clustered distribution in kinematic or integrals-of-motion spaces are not necessarily accreted. The probability that these clumps lie in regions dominated by in-situ stars is indeed very high (see Tables~\ref{tablefrac1}, \ref{tablefrac2}, \ref{tablefrac4}, \ref{tablefrachelmi1}, \ref{tablefrachelmi2}, \ref{tablefrachelmi4}).
\item\emph{heated halo stars rotate:}  this result is fundamental for all studies that look for accreted streams in the $L_z-L_\perp$ space and discriminate  between the in-situ and the accreted populations by assuming that the in-situ halo should not rotate. Our models indeed show that this assumption -- controversial from the observational point of view \citep[see, for example, ][]{chiba00, carollo07, kepley07, smith09, an15} -- is not valid at least for the part of  in-situ halo made of heated  thin/thick disc stars. As we show in Fig.~\ref{LzLperp1satRsun8part1}, indeed, the distribution of in-situ halo stars in the $L_z-L_\perp$ space presents an excess of prograde orbits (i.e. positive $L_z$). As a consequence the distribution of those in-situ halo stars is not symmetric with respect to the $L_z=0$ axis. This asymmetry has been found in observation samples of stars at few kpc from the Sun, for prograde orbits with high values of $L_\perp$. Stars in these regions have been interpreted as accreted stars exactly because it has been assumed that any in-situ population should be not-rotating and thus symmetric with respect to the $L_z=0$ axis. Contrary to this common assumption, here we show that in-situ heated halo stars are expected to rotate. The amount of rotation depends on the number of accretion events and on the total accreted mass: the larger the accreted mass, the slower the in-situ halo rotates. 
\item \emph{heated stars overlap with accreted stars:} this finding affects all the spaces studied in this paper -- $E-L_z$, $L_\perp-L_z$, $APL$ -- and the proportion of in-situ stars is so important in each of these spaces ( see, for example, Tables~\ref{tablefrac1}, \ref{tablefrac2}, \ref{tablefrac4}, \ref{tablefrachelmi1}, \ref{tablefrachelmi2}, \ref{tablefrachelmi4} and Figs.~\ref{ElzGLOB}, \ref{ELz1satRsun8part1}, \ref{ELz1satRsun8part1_halo}, \ref{ELz2satRsun8part1_halo}, \ref{ELz4satRsun8part1_halo}, \ref{LzLperp1satRsun8part1}, \ref{velhelmi2}, \ref{velhelmi3}, \ref{velhelmi4}, \ref{MW1APL}, \ref{MW2APL}) to lead us to conclude that the search for accreted streams in those spaces is mostly inefficient, simply because the probability to find in-situ stars in a given region of those spaces is significant everywhere and it is not possible to define \emph{a priori}  where the chance to find accreted stars is the highest. The extension and location of the region dominated by accreted stars depend indeed on a number of parameters not known -- parameters that indeed we would like to constraint by a search in kinematic spaces -- as the number of accreted satellites, their masses and their orbital properties. 
\end{itemize}

To the difficulty in separating in-situ and accreted stars, we need to add also the difficulty in recovering the properties of the accreted satellites (in particular their number and their masses) from their clumpy distribution in kinematic and integral-of-motion spaces. For example, Fig.~\ref{ElzsatGLOB} shows that -- even in the ideal case where we are able to separate in-situ from accreted stars -- the number of ripples and clumps found for the accreted population in the $E-L_z$ space does not reveal anything  about the number of progenitors beyond this distribution: a simple 1:10 merger is able to produce tens of distinct structures in this space, each of these with sizes and densities similar to those generated by a 1:100 accretion (see following of this Section). This finding is common to all the analyzed spaces. Another example is given by the $v_r-v_\phi$, $v_z-v_\phi$ and $v_r-v_z$ spaces: on the basis of the velocities found for accreted stars at the solar vicinity (their values and characteristics, see Figs.~\ref{velhelmi2}, \ref{velhelmi3}, \ref{velhelmi4}), it is not possible to derive the number of accretion events which contributed to shape the accreted population. 

\subsection{Chemistry and ages, the key for differentiating the origin of stars in the Galaxy}

\begin{figure}
\begin{center}
\includegraphics[clip=true, trim = 0mm 0mm 0mm 0mm, width=1\columnwidth]{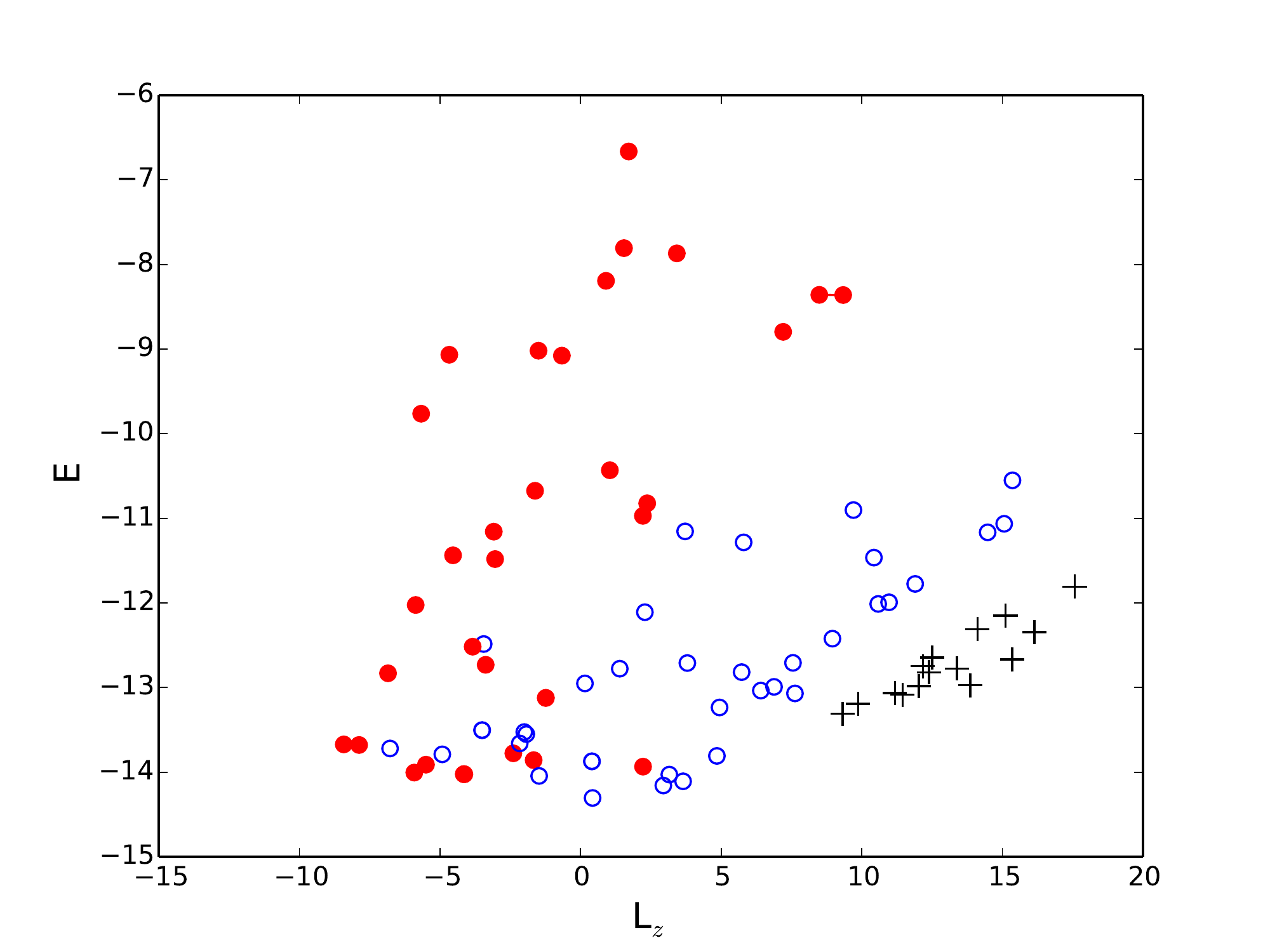}
\includegraphics[clip=true, trim = 0mm 0mm 0mm 0mm, width=1\columnwidth]{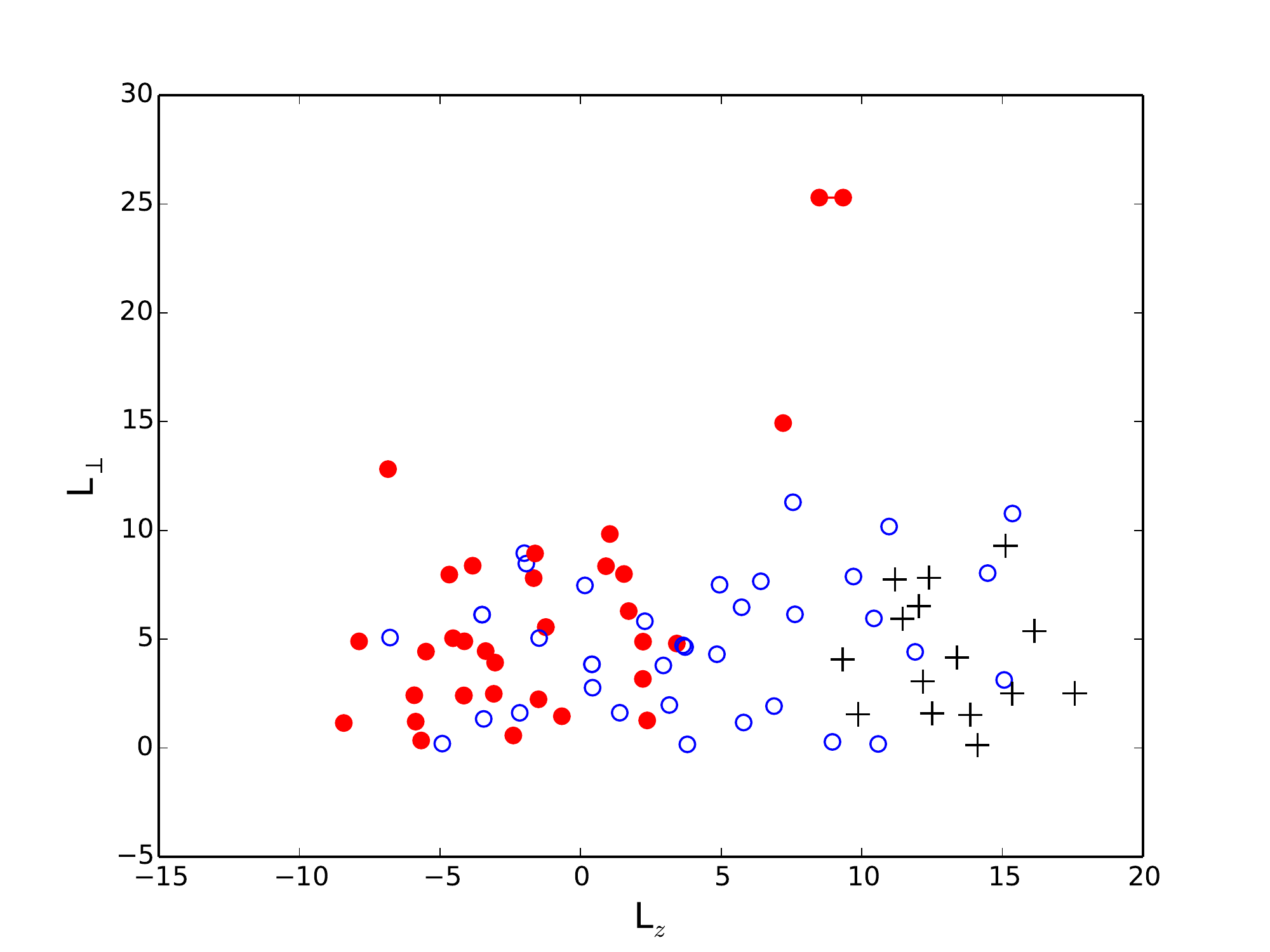}
\includegraphics[clip=true, trim = 0mm 0mm 0mm 0mm, width=1\columnwidth]{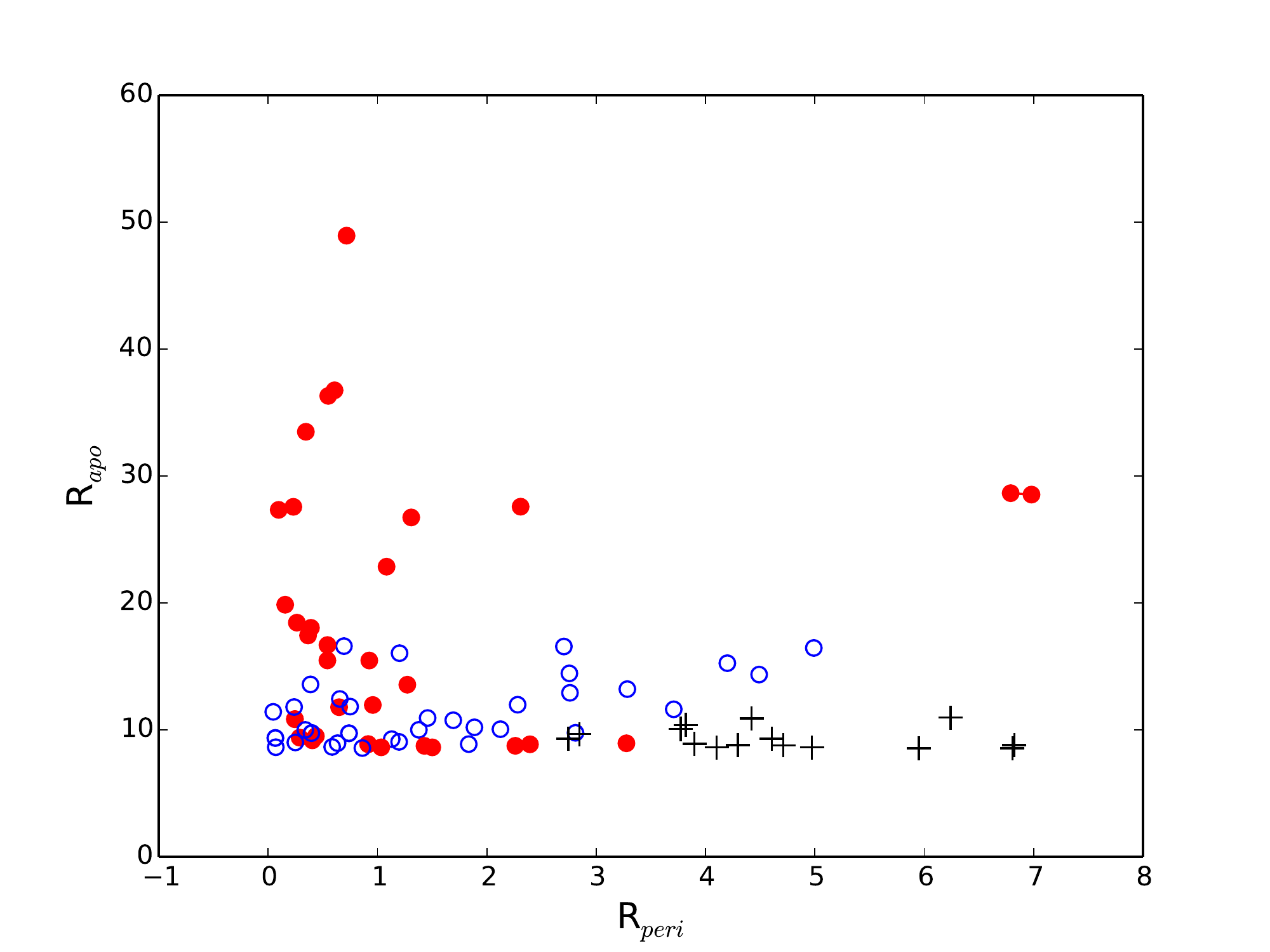}
\caption{\emph{From top to bottom:} Distribution in the $E-L_z$, $L_\perp-L_z$ and $R_{min}-R_{max}$ spaces of halo and thick disc stars from the \citet{nissen10} sample.  As in \citet{nissen10}, thick disc stars are indicated with black crosses,  ``high-$\alpha$" halo stars with empty blue circles and ``low-$\alpha$" halo stars with red filled circles. Stars with an uncertain classification (see Tables 3 and 4 in \citet{nissen10}) have not been included. As in \citet{nissen10},  the components of the visual binary star, G 112-43 and G 112-44, are connected by a straight line.  Energies are expressed in units of $10^4$~km$^2$/s$^2$, angular momenta in units of 100~km/s/kpc, distances in kpc. }\label{nissen}
\end{center}
\end{figure}

How to overcome this degeneracy ? One may be tempted to think that a simple cut in metallicity (for example, selecting only stars with [Fe/H] $\lesssim -1 $~dex) would be sufficient to eliminate the contamination of in-situ stars. The reasoning behind this choice would be to think that heated halo stars -- previously in the disc -- should have metallicities typical of the Milky Way disc, that is [Fe/H]$\ge -1$~dex. Following this reasoning, a cut in metallicity that removes the metallicities of current disc stars would allow  to remove the in-situ contamination. However this reasoning risks to be incorrect, because the metallicities expected for the heated halo stars would be the metallicities of the disc \emph{at the time the accretion events took place and not the metallicities of the current (i.e. redshift $z=0$) Milky Way disc}. Thus the metallicities of the in-situ heated halo population critically depend on the time relatively massive mergers (mass ratio $\lesssim$ 1:10) took place. In the cosmological simulations analyzed by \citet{tissera12}, for example, in-situ and accreted inner halo stars   have similar metallicities (see Table~5, in their paper), generally below [Fe/H]$=-1$~dex. Any cut on the basis of metallicity only would thus not be able to remove the in-situ population, simply because part of this population has been possibly kinematically heated in the halo at early times -- when the most massive mergers took place -- and thus has metallicities typical of current halo stars. 

To solve the question of the origin of stars in the Galaxy, their site and mode of formation, we need to look for detailed chemical abundances and/or ages. $\Lambda$CDM models indeed predict that several chemical and chemo/age patterns should be expected in Milky Way-like galaxies, for stars originating in different environments \citep[see][for some recent  works]{gibson13, few14, snaith16}. However, the ability to find and differentiate those patterns in future spectroscopic samples of stars crucially depends on the precision on the abundance estimates that future surveys will be able to reach  \citep[see Figs.~12 and 13 in][]{snaith16}. From solar vicinity sample of stars, where abundances with precisions below 0.1~dex have been attained, it has been  possible to reveal a number of chemical (or chemical/age) patterns.  The dominant patterns seem characteristic of the evolution of the in-situ galactic stellar populations \citep{haywood13, spina16, nissen16}, with possible dilution from gas accretion in the outer disc \citep{haywood13, snaith15}, while some others possibly have an extragalactic origin \citep{nissen10}. This approach seems very promising and its predictive power -- in terms of ability to separate populations of different origins -- seems to surpass that of kinematic diagnostics, as it can be evinced already from stellar samples at the solar vicinity (see next). 

\subsection{An illustrative example: the mapping of a chemically selected, local stream into kinematic spaces}

To illustrate the potentiality of the search for accreted/in-situ populations when detailed chemical abundances are available, in Fig.~\ref{nissen} we show, respectively, the distribution in the $E-L_z$, $L_\perp-L_z$ and $R_{apo}-R_{peri}$ spaces of halo and thick disc stars at the solar vicinity from  the sample studied by \citet{nissen10} and subsequently in \citet{nissen11, nissen12, schuster12}.Why did we choose this sample ? Because by studying  detailed chemical abundances, \citet{nissen10} show that the halo stars that make it can be separated into two distinct populations, on the basis of their [$\alpha$/Fe] content \citep[see Fig.~1, ][]{nissen10}. The ``high-$\alpha$" sequence appears in chemical continuity (with some overlap)  with the thick disc and -- as proposed by \citet{nissen10} -- it may be made of  stars originally in the disc or bulge of the Galaxy and subsequently heated by merging satellites.  The currently favored hypothesis for the  ``low-$\alpha$" sequence is that it is made of  stars accreted from dwarf galaxies, with some of them possibly associated to $\omega$ Cen progenitor. This sample thus very probably contains both an in-situ and an accreted halo population and since parallaxes, radial velocities and proper motions are  available \citep[see][for details]{nissen10, schuster12}, we can derive orbital properties and  make use of them as a natural testbed to probe the efficiency of the detection  of accreted streams by kinematic and integrals-of-motion diagnostics. 

To this aim, we have integrated the orbits of all stars in this sample in the \citet{allen91} Galactic potential for 7~Gyr.  We show the value of their energies, angular momenta, $R_{apo}$ and $R_{peri}$ in Fig.~\ref{nissen}.  This figure clearly shows that \emph{in none of those kinematic spaces the sequence of ``low-$\alpha$" accreted halo stars shows up as a cluster or substructure distinct and separated from the whole sample}. As already evidenced by \citet{schuster12}, it is true that the ``low-$\alpha$" sequence shows, on average, larger values of energies and $R_{apo}$ than those reached by the ``high-$\alpha$", in-situ, sequence, but while the kinematics can be used to study differences among those (chemically identified) populations, it is clear that it cannot be used to identify the accreted stream and separate it from the bulk of the in-situ population. Without marking the different populations with different colors, in none of the spaces shown in Fig.~\ref{nissen} would it be possible to recognize and separate the two halo populations. There is a not negligible overlap between the accreted and in-situ populations: this is particularly striking in the $L_\perp-L_z$ plane, where all the accreted stars overlap with the in-situ population, except for two stars  \citep[i.e. G112-43 and G112-44, a couple of visual binaries, see][for their chemical properties and orbit]{nissen10, allen00} which lie in the upper-right part of the diagram, at values of $L_z$ and $L_\perp$ comparable to those of the Helmi region.\\
This example thus illustrates  the (general) inefficient power of kinematic diagnostics in finding accreted streams and, at the same time, how promising the search for accreted streams in chemical abundance spaces can be\footnote{But note that for this, we need very high quality spectroscopic data, the separation in [$\alpha/Fe$] between the two halo sequences found by \citet{nissen10} being typically less than  0.1~dex.} when compared to kinematic detections (cf. Fig.~1 in \citet{nissen10} with Fig.~\ref{nissen} in this paper). The power of this approach is considerable and it is expected to increase  once ages and detailed chemical abundances of several hundreds thousand stars will be available.

\subsection{The kinematic signature of mergers with a mass ratio $\sim$1:100}\label{1su100sect}

\begin{figure*}
\begin{center}
\includegraphics[clip=true, trim = 0mm 0mm 10mm 9mm, angle=270,width=2.\columnwidth]{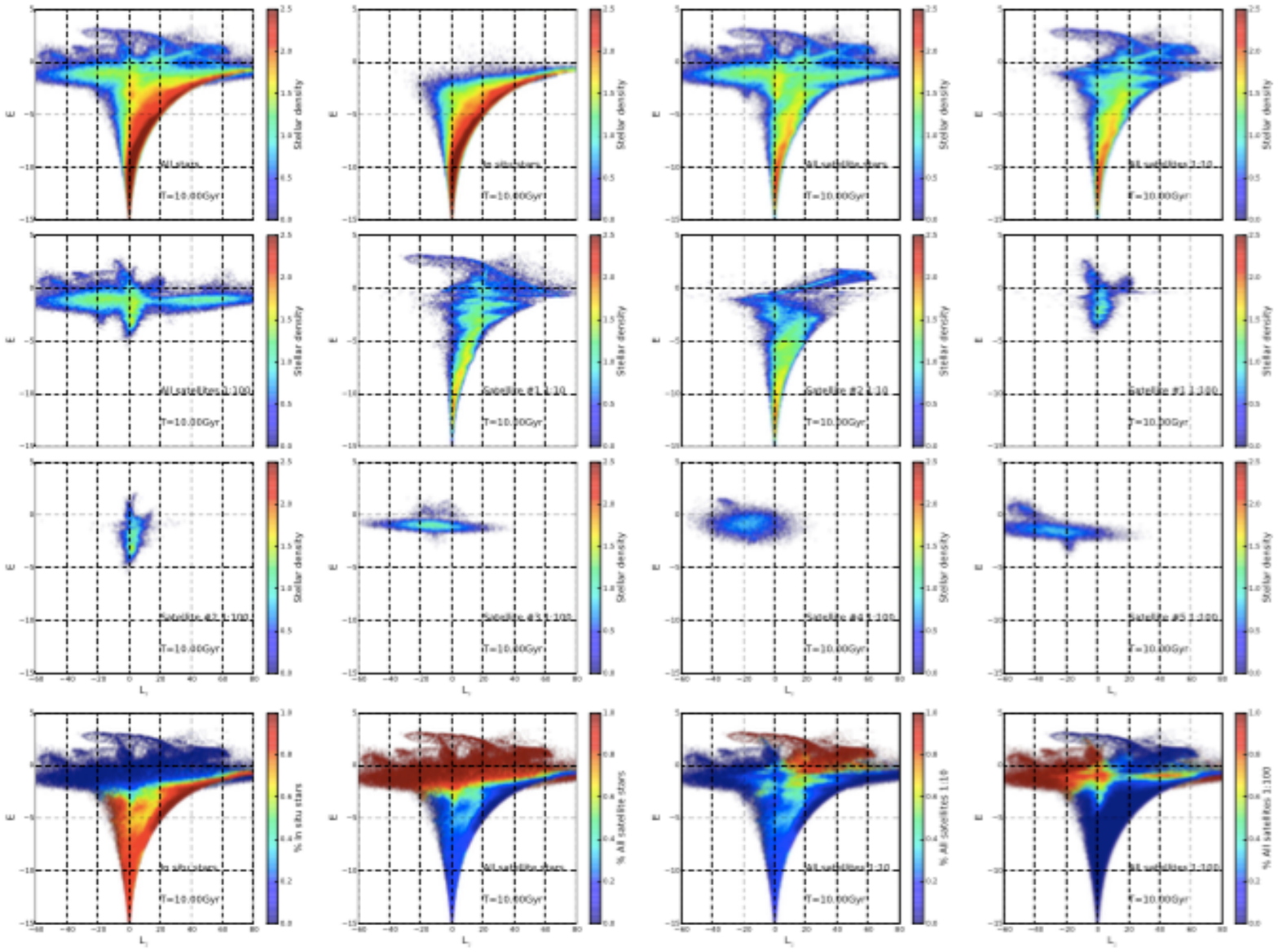}
\caption{\emph{Mergers with a 1:100 mass ratio.} Distribution in the $E-L_z$ space of stars for the simulation of a $2\times(1:10)+8\times(1:100)$ merger.  In the first row, all stars (\emph{first column}), in-situ stars (\emph{second column}), all accreted stars (\emph{third column}) and all accreted stars from the 1:10 mass ratio mergers (\emph{last column}) are shown. In the second row, the distribution of all accreted stars from 1:100 mass ratio mergers is shown (\emph{first column}), followed by the distribution of the two 1:10 mass ratio satellites, separately (\emph{second and third columns}). The distribution in the $E-L_z$ space of some of the 1:100 mass ratio satellites, separately is also shown (\emph{second row, last column and all panels in the third row}). Finally, in the last row, the fractional contribution to the total distribution of in-situ stars (\emph{first column}), all accreted stars (\emph{second column}), all accreted stars from 1:10 mergers (\emph{third column}) and all accreted stars from 1:100 mergers (\emph{last column}) is shown. In all panels, the $E-L_z$ distributions are shown after evolving the 8$\times$(1:100) satellites for 10~Gyr, interacting with the remnant of the 2$\times$(1:10) merger (see Sect.~\ref{1su100sect}).}\label{1su100}
\end{center}
\end{figure*}

Mergers with mass ratios similar to those analyzed in Sect.~\ref{results} appear inevitable for Milky Way-type galaxies in  $\Lambda$CDM simulations \citep{read08, stewart08, delucia08, deason16}. However, these simulations also predict that accretions of satellites with smaller masses should also take place. Would it be possible to recover these low-mass accretions in kinematic or integrals-of-motion spaces more easily than what our models suggest for more massive satellites ? In the $E-L_z$ space,  for example, we have seen that the non-conservation of energy and $L_z$ is mostly driven by dynamical friction. Because the latter depends on the mass of the accreted galaxy, one would expect that low mass satellites should conserve their initial location in the $E-L_z$ space, without experiencing the significant changes in energy and angular momentum quantified for the more massive systems in Sect.~\ref{ELzsect}. Motivated by this question and by the need to quantify the ability to recover accretions with low mass ratios in a Milky Way-type galaxy by means of  kinematics alone, we have run some supplementary simulations, where the Milky Way type galaxy -- after experiencing a 1x(1:10) or a 2x(1:10) merger -- accrete four or eight satellites with mass ratio 1:100. More in detail, we have taken the final snapshot of the simulations 1x(1:10) and 2x(1:10) analyzed in Sect.~\ref{results} and for each of these two new ``initial conditions" we have run two simulations, accreting respectively, 4 or 8 satellites with masses which are 1/100 smaller than that the Milky Way-type galaxy initially (before the 1:10 mergers) had. The low mass satellites are rescaled version of the Milky Way-type galaxy: they masses are a factor 100 smaller, their sizes have been divided by a factor of 10. For computing time reasons, for each of these simulations we have reduced the number of particles of both the Milky Way type galaxy and its satellites by a factor of 10 and rescaled the masses of the single particles accordingly. This means for example that the 1x(10)+4x(1:100) merger contains about 2.8 million particles in total, instead of 28 million particles, expected if we had used the resolution adopted for all simulations described in Sect.~\ref{results}.   Each low mass satellite, in particular, is made of 25000 particles and does not contain any  population of globular clusters. 
The lower resolution employed to run these simulations does not allow to appreciate and study the fine structure observed in the integrals-of-motions spaces discussed in  Sect.~\ref{results}. However, it is still possible to quantify the contamination and overlap between in-situ stars and the different accreted stellar populations. An example of the difficulty in deciphering the signature left by low mass mergers  is given in Fig.~\ref{1su100}. In this example, the Milky Way type galaxy, after experiencing two mergers with a mass ratio 1:10, accretes eight satellites with mass ratios 1:100. The distribution of all stars in the $E-L_z$ space is shown here, independently of their location in the galaxy. The points that we want to emphasize from the analysis of these supplementary simulations are the following:
\begin{itemize}
\item as found in the previous section, even in the case of a 2$\times$(1:10)+8$\times$(1:100) merger, as that shown in Fig.~\ref{1su100}, most of the $E-L_z$ space is dominated by in-situ stars. The only region where the contribution of satellites becomes dominant is that at energies $E \gtrsim -2$. Note however that this region (that dominated by accreted stars, i.e. $E\gtrsim -2$) would be inaccessible if we had restricted the search to a sphere of $\sim 10$~kpc from the Sun.  
\item Low mass satellites are characterized by structures, in the $E-L_z$ plane, comparable to one of the several generated by a more massive accretion. Indeed, if we compare the overdensities generated by 1:100 mass ratio mergers, they are similar in sizes and densities to one of tens of clumps generated by a 1:10 merger (cf., for example the distribution of the satellites 1 and 2 with mass ratio 1/100 to part of the distribution generated by a 1:10 merger). 
\end{itemize}
To conclude, even if in general mergers with mass ratios $\sim$ 1:100 tend to better preserve their initial distribution in the $E-L_z$ space, it is enough that the Milky Way had experienced one of few 1:10 mergers in its past to render the $E-L_z$ space hardly decipherable. Mergers with such mass ratios indeed erase the signature of the less massive ones, because they dominate in mass and  because each of the overdensities they generate in the $E-L_z$ space are indistinguishable from those generated from less massive accretions.

\subsection{Limitations of the modelling presented in this work}\label{our limits}

Before moving to the main conclusions of this work, it is necessary to remind its main limitations that we wish to address in future studies and discuss some possible avenues to  overcome them. \\

A limitation comes from the \emph{limited range of orbital parameters and internal parameters} (such as : satellite to MW-type galaxy mass ratios, M/L ratios and relative number of particles for the baryonic and dark matter components, etc) explored. Even if a more extended parameters space is suitable, with initial conditions possibly taken from cosmological simulations, we think the current choice does not impact the main results of this work, that is the inner halo formation by disc heating and the overlap of in-situ and accreted stars in kinematic and integrals-of-motion spaces. Even if not specifically focused on integrals-of-motion spaces, several works have been investigating  the impact of satellite accretions in heating pre-existing stellar discs and the resulting kinematic properties of in-situ and accreted populations (a not exhaustive list is given in Sect.~\ref{overlap}). The results presented in those papers have explored a large interval of internal and orbital parameters, showing in particular that the fraction of satellite stars over the total number of stars
found at a given height above the plane of the remnant galaxy 
depends very weakly on the orbital conditions of the merger \citep[see Fig.~16 in][]{villalobos08}. This gives us confidence that even if we have simulated
only few orbits, our results do not depend strongly on the choice of the
orbital parameters. Note that also the thickness of the final disc is
mostly independent on the prograde/retrograde orbit \citep[again, see][]{villalobos08}. On the basis of these results, we fell confident that this point does not constitute a main limitation to our conclusions. \\
A second limitation of our modelling consists in the \emph{absence of an in-situ halo population in the Milky Way-type galaxy that pre-exists to the accretion event(s)}.  Indeed, as we have recalled several times in this paper, in our simulations the in-situ halo is exclusively made of  in-situ disc stars heated by the interaction and  no in-situ halo component pre-exists to the accretion(s). It will be thus important to introduce it in future simulations, to quantify its response to the accretion(s), its distribution in integrals-of-motion spaces and the degree of substructures it can show in those spaces. The addition of this component to the modelling will naturally require to explore a number of different parameters (flattening, rotation, concentration among others), to understand their impact on the final result. \\
A third limitation, which is probably the most urgent to investigate, is the \emph{lack of a gaseous component both in the Milky Way disc and in the satellite(s)}. It is not easy to anticipate how the inclusion of gas in the simulations can impact the results. On the one side, it has been shown \citep{moster10, qu11} that the presence of a thin gaseous disc can limit the amount of disc heating: for gas fractions of the order of 20\%(40\%), the heating may be reduced by 25 per cent (40 per cent). For accretions occurred in the early history of the Galaxy, typically before redshift $z\sim1$, the fractions of gas  in the Milky Way disc may have been significantly higher (with $f_{gas}\ge$50\%) than those simulated in the above cited works. Moreover, in those early phases of the Galactic evolution it is not even clear how thin the gaseous disc may have been.  While in nearby disc galaxies indeed gaseous discs have scale heights between few tens and 100~pc and velocity dispersions of the order of $\sim$10~km/s, at higher redshift  gaseous discs of about 1~kpc (corresponding to velocity dispersions of the order of 100~km/s) have been observed in the continuum and in the gas component \citep{elmegreen06, epinat12}. In the N-body models cited above, the typical scale heights employed for the gaseous component are between 200 and 400~pc. It is thus crucial to quantify the heating of stellar discs and the efficiency in forming an in-situ halo  also in more  extreme conditions than those modeled so far, also because galactic stellar archeology seems to suggest that the Milky Way may have experienced an intense phase of star formation and significant gas turbulence at those epochs \citep{haywood13, snaith14, lehnert14}.  If the inclusion of a gaseous component in the Milky Way disc seems inevitable to simulate accretion episodes that occurred at early times, the same is valid for modelling the accreted satellites. In the case of gas-rich satellites, the stellar fraction of the baryonic mass would be lower than 100\%, as assumed in this paper and a certain  fraction of the gaseous mass may  be lost by tidal effects/ram pressure stripping before being accreted and converted into new stars. This would naturally lead the accretions to be less ``stellar-rich" that those simulated in the present paper, thus effectively reducing the amount of stars of extra-galactic origin that a satellite of a given mass can bring into the inner halo (i.e. inside 20~kpc) of a Milky Way-type galaxy.
Finally, it is worth emphasizing that we are using idealized simulations of local mergers to quantify the response of a stellar disc to merger events that may have occurred several Gyrs ago, when both the stellar masses of the Milky Way and of its satellite(s) were significantly different from those employed here.  Because the properties of the Milky Way at higher z are still largely unknown (in terms of mass, effective radius, etc ..) and similarly those of the satellite galaxies, idealized simulations of local mergers are currently still one of the only viable ways to quantify the response of a stellar disc to accretion events of a given mass ratio, even for those occurring at higher z (with the limitations previously discussed, like the lack of a gaseous component). Some N-body models similar to those analyzed in this paper have tried to mimic mergers at higher redshift, finding that the evolution and thickening of the disc, and its dependence on orbital parameters, do not depend on the redshift, but only on the mass ratio of the merger \citep[cf, for example, Figs~10 and 14 in][]{villalobos08}.  This is a robust
    evidence of the fact that what matters is the relative mass ratio
    of the interacting galaxies \citep[see also][]{quinn93, bournaud05, bournaud09, qu10}, and the relative
    variation of their parameters, rather than the absolute values. As
    a consequence, at first approximation local mergers can be rescaled
    to mimic also mergers at higher $z$. This - again - can be done
    warning the reader of the limitations of these models, like for
    example the lack of gas that can play an important role in the
    process.

\section{Conclusions}\label{conclusions}

In this paper, we have analyzed high resolution, dissipationless simulations of a Milky Way-type galaxy accreting one or several (up to four) satellites with mass ratio 1:10. These simulations have been complemented by four simulations, at lower spatial resolution, of a Milky Way-type galaxy, accreting several (from four to eight) satellites with mass ratios 1:100, after experiencing one or two mergers with mass ratios 1:10. The novelty of this work with respect to the majority of those already available in the literature, is to analyze fully consistent models, where both the satellite(s) and the Milky Way galaxy are ``live" systems, which can react to the interaction, experience kinematical heating, tidal effects and dynamical friction. We have analyzed this set of simulations to investigate the possibility to make use of kinematics information only to find accreted stars  in the Galaxy, remnants of past accretion events and which have lost their spatial coherence. In particular, we have investigated integrals--of--motion spaces, like the $E-L_z$ and the $L_\perp-L_z$ spaces and kinematic spaces like the $R_{apo}-R_{peri}$  and velocity  spaces, to understand if we can realistically make use of those spaces to search for accreted streams and if this search is really efficient and meaningful.
Our main conclusions are the following. \\
In the $E-L_z$ space:
\begin{itemize}
\item  because the energy and angular momentum of a satellite are not conserved quantities during an interaction, each satellite gives origin to several independent overdensities. This is particularly true for satellites with mass ratios $\sim$ 1:10, which can experience severe dynamical friction during the merger event. In some cases, also satellites with a mass ratio 1:100 can give rise to several, independent, overdensities; 
\item multiple satellites overlap; 
\item in-situ stars affected by the interaction(s) tend to progressively populate a region of lower angular momentum and higher energy that the one initially (i.e. before the interaction) populated; 
\item most of the accreted stars overlap with in-situ stars. This point is valid even if the search is restricted to halo stars only, because in our simulations a substantial part of the inner stellar halo (distances from the galaxy centre less than 10-20~kpc) is made of in-situ stars, originally in the disc and then heated by the interaction(s).
\end{itemize}
In the $L_\perp-L_z$ space:
\begin{itemize}
\item the overlap between the in-situ and accreted population is considerable everywhere. There is no particular region where the contribution of accreted stars appears dominant; 
\item  the distribution is asymmetric with respect to the axis $L_z = 0$, skewed towards positive $L_z$ values (i.e. prograde motions). In particular  in-situ halo stars,  kinematically heated by the interaction(s), preserve part of their initial rotation. In the $L_\perp-L_z$ plane, this rotation reveals itself  as an excess of stars on prograde orbits. The distribution of in-situ halo stars is thus asymmetric in the $L_\perp-L_z$ space, similarly to the distribution of accreted stars. This finding is fundamental: it implies that any observational evidence of an asymmetric distribution in the $L_\perp-L_z$ space skewed towards prograde, inclined orbits is not in itself an indication that this region of the space is dominated by accreted stars. Any accretion event indeed generates an in-situ halo population (made of pre-existing disc stars heated by the interaction), whose distribution, at large values of $L_\perp$, is skewed towards large values of $L_z$.
\end{itemize}
In velocity spaces:
\begin{itemize}
\item for a selected spatial volume around the Sun and for similar values of $L_z$ and $L_\perp$, it is not possible to differentiate in-situ from accreted stars;
\item in particular, the kinematic characteristics observed for stars at the solar vicinity in the ``Helmi region" can be reproduced by accreted stars and by in-situ stars as well. The finding that, under certain conditions, N-body models can reproduce the velocities observed in the ``Helmi region" by means of an accreted stream is not a probe in itself that the stars observed in that region are accreted: once the spatial volume is fixed, all stars in that volume with similar angular momenta will have similar velocities, independent on their accreted or in-situ nature;
\item not only our models suggest that the use of these velocity spaces cannot solve the question of the in-situ or accreted origin of the stars that make them, but also that it is not possible to recover the mass of the progenitor satellite  -- if any -- from them. Our models show that the kinematic properties observed for the Helmi stream can be reproduced with one (or multiple) satellite(s) significantly more massive than that suggested by \citet{helmi99, kepley07} as the progenitor of the stream. The problem of reconstructing the orbital and internal properties of the accreted satellite(s) in the ``Helmi region" by kinematics alone is thus degenerate. 
\end{itemize}
In the apocentre-pericentre-angular momentum space, the APL space:
\begin{itemize}
\item as for all the other spaces investigated in this paper, the overlap of in-situ and accreted stars is substantial also in the $R_{apo}-R_{peri}$ space, at the point that no clear distinction can be made among in-situ and accreted stars on the basis of this space, only. 
\item The global distribution of all the stars in the volume in the apocentre-pericentre space reveals the presence of several streaks, already noticed by \citet{helmi06} in their models. Each of these streaks spans a large range in eccentricities, from $e\le 0.1$ to $e\ge 0.7$. They are visible both in the accreted and in-situ component and -- among in-situ stars -- not only in the thin and thick discs, as already pointed out by \citet{gomez12}, but also in the halo.  
\item all the main characteristics observed among solar vicinity stars in the $R_{apo}-R_{peri}$ space -- namely their diagonal distribution in the  $R_{apo}-R_{peri}$ space, the presence of streakes or overdensities, the trend observed for the angular momentum $L_z$, with $L_z$ increasing with $R_{apo}$ and $R_{peri}$ -- all these features are common both to  the accreted and to the in-situ populations.
\end{itemize}

In agreement with previous works, we find that all these spaces are rich in substructures, but that the origin of these substructures cannot be determined with kinematics. The in-situ stellar halo, formed as a result of the interaction, is neither smooth or non-rotating. As a consequence, an extreme caution must  be employed before interpreting overdensities in any of those spaces as evidence of relics of accreted satellites \citep[see, for example][]{helmi99, gould03, helmi06, kepley07, morrison09, refiorentin15}. \\ We consider this work to constitute a first step towards a more realistic modelling of accreted--in-situ populations in kinematic spaces in view of Gaia and its first data releases. But it is also intended as a cautionary remark about the interpretation of current kinematic data where substructures have been detected and an extra-galactic origin is currently favored  for them: all spaces and regions where stellar streams have been detected  by means of kinematics are indeed -- accordingly to our models -- severely dominated by in-situ stars.  \\ 

On the basis of these simulations, our conclusion is that the kinematic detection of stellar streams in the Galaxy -- streams made of stars accreted long enough ago to have lost their spatial coherence --  is mostly inefficient.  Detailed chemical abundances and/or ages are critically needed and will be definitely necessary to identify accreted populations, that should appear in those spaces as distinct patterns from those described by  in-situ stars.

\section*{Acknowledgments}
This work has been supported by the Ile-de-France Region and the DIM-ACAV, through the grant ``Reconstructing the accretion history of the Milky Way through its globular clusters system" and  by the ANR (Agence Nationale de la Recherche)  through the MOD4Gaia project (ANR-15-CE31-0007, P.I.: P. Di Matteo). The authors also thank the CNRS, for its financial support through the MASTODONS project ``The origin and evolution of our Galaxy: data validation" (P.~I.~: F. Arenou). MM acknowledges support from the CNR short-term mobility (STM) programme, 2015. We warmly thank: P.~E. Nissen and W.~J. Schuster, for providing us with a tabular form of the data used in Sect.~4.5, and for their comments on a first version of this manuscript;    R.~Capuzzo-Dolcetta, B.~Famaey, V.~Hill,  R.~Ibata, D.~Katz, N.~Martin, A.~Mastrobuono-Battisti, M.~D.~Lehnert,  Y.~Revaz, and O.~N. Snaith, for stimulating discussions and remarks. This work was granted access to the HPC resources of TGCC and CINES under the
allocations 2014-040507 and 2016-040507 made by GENCI. Finally, we wish to thank the referees,  for their prompt reports and for their comments which helped improving the clarity of this paper.

\bibliographystyle{aa}
\bibliography{biblio}
\begin{appendix}
\section{The ``isolated" Milky Way}\label{isoMW}

\begin{figure*}
\begin{center}
\includegraphics[clip=true, trim = 50mm 0mm 50mm 0mm, angle=270,width=2.2\columnwidth]{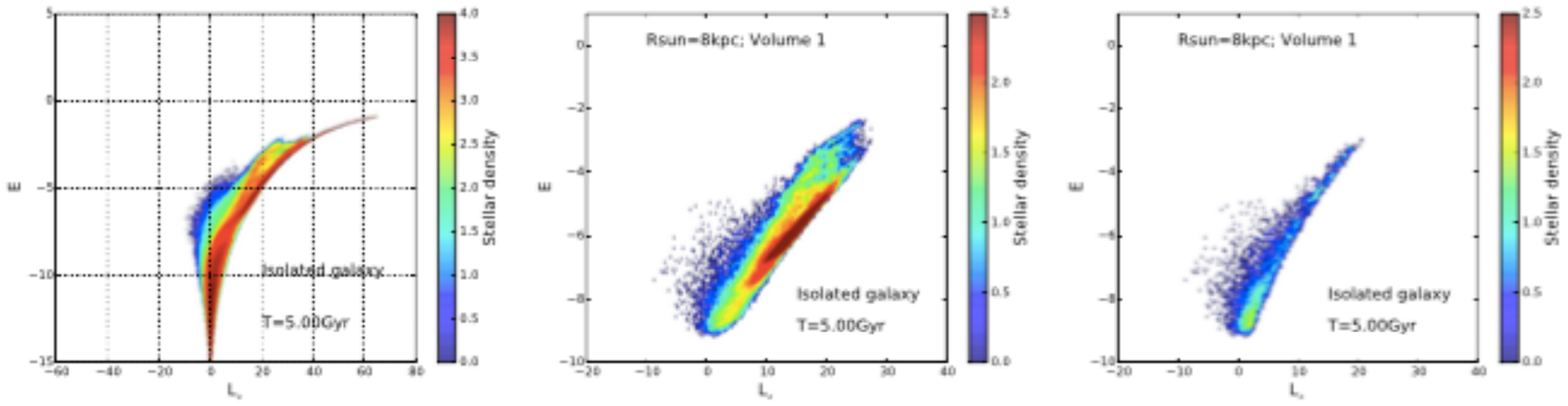}
\caption{Distribution  in the $E-L_z$ space of stars in the isolated galaxy. \emph{Left panel: } all stars in the simulations are shown; \emph{middle panel: } only stars in a ``solar vicinity" volume are shown; \emph{right panel}: only halo stars of the ``solar vicinity" volume shown in the middle panel are plotted. Note that the range of $E$ and $L_z$ values shown in the right panel is not the same adopted in the remaining two panels. }\label{ELziso}
\end{center}
\end{figure*}

\begin{figure*}
\begin{center}
\includegraphics[clip=true, trim = 30mm 0mm 50mm 0mm, angle=270,width=2.\columnwidth]{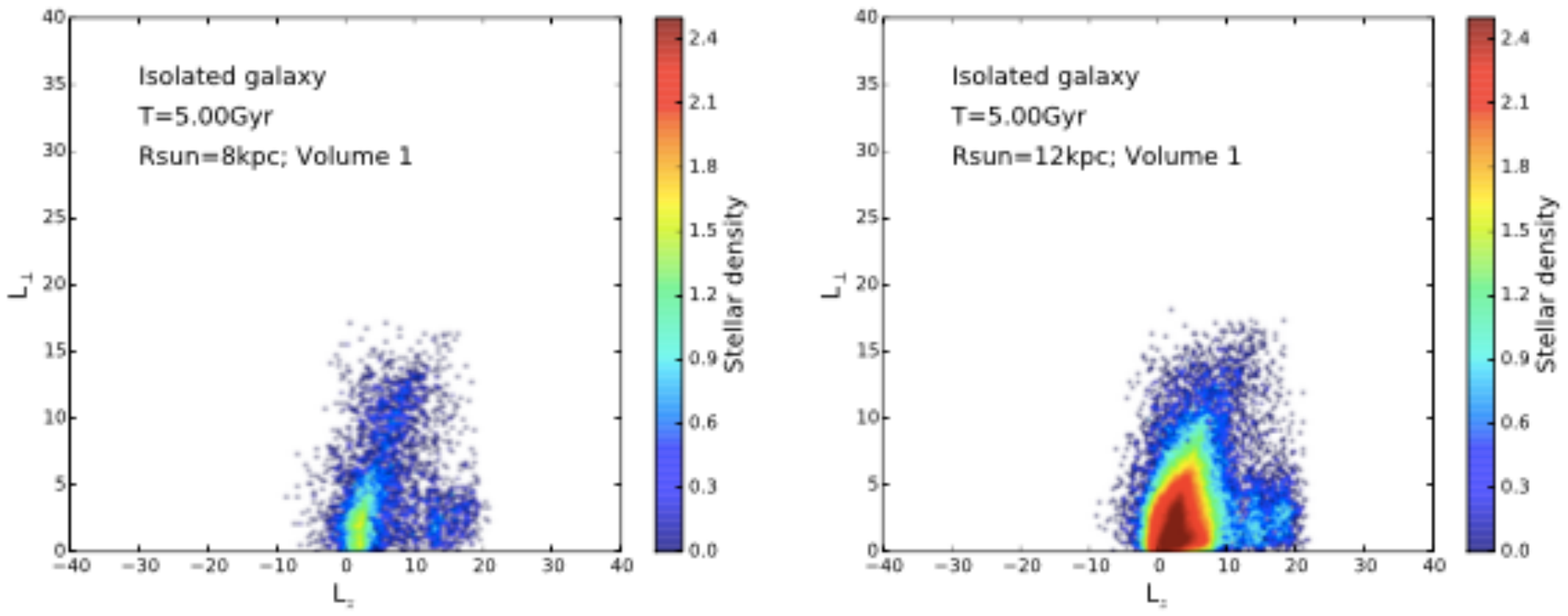}
\caption{Distribution  in the $L_\perp-L_z$ space of stars in the isolated galaxy. \emph{Left panel: } only halo stars in a ``solar vicinity" volume  at R=8~kpc are shown; \emph{right panel}: only halo stars in a ``solar vicinity" volume  at R=12~kpc are shown.}\label{LzLperpiso}
\end{center}
\end{figure*}

\begin{figure*}
\begin{center}
\includegraphics[clip=true, trim = 30mm 0mm 50mm 0mm, angle=270,width=1.7\columnwidth]{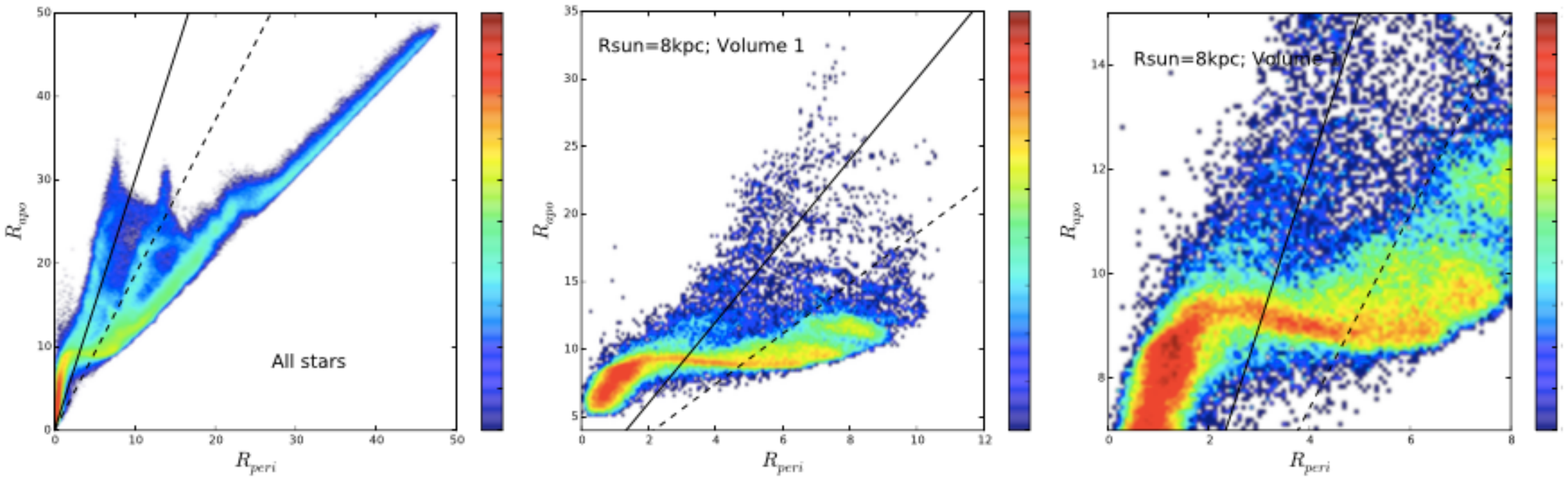}
\caption{Distribution  in the $R_{apo}-R_{peri}$ space of stars of the isolated galaxy. \emph{Left panel: } all stars in the simulations are shown; \emph{middle panel: } only stars in a ``solar vicinity" volume are shown; \emph{right panel}: only  stars of the ``solar vicinity" volume shown in the middle panel and with $0 \le R_{peri} \le 8$~kpc and $0\le R_{apo} \le 15$~ kpc are plotted. Note that the range of $R_{apo}$ and $R_{peri}$ values is not the same in the three panels. }\label{Rapoperiiso}
\end{center}
\end{figure*}

To understand how one or several mergers can impact the distribution of stars in integrals-of-motion and kinematic spaces,  it is interesting to compare with the distribution one would obtain for a Milky Way-type galaxy evolved in isolation. To this aim, we have run a simulation of the Milky Way-type galaxy, evolved in isolation for 5~Gyr and which initially has the same internal parameters as those adopted for the minor merger simulations (see Table~\ref{galparamtable}).  in Figs.~\ref{ELziso}, \ref{LzLperpiso} and \ref{Rapoperiiso} the corresponding distribution of stars  in the $E-L_z$ space, in the $L_\perp-L_z$ space and in the $R_{apo}-R_{peri}$ space are shown.  Comparing these figures with those presented in Sect.~\ref{results}, it is clear that interactions tend to redistribute stars over a much larger portion of all the spaces considered. In the $E-L_z$ plane, halo stars in the isolated galaxy -- which constitute the tail of the thick disc distribution -- are few and tend to redistribute over a thin region in this space.  In the $L_\perp-L_z$ space, halo stars at ``solar vicinity" volumes redistribute over a considerably less extended region of the space, than that occupied in the case of one or several accretion events (cfr Figs.~\ref{LzLperpiso} and \ref{LzLperp1satRsun8part1}). A similar conclusion can be reached for the $R_{apo}-R_{peri}$ space: looking at the distribution of all stellar particles in the simulation, three main structures are observed, corresponding to thin disc, intermediate disc and thick disc stars (see left panel, Fig.~\ref{Rapoperiiso}).  When the analysis is restricted to the solar vicinity volume (middle and right panels, Fig.~\ref{Rapoperiiso}), one sees how thinner the distribution is with respect to those shown in Figs.~\ref{MW1APL} and \ref{MW2APL}. Note in particular the presence of stars belonging to the stellar bar, which show up as an overdensity at high eccentricity in this plane ($R_{peri}< 2$~kpc, $R_{apo}< 10$~kpc).

\end{appendix}

\begin{appendix}
\section{The spatial distribution of clumps in the $E-L_z$ space}\label{spatialclumps}

\begin{figure*}
\begin{center}
\includegraphics[clip=true, trim = 0mm 40mm 20mm 30mm, width=1.8\columnwidth]{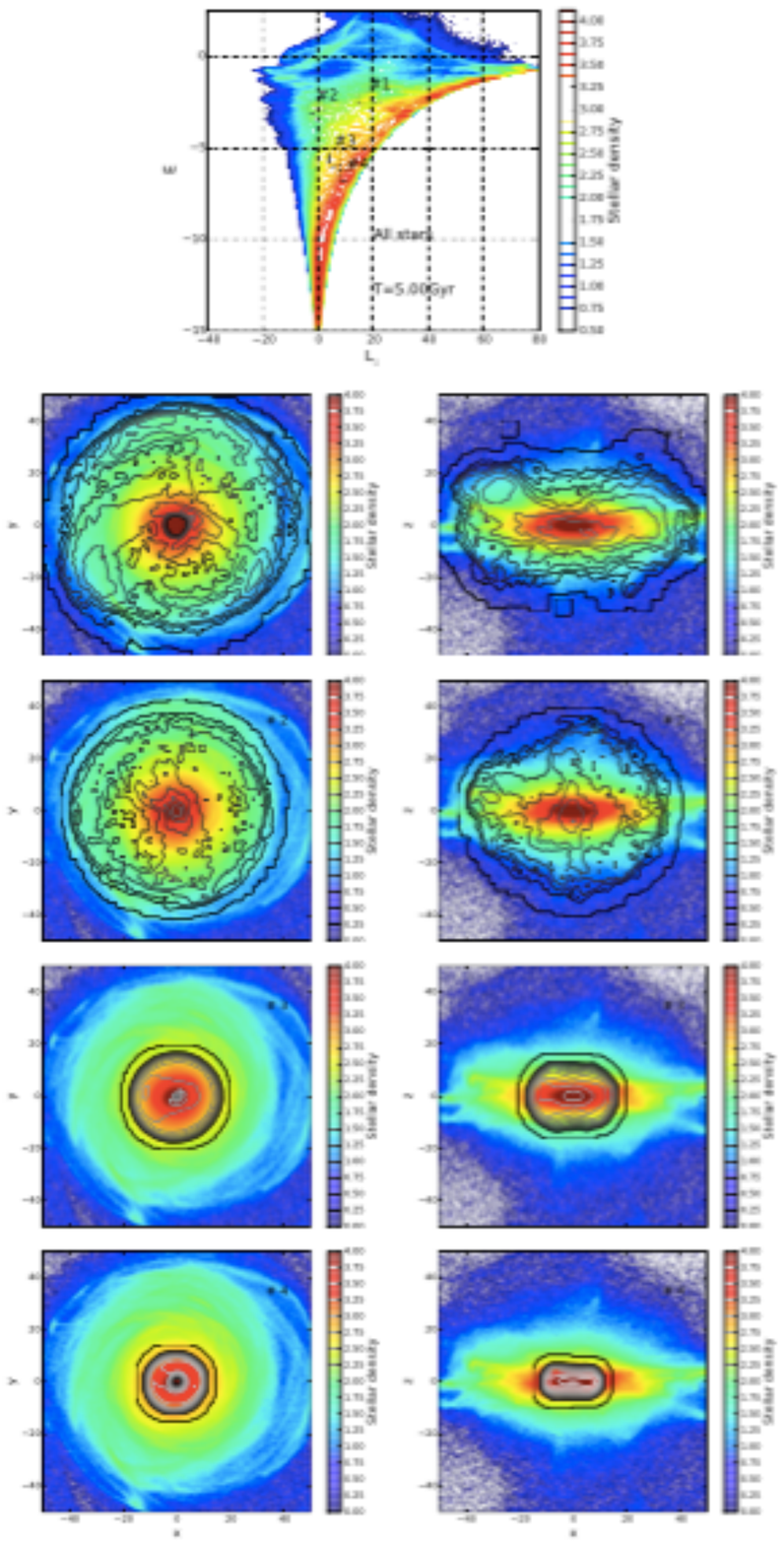}
\caption{\emph{Spatial distribution of some of the stellar clumps visible in the $E-L_z$ space for the 2$\times$(1:10) merger. All stellar particles are included in this plot. Top panel:} distribution in the $E-L_z$ space, with isodensity contours indicated by colored lines. Four clumps have been identified by eye, their are indicated by the symbols \#1, \#2, \#3 and \#4. \emph{From the second to the last row: } Projection on the $x-y$ plane (left column) and on the $x-z$ plane (right column) of stars belonging to each of these clumps, in order of decreasing energy, that is from \#1 (second row), to \#4 (last row). Their distribution is shown by black contours and it is superimposed to the whole distribution of stars.}\label{ALLxyzclumps}
\end{center}
\end{figure*}

\begin{figure*}
\begin{center}
\includegraphics[clip=true, trim = 0mm 40mm 20mm 30mm, width=1.8\columnwidth]{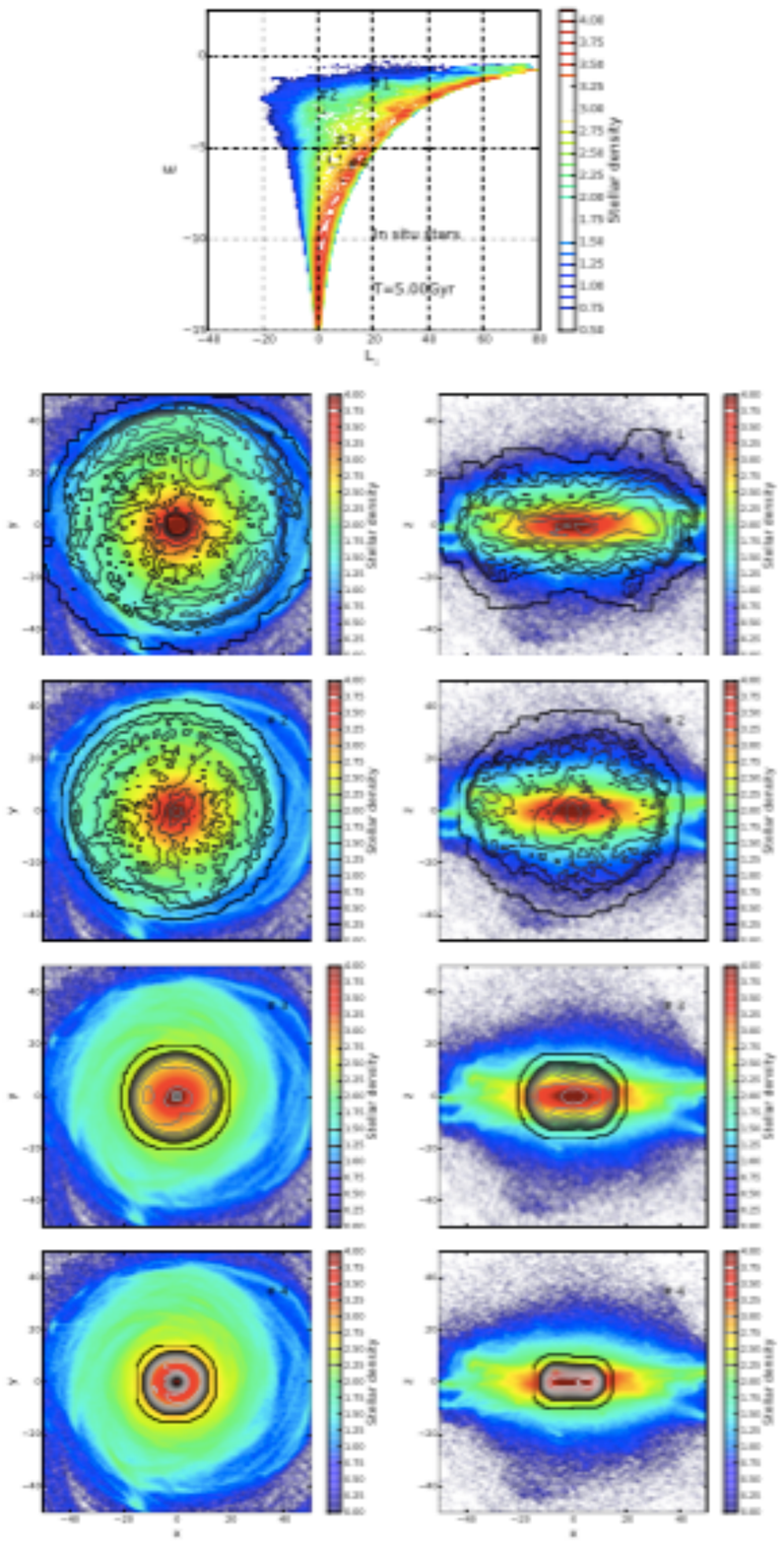}
\caption{Same as Fig.~\ref{ALLxyzclumps}, but for in-situ stars only.}\label{MWxyzclumps}
\end{center}
\end{figure*}

\begin{figure*}
\begin{center}
\includegraphics[clip=true, trim = 0mm 40mm 20mm 30mm, width=1.8\columnwidth]{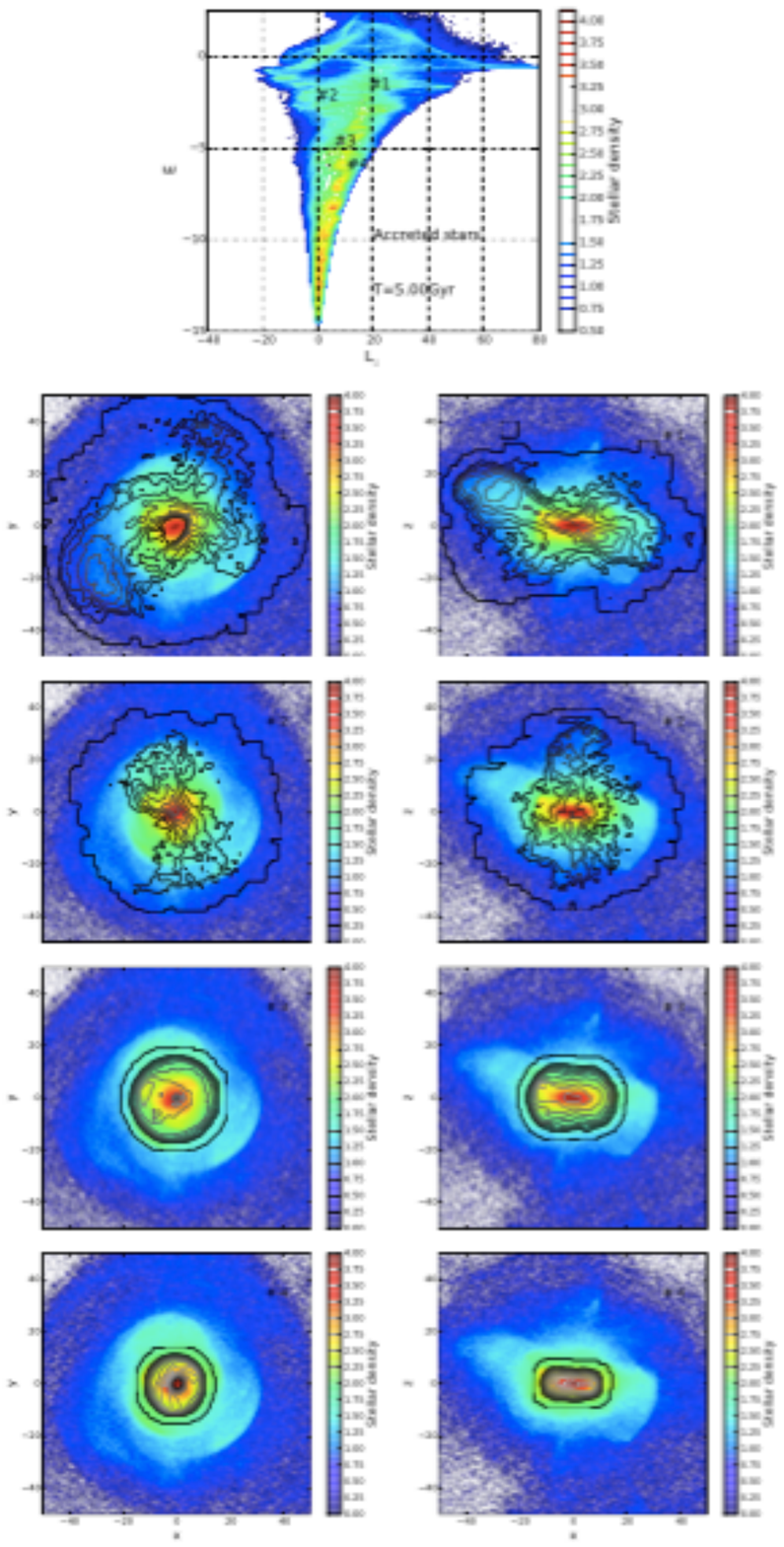}
\caption{Same as Fig.~\ref{ALLxyzclumps}, but for accreted stars only.}\label{allbigsatxyzclumps}
\end{center}
\end{figure*}

In this section we show the spatial distribution of some of the stellar clumps observed in the $E-L_z$ space. To this aim, in the top panels of Figs.~\ref{ALLxyzclumps}, \ref{MWxyzclumps} and \ref{allbigsatxyzclumps} we report the distribution in this space, respectively,  for all stars, in-situ stars only and accreted stars only, in the case of the 2$\times$(1:10) merger. From this distribution we have selected, by eye, four clumps -- which correspond to local maxima in the $E-L_z$ space -- which are indicated with the symbols \#1, \#2, \#3 and \#4 in all the figures. For all stars belonging to each of these clumps, we report in the second, third, fourth and fifth rows of  Figs.~\ref{ALLxyzclumps}, \ref{MWxyzclumps} and \ref{allbigsatxyzclumps} -- in order of decreasing energy -- their spatial distribution in the $x-y$ and in the $x-z$ planes. \\ As expected, for all populations (in-situ, accreted and the two combined), the spatial distribution of stars in a given clump depends on its energy, with stars in the less bound clumps  having a more extended radial and vertical extent. Also, stars belonging to less bound clumps show a more inhomogeneous (and unrelaxed) spatial distribution than stars belonging to clumps of lower energy, independent on their in-situ or accreted origin. Finally we emphasize that both in-situ and accreted stars show a very disturbed spatial distribution in the outer disc: stellar streams,  plumes, are indeed visible in both populations, as it is evident by comparing the $x-z$ distribution in Figs.~\ref{MWxyzclumps} and \ref{allbigsatxyzclumps}.
\end{appendix}

\end{document}